\documentclass[preprint]{aastex}
\usepackage{graphicx}   \usepackage{verbatim}
\graphicspath{{figures/}}
\usepackage{bm}
\newcommand{\Ks}{{\ensuremath K_s}}
\newcommand{\JHKs}{{\ensuremath JH$K_s$}}
\newcommand{\snia}{SN~Ia}
\newcommand{\sneia}{SNe~Ia}

\newcommand{\file}[1]{\textit{#1}}
\newcommand{\code}[1]{\texttt{#1}}

\begin{document}

\title{The First Data Release from SweetSpot: 74 Supernovae in 36 Nights on WIYN+WHIRC}

\author{
Anja~Weyant\altaffilmark{1,2}, W.~M.~Wood-Vasey\altaffilmark{1},
Richard~Joyce\altaffilmark{3},
Lori~Allen\altaffilmark{3},
Peter~Garnavich\altaffilmark{4},
Saurabh~W.~Jha\altaffilmark{5}
Jessica~R.~Kroboth\altaffilmark{1},
Thomas~Matheson\altaffilmark{3},
Kara~A.~Ponder\altaffilmark{1,2}
}

\altaffiltext{1}{
Pittsburgh Particle Physics, Astrophysics, and Cosmology Center (PITT PACC).
Physics and Astronomy Department, University of Pittsburgh,
Pittsburgh, PA 15260, USA
}
\altaffiltext{2}{
Visiting astronomer, Kitt Peak National Observatory, National Optical Astronomy Observatory, which is operated by the Association of Universities for Research in Astronomy (AURA) under a cooperative agreement with the National Science Foundation.
}
\altaffiltext{3}{
National Optical Astronomy Observatory,
950 North Cherry Avenue,
Tucson, AZ 85719, USA
}
\altaffiltext{4}{
Physics Department
University of Notre Dame
Notre Dame, IN, 46556, USA
}
\altaffiltext{5}{
Department of Physics and Astronomy,
Rutgers, the State University of New Jersey,
136 Frelinghuysen Road,
Piscataway, NJ 08854, USA
}

\email{anw19@pitt.edu, wmwv@pitt.edu}


\date{\today}

\newcommand{\numLc}{33}
\newcommand{\numLcPlateau}{25}
\newcommand{\numLcPreBMax}{4}
\newcommand{\numLcPreTwentyDaysBMax}{27}
\newcommand{\numLcPoints}{186}
\newcommand{\avgLcPoints}{5.6}
\newcommand{\medLcPoints}{5.0}
\newcommand{\numLcPointsPlateau}{75}
\newcommand{\numLcPointsPreBMax}{9}
\newcommand{\numLcPointsPreTwentyDaysBMax}{84}
\newcommand{\numLcJ}{31}
\newcommand{\numLcPlateauJ}{22}
\newcommand{\numLcPreBMaxJ}{4}
\newcommand{\numLcPreTwentyDaysBMaxJ}{37}
\newcommand{\numLcPointsJ}{79}
\newcommand{\avgLcPointsJ}{2.5}
\newcommand{\medLcPointsJ}{2.0}
\newcommand{\numLcPointsPlateauJ}{33}
\newcommand{\numLcPointsPreBMaxJ}{4}
\newcommand{\numLcPointsPreTwentyDaysBMaxJ}{37}
\newcommand{\numLcH}{33}
\newcommand{\numLcPlateauH}{25}
\newcommand{\numLcPreBMaxH}{4}
\newcommand{\numLcPreTwentyDaysBMaxH}{43}
\newcommand{\numLcPointsH}{94}
\newcommand{\avgLcPointsH}{2.8}
\newcommand{\medLcPointsH}{3.0}
\newcommand{\numLcPointsPlateauH}{39}
\newcommand{\numLcPointsPreBMaxH}{4}
\newcommand{\numLcPointsPreTwentyDaysBMaxH}{43}
\newcommand{\numLcKs}{5}
\newcommand{\numLcPlateauKs}{2}
\newcommand{\numLcPreBMaxKs}{1}
\newcommand{\numLcPreTwentyDaysBMaxKs}{4}
\newcommand{\numLcPointsKs}{13}
\newcommand{\avgLcPointsKs}{2.6}
\newcommand{\medLcPointsKs}{2.0}
\newcommand{\numLcPointsPlateauKs}{3}
\newcommand{\numLcPointsPreBMaxKs}{1}
\newcommand{\numLcPointsPreTwentyDaysBMaxKs}{4}
\newcommand{\numLcThreeH}{20}
\newcommand{\numLcThreeJ}{12}
\newcommand{\numLcThreeK}{2}
\newcommand{\numLcOneJ}{31}
\newcommand{\numLcOneH}{33}
\newcommand{\numLcOneK}{5}

\newcommand{\numTwomassLc}{786}
\newcommand{\numTwomassLcPoints}{6287}
\newcommand{\avgTwomassLcPoints}{8.0}
\newcommand{\medTwomassLcPoints}{6.0}
\newcommand{\numTwomassLcJ}{705}
\newcommand{\numTwomassLcPointsJ}{2663}
\newcommand{\avgTwomassLcPointsJ}{3.8}
\newcommand{\medTwomassLcPointsJ}{3.0}
\newcommand{\numTwomassLcH}{786}
\newcommand{\numTwomassLcPointsH}{3144}
\newcommand{\avgTwomassLcPointsH}{4.0}
\newcommand{\medTwomassLcPointsH}{3.0}
\newcommand{\numTwomassLcKs}{141}
\newcommand{\numTwomassLcPointsKs}{480}
\newcommand{\avgTwomassLcPointsKs}{3.4}
\newcommand{\medTwomassLcPointsKs}{2.0}
\newcommand{\numTwomassLcThreeH}{541}
\newcommand{\numTwomassLcThreeJ}{425}
\newcommand{\numTwomassLcThreeK}{64}
\newcommand{\numTwomassLcOneJ}{705}
\newcommand{\numTwomassLcOneH}{786}
\newcommand{\numTwomassLcOneK}{141}

\newcommand{\numObjects}{80}
\newcommand{\numObjectsJ}{75}
\newcommand{\numObjectsH}{80}
\newcommand{\numObjectsKs}{15}
\newcommand{\numSnObjectsPlateau}{59}
\newcommand{\numSnObjectsPlateauJ}{55}
\newcommand{\numSnObjectsPlateauH}{59}
\newcommand{\numSnObjectsPlateauKs}{7}
\newcommand{\numSnObjectsPreBMax}{10}
\newcommand{\numSnObjectsPreBMaxJ}{10}
\newcommand{\numSnObjectsPreBMaxH}{10}
\newcommand{\numSnObjectsPreBMaxKs}{3}
\newcommand{\numSnObjectsPreTwentyDaysBMax}{63}
\newcommand{\numSnObjectsPreTwentyDaysBMaxJ}{59}
\newcommand{\numSnObjectsPreTwentyDaysBMaxH}{63}
\newcommand{\numSnObjectsPreTwentyDaysBMaxKs}{7}
\newcommand{\numObjObservations}{710}
\newcommand{\numObjObservationsJ}{291}
\newcommand{\numObjObservationsH}{326}
\newcommand{\numObjObservationsKs}{93}
\newcommand{\avgObjObservations}{8.9}
\newcommand{\avgObjObservationsJ}{3.9}
\newcommand{\avgObjObservationsH}{4.1}
\newcommand{\avgObjObservationsKs}{6.2}
\newcommand{\numSne}{74}
\newcommand{\numSneJ}{69}
\newcommand{\numSneH}{74}
\newcommand{\numSneKs}{9}
\newcommand{\numSnObservations}{493}
\newcommand{\numSnObservationsJ}{219}
\newcommand{\numSnObservationsH}{250}
\newcommand{\numSnObservationsKs}{24}
\newcommand{\avgSnObservations}{6.7}
\newcommand{\avgSnObservationsJ}{3.2}
\newcommand{\avgSnObservationsH}{3.4}
\newcommand{\avgSnObservationsKs}{2.7}
\newcommand{\medSnObservations}{6.0}
\newcommand{\medSnObservationsJ}{3.0}
\newcommand{\medSnObservationsH}{3.0}
\newcommand{\medSnObservationsKs}{2.0}
\newcommand{\numSnObservationsPlateau}{185}
\newcommand{\numSnObservationsPlateauJ}{84}
\newcommand{\numSnObservationsPlateauH}{91}
\newcommand{\numSnObservationsPlateauKs}{10}
\newcommand{\numSnObservationsPreBMax}{27}
\newcommand{\numSnObservationsPreBMaxJ}{12}
\newcommand{\numSnObservationsPreBMaxH}{12}
\newcommand{\numSnObservationsPreBMaxKs}{3}
\newcommand{\numSnObservationsPreTwentyDaysBMax}{216}
\newcommand{\numSnObservationsPreTwentyDaysBMaxJ}{98}
\newcommand{\numSnObservationsPreTwentyDaysBMaxH}{105}
\newcommand{\numSnObservationsPreTwentyDaysBMaxKs}{13}
\newcommand{\numSnThreeH}{57}
\newcommand{\numSnThreeJ}{47}
\newcommand{\numSnThreeK}{3}
\newcommand{\numSnOneJ}{69}
\newcommand{\numSnOneH}{74}
\newcommand{\numSnOneK}{9}

\begin{abstract}

SweetSpot is a three-year National Optical Astronomy Observatory (NOAO) Survey program to observe Type Ia supernovae (\sneia) in the smooth Hubble flow with
the WIYN High-resolution Infrared Camera (WHIRC) on the WIYN 3.5-m telescope.
We here present data from the first half of this survey, covering the 2011B--2013B NOAO semesters, and consisting of
\numSnObservations\ calibrated images of \numSne\ \sneia\ observed in the rest-frame near-infrared (NIR) from $0.02<z<0.09$.
Because many observed supernovae require host-galaxy subtraction from templates taken in later semesters, this release contains only the \numLcPoints\ NIR ($JHK_s$) data points for the \numLc\ \sneia\ that do not require host-galaxy subtraction.
The sample includes \numLcPreBMax\ objects with coverage beginning before the epoch of $B$-band maximum and \numLcPreTwentyDaysBMax\ beginning within 20 days of $B$-band maximum.
We also provide photometric calibration between the WIYN+WHIRC and Two-Micron All Sky Survey (2MASS) systems
along with lightcurves for \numTwomassLc\ 2MASS stars observed alongside the \sneia.

This work is the first in a planned series of three SweetSpot Data Releases.
Future releases will include the full set of images from all 3 years of the survey,
including host-galaxy reference images and updated data processing with host-galaxy reference subtraction.
SweetSpot will provide a well-calibrated sample that will 
help improve our ability to standardize distance measurements to \sneia,
examine the intrinsic optical-NIR colors of \sneia\ at different epochs, 
explore nature of dust in other galaxies,
and act as a stepping stone for more distant, potentially space-based surveys. 

\end{abstract}

\section{Introduction}\label{sec:introduction}

Observations of Type Ia supernovae (\sneia) indicate that
 the Universe is accelerating in its expansion \citep{Riess98, Perlmutter99, Astier06, Riess07, Wood-Vasey07, Kessler09, Conley11, Betoule14, Rest14, Scolnic14, Scolnic17}.
These results imply that a form of energy coined ``dark energy'' permeates the universe driving this acceleration or that the theory of general relativity is invalid on cosmological scales.

Learning more about the nature of dark energy requires the study of both nearby and distant supernovae.  This range is necessary both for the comparison of luminosity distance across a range of redshifts and because studying nearby supernovae offers opportunity for much more detailed and higher signal-to-noise ratio studies of the detailed properties of the supernovae and their host environments.

A well-established systematic affecting derived cosmological parameters from SNe~Ia is reddening and extinction due to dust~\citep[see, e.g.,][]{Jha07,Conley07,WangX06,Goobar08,Hicken09b,Wang09,Folatelli10,Foley11,Chotard11,Scolnic13}.
Reddening resulting from dust is difficult to separate from variations in the intrinsic colors of \sneia~\citep{Mandel16}.
Near-infrared (NIR) studies offer the opportunity to study \sneia\ with less confusion from reddening and extinction due to dust than in the restframe optical passbands.  For supernovae observed in both the optical and NIR, much more can be learned about the nature of dust through the comparison across wavelengths \citep[e.g.,][]{Amanullah15}.

\sneia\ are likely to be superior distance indicators in the $H$ and $K_s$ \citep{Folatelli10,Kattner12, Wood-Vasey08, Barone-Nugent12}.
The intrinsic dispersion in the NIR {\em uncorrected} brightness is comparable to 
that of {\em corrected} brightness in the optical.
This provokes optimism that if relationships similar to the optical width-luminosity and color-luminosity relationships exist in the NIR, then it may be possible to determine corrected NIR brightness to an intrinsic dispersion smaller than in the optical \citep[e.g.,][]{Kattner12, Dhawan15}.
This property was explained theoretically by \citet{Kasen06} whose synthetic lightcurve calculations predict that \sneia\ should be excellent standard candles in the NIR, particularly in $H$.
These benefits of studying \sneia\ in the NIR motivate the use of large-aperture telescopes to overcome the significant background of the night sky in the NIR to further examine the nature of \sneia\ in this promising wavelength regime.

SweetSpot was an approved 3-year 72-night National Optical Astronomy Observatory (NOAO) survey to image $\sim 150$ nearby \sneia\ with redshift $z<0.1$ in the NIR.
This work was carried out at the WIYN 3.5-m Observatory\footnote{At the time of these observations, the WIYN Observatory was a joint facility of the University of Wisconsin-Madison, Indiana University, Yale University, and the National Optical Astronomy Observatory.  \url{http://www.wiyn.org}}
 at Kitt Peak using the WIYN High Resolution Infrared Camera \citep[WHIRC;][]{Meixner10,Smee11}.
The concept of the survey was to build a sample with 3--6 observations per lightcurve in $JH$ with a nearby subset of $\sim 25$ observed in $JHK_s$ out to late phases ($\ge30$~days) with 6--10 observations per lightcurve.
The goals of the program are to standardize the luminosity of \sneia\ by populating the NIR Hubble diagram in a regime less affected by peculiar motions than previous NIR work, quantify any relation between NIR Hubble residual and properties of the host galaxy, improve our understanding of the intrinsic colors of \sneia, and learn more about dust in the host galaxies of \sneia.

In this first data release from the main SweetSpot survey we present lightcurves for \numLc\ \sneia\ from the first half of the survey that are on such low surface-brightness regions of their host galaxies that they did not require host-galaxy subtraction to extract useful lightcurves.
These \sneia\ were observed with a median
of \medSnObservations\ combined observations in $J$ and $H$.
Our second data release will add host-galaxy image subtraction based on final reference images obtained during the second half of the survey.  Explorations of intrinsic color and dust will await the addition of optical data for these supernovae, which is planned for our third data release.
These present NIR lightcurves are sufficient to increase the sample of \sneia\ available to quantify the dispersions in the NIR brightness or \sneia\ and to the explore the relationship between NIR brightness and host-galaxy properties \citep{Ponder17,Ponder18}.

We begin with a description of the telescope and camera followed by a description of the observations and data in Section~\ref{sec:observations}.
The data processing and photometry are discussed in Section~\ref{sec:processing}.
In Section~\ref{sec:calibration} we describe the calibration of our system to the Two Micron All-Sky Survey \citep[2MASS;][]{2MASS}.
Section~\ref{sec:data} presents and discusses the image and catalog properties and the lightcurves for \numLc\ \sneia.
Instructions to access these data are presented in Section~\ref{sec:access}.
We discuss the utility of this present data release and the path to improvements for future data releases in Section~\ref{sec:discussion}.
Section~\ref{sec:conclusion} summarizes the data release and outlines the future data releases from SweetSpot.
\section{Observations}\label{sec:observations}

\subsection{Description of Telescope and Camera}

The WIYN Observatory hosts a 3.5-meter telescope at Kitt Peak National Observatory (KPNO).
The Ritchey-Chretien designed telescope has two Nasmyth ports each of which can house one of five instruments currently available, including WHIRC.

WHIRC is a NIR (0.9--2.5~$\mu$m) imager with a 3.3\arcmin\ field of view and a 0.1\arcsec\ pixel scale.
The instrument features a 2048x2048 Raytheon Virgo HgCdTe detector.
WHIRC is equipped with $JHK_s$ broadband filters and 10 narrow-band filters --- for this work we only utilize the broadband filters.\footnote{\url{http://www.noao.edu/kpno/manuals/whirc/whirc.user.html}}

The WHIRC camera is installed on the WIYN Tip/Tilt module (WTTM) \citep{Claver03},
which can provide improved seeing in good conditions.
In our observations under the SweetSpot program we have found an improvement to the full-width at half-maximum (FWHM) of the image from the tip-tilt correction of 0.1--0.2\arcsec.
To successfully use WTTM requires a suitable guide star ($R\lesssim 15$~mag) that remains in the WTTM field of view while the dither offsets necessary in NIR observations are executed.
For most of the nights in this first data release, WTTM was unavailable due to hardware failures.
We were able to use it on the following nights:\footnote{All dates in this paper are with respect to the local observatory time, MST, at the beginning of the observing night -- even if we didn't start until the second half of the night.} 2011-10-25, 2011-11-15, 2011-11-21, 2012-11-22, 2013-12-09, 2013-12-13.

\subsection{Description of Supernova Sample}

The SweetSpot data presented in this work represent the first half of the 3-year survey plus the original pilot data from \citet{Weyant14}.  Specifically we publish data from the 2011B, 2012A, 2012B, 2013A, and 2013B semesters.
We were awarded 37 effective WIYN nights over 39 distinct calendar nights (there were 4 half nights).
We obtained usable on-sky data during 29.5 effective nights across 36 distinct calendar nights (we accounted usable time in half-night chunks).
There were 7.5 equivalent nights of lost time due to weather and occasional instrument trouble.
Table~\ref{table:nights_awarded} presents a summary of the number of nights per semester and number of usable nights per semester.
Table~\ref{table:nights_observed} summarizes the number of supernovae, number of Persson standard stars~\citep{Persson98}, and whether or not the night was deemed to be photometric.

We selected \sneia\ for follow-up based on IAU Central Bureau for Astronomical Telegrams\footnote{\url{http://www.cbat.eps.harvard.edu/cbet/RecentCBETs.html}} (CBET) and The Astronomers Telegram.\footnote{\url{http://www.astronomerstelegram.org}}
To meet our program goals, the criteria for selection were:
(1) spectroscopically confirmed as Type Ia;
(2) at a phase near $B$-band maximum;
and
(3) in our preferred redshift range of $0.02<z<0.08$.
We used the ``Latest Supernovae'' website\footnote{\url{http://www.rochesterastronomy.org/supernova.html}}
 maintained by David Bishop~\citep{Gal-Yam13} to track available information about our supernovae,
 and generated finding charts using data from IRSA\footnote{\url{http://irsa.ipac.caltech.edu/applications/FinderChart/}}.
Table~\ref{table:snsummary} presents a complete list of the basic information and the discovery and classification sources for the \sneia\ observed in this data release.

\begin{figure}
\plotone{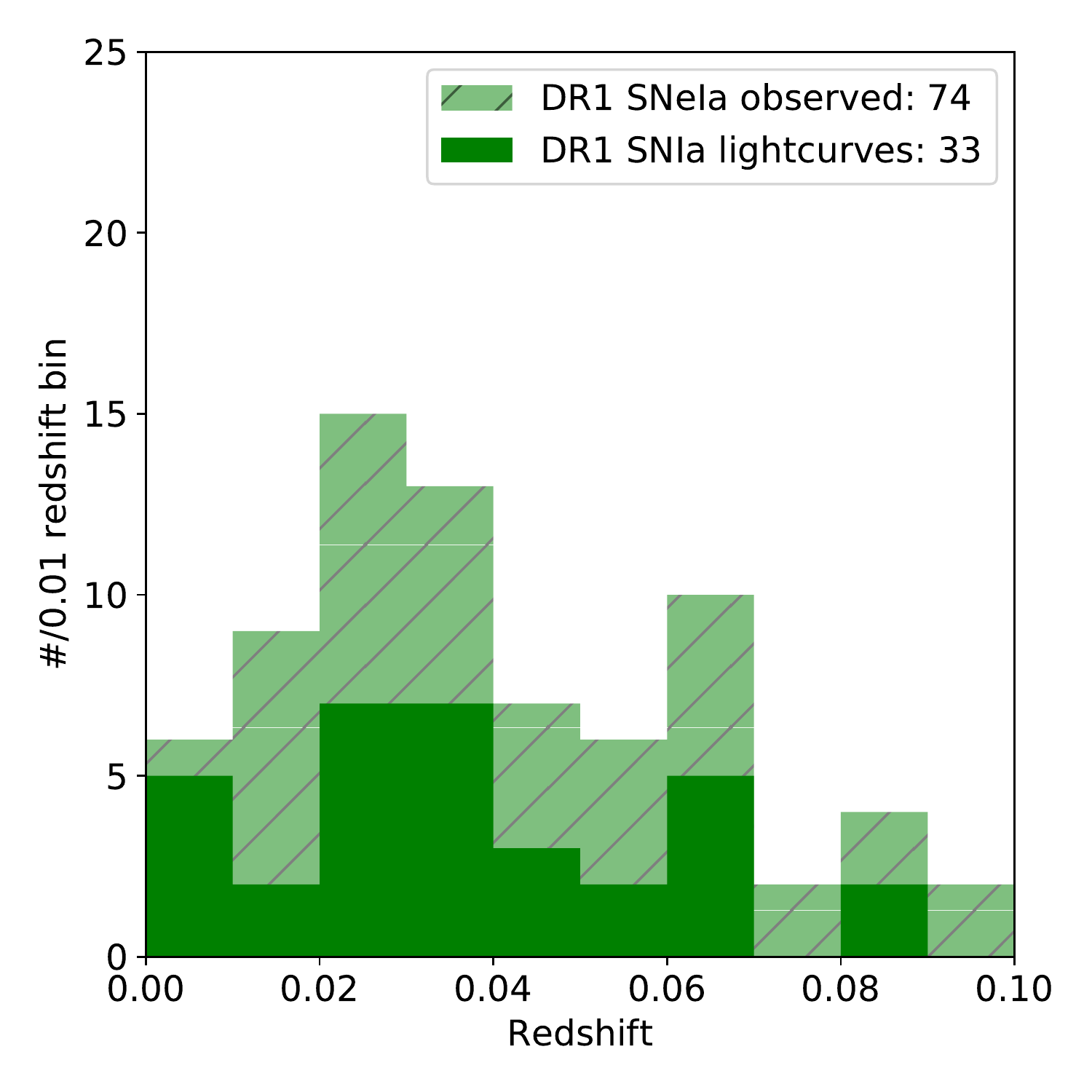}
\caption{Distribution in redshift of the supernova presented here.  The SweetSpot survey is significantly increasing the number of \sneia\ above $z>0.03$ with observations in the NIR where measurements of the distance-redshift relation are less affected by peculiar velocities.
\numSne\ supernovae were observed in DR1 (hatched light green).
\numLc\ have lightcurves in DR1 (solid green).
Green was chosen because all supernovae were observed at least in H-band -- which will be represented by green throughout the rest of this paper.
}
\label{fig:redshiftdistribution}
\end{figure}

Figure~\ref{fig:redshiftdistribution} presents the redshift distribution of the \sneia\ observed in DR1.
The median redshift of the full sample is 0.034.
 At this redshift, a peculiar velocity of 150 km/s is only 1.5\% of the motion from the Hubble flow of $cz=10,200$~km/s.
This is also higher than the median redshift of the current published data set which is 0.019~\citep{Friedman15}.

Our goal was to get the first observation in the lightcurve at least as early as 10--20 days after the time of $B$-maximum, where we expect \sneia\ to be most standard in $H$.
Due to constraints of when WHIRC was mounted on the telescope, a lack of targets at the beginning of the fall season, and the perpetual vagaries of weather, we were not always able to meet this goal.
For the \numSne\ \sneia\ in this data release, (\numSnObjectsPreBMax, \numSnObjectsPreTwentyDaysBMax) have their first observations before (0, 20) days after $B$-band maximum.
For the \numLc\ lightcurves published in this data release, (\numLcPreBMax, \numLcPreTwentyDaysBMax) have their first observations before (0, 20) days after $B$-band maximum.

\begin{figure}
\plotone{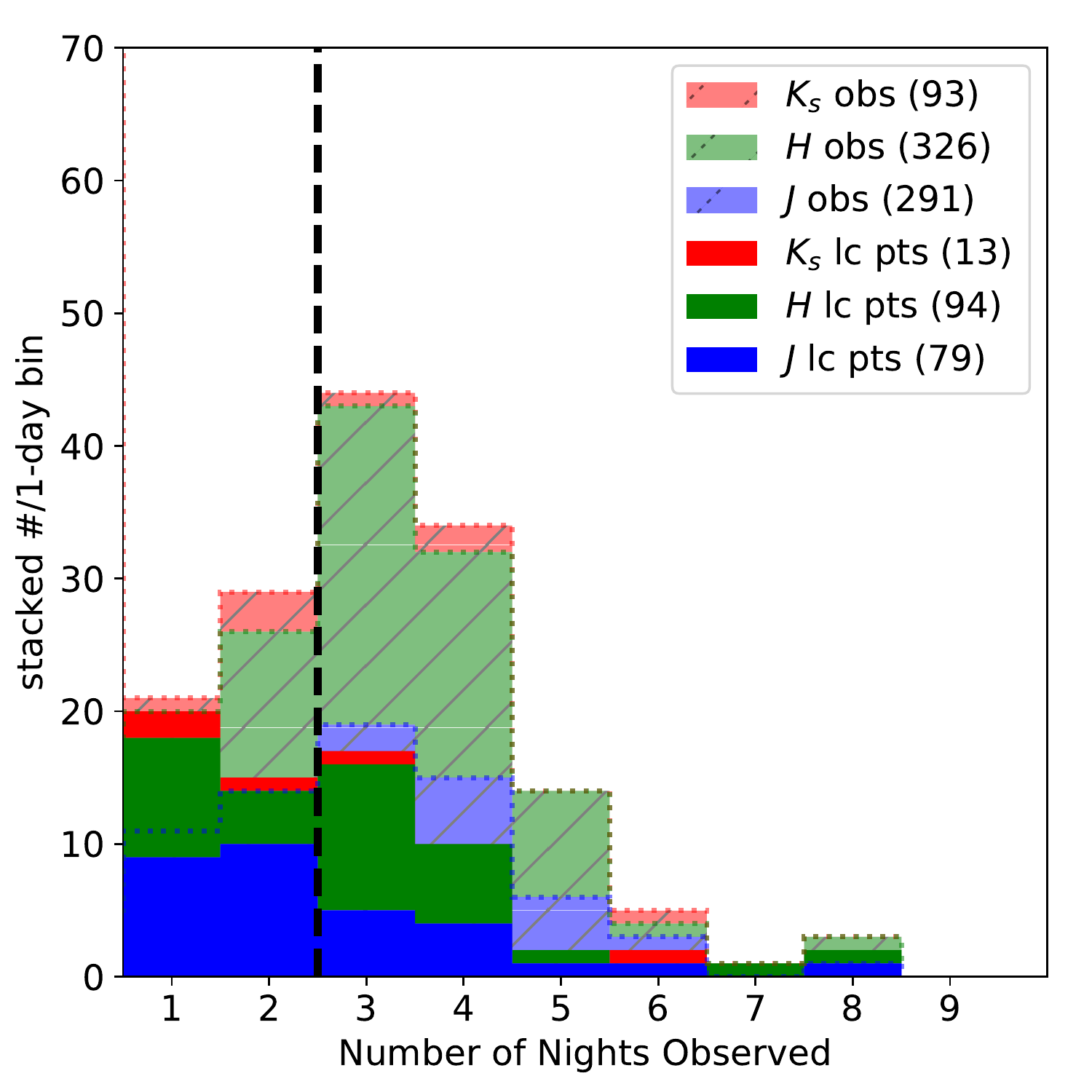}
\caption{Number of observations for each supernovae (dotted) and filter.
Number of points in each lightcurve being presented in this paper (solid).
Stacked histograms for each of $J$ (blue), $H$ (green), and $K_s$ (red).
}
\label{fig:npoints}
\end{figure}

\begin{figure}
\plottwo{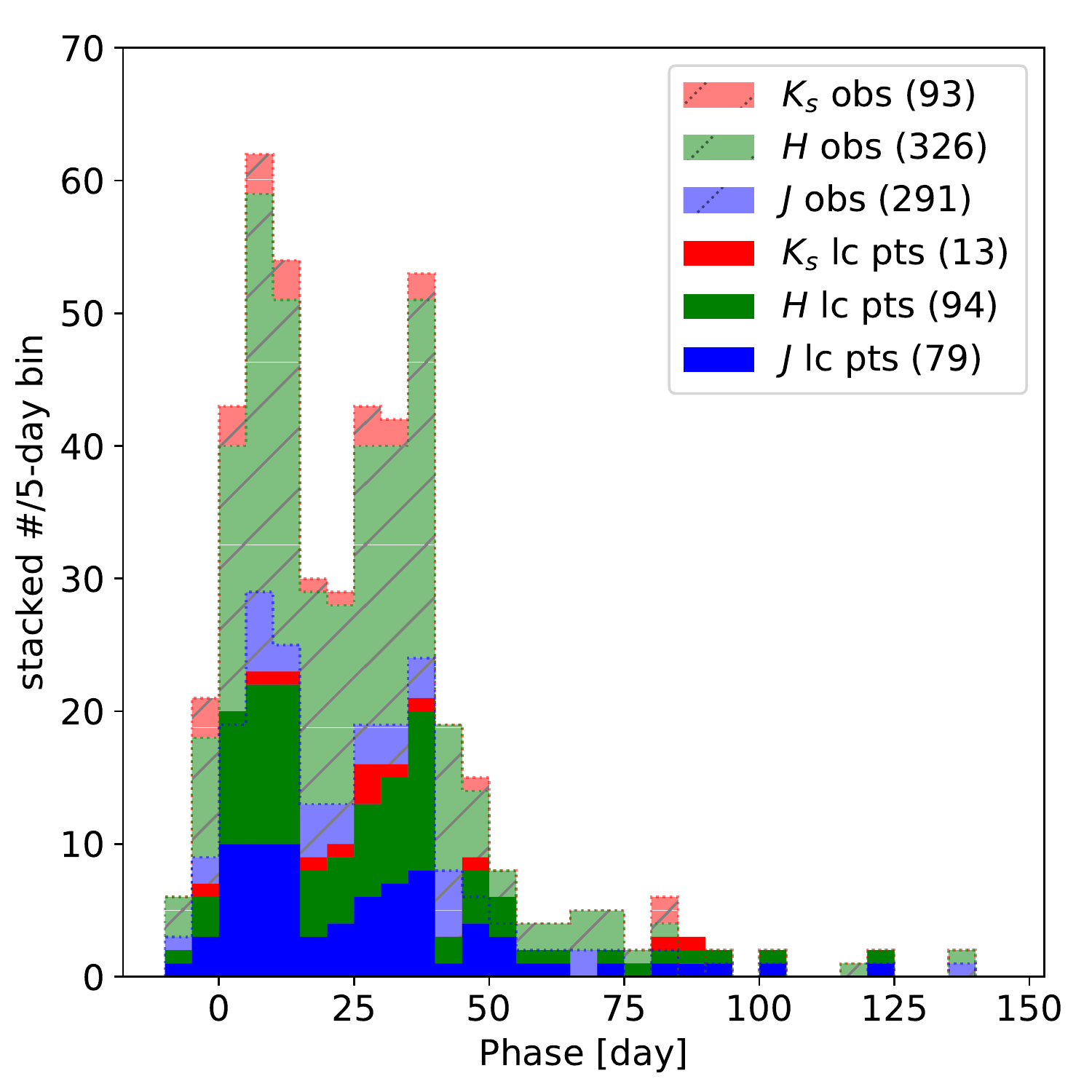}{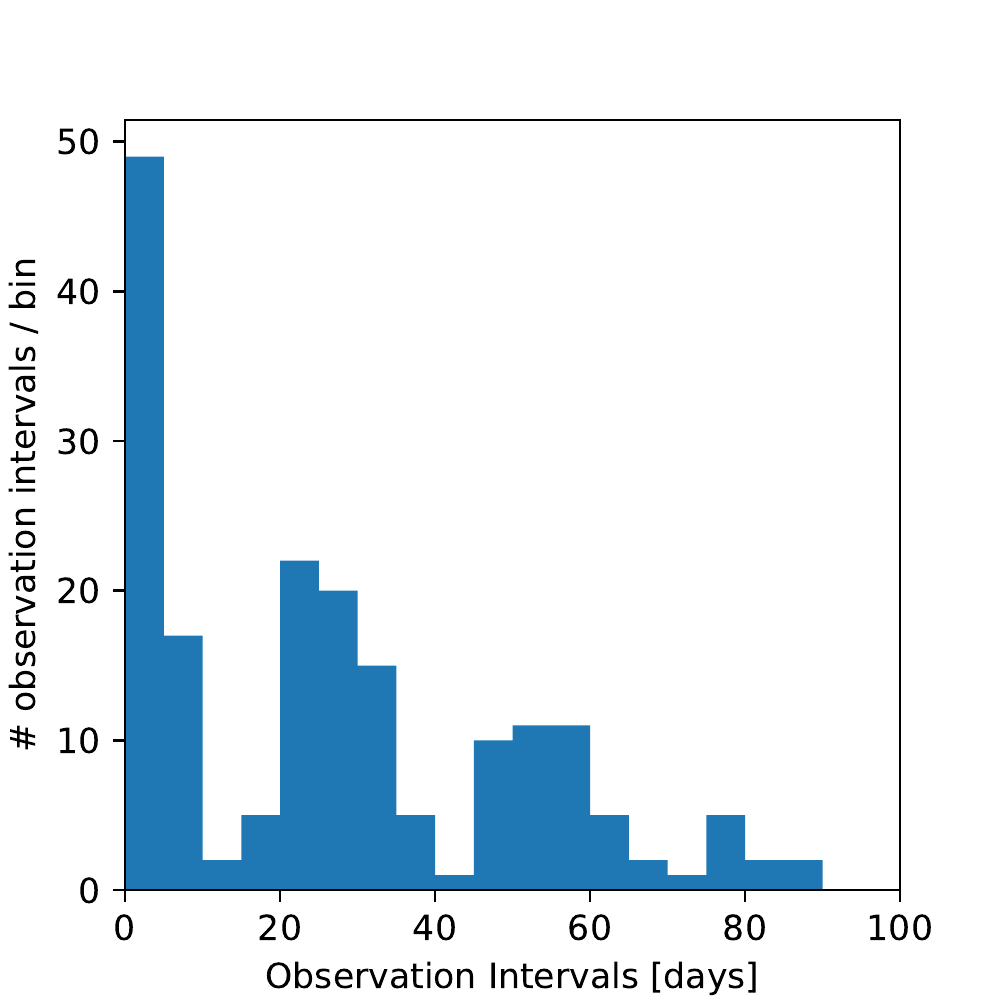}
\caption{(left) Phase of \numSnObservations\ observations (dotted) and \numLcPoints\ lightcurve points (solid) stacked for each of $J$ (blue), $H$ (green), and $K_s$ (red).  The two-peaked structure is primarily due to the scheduling of our observed nights in grey and bright time.
(right) The distribution of intervals between observations -- truncated at 100 days for easy of comparison because we follow very few supernovae past 100 days.
}
\label{fig:nphase}
\end{figure}

Figure~\ref{fig:npoints} shows the distribution of the number of observations of each SN~Ia.
The significantly higher sky background in $K_s$
 limited us to obtaining $K_s$ observations only for \sneia\ at $z<0.03$.
Figure~\ref{fig:nphase} shows the distribution of observed phase for the observations and lightcurve points for \sneia\ in DR1.
This observed \snia\ phase distribution is largely determined by the the phase distribution of our awarded nights.

\subsection{Description of Observations}

\begin{figure}
\gridline{\fig{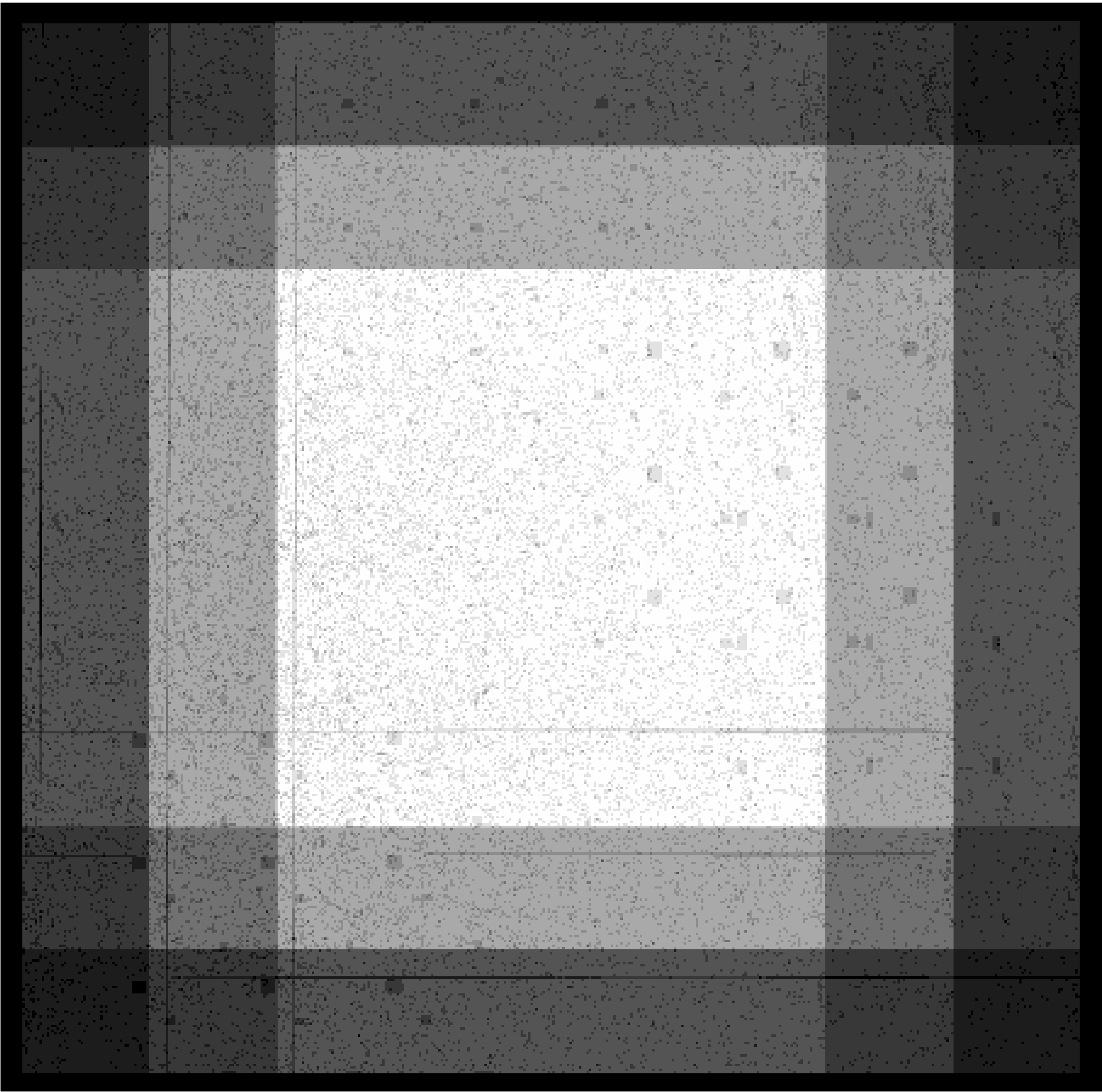}{0.3\textwidth}{(a)}\fig{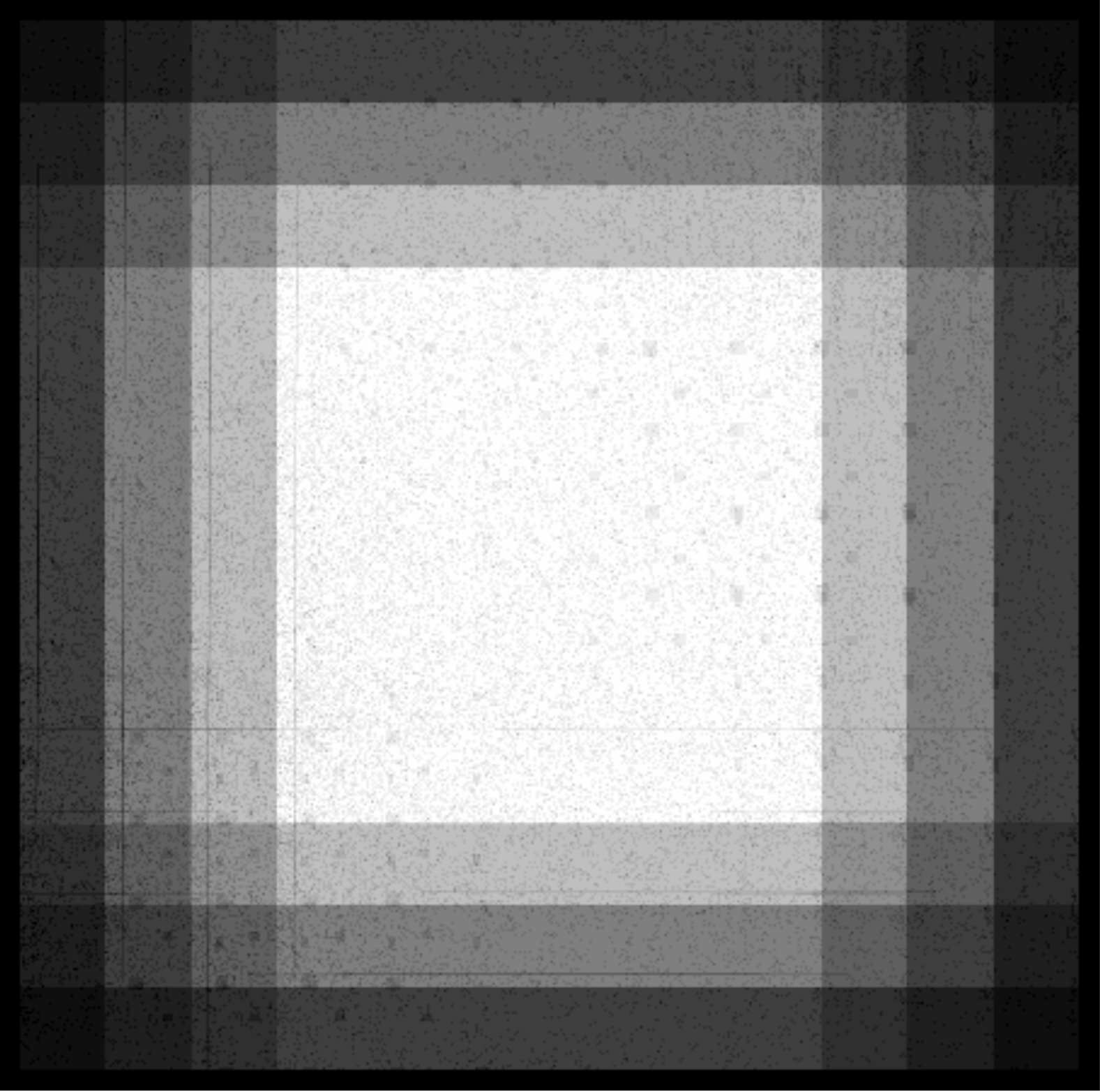}{0.3\textwidth}{(b)}\fig{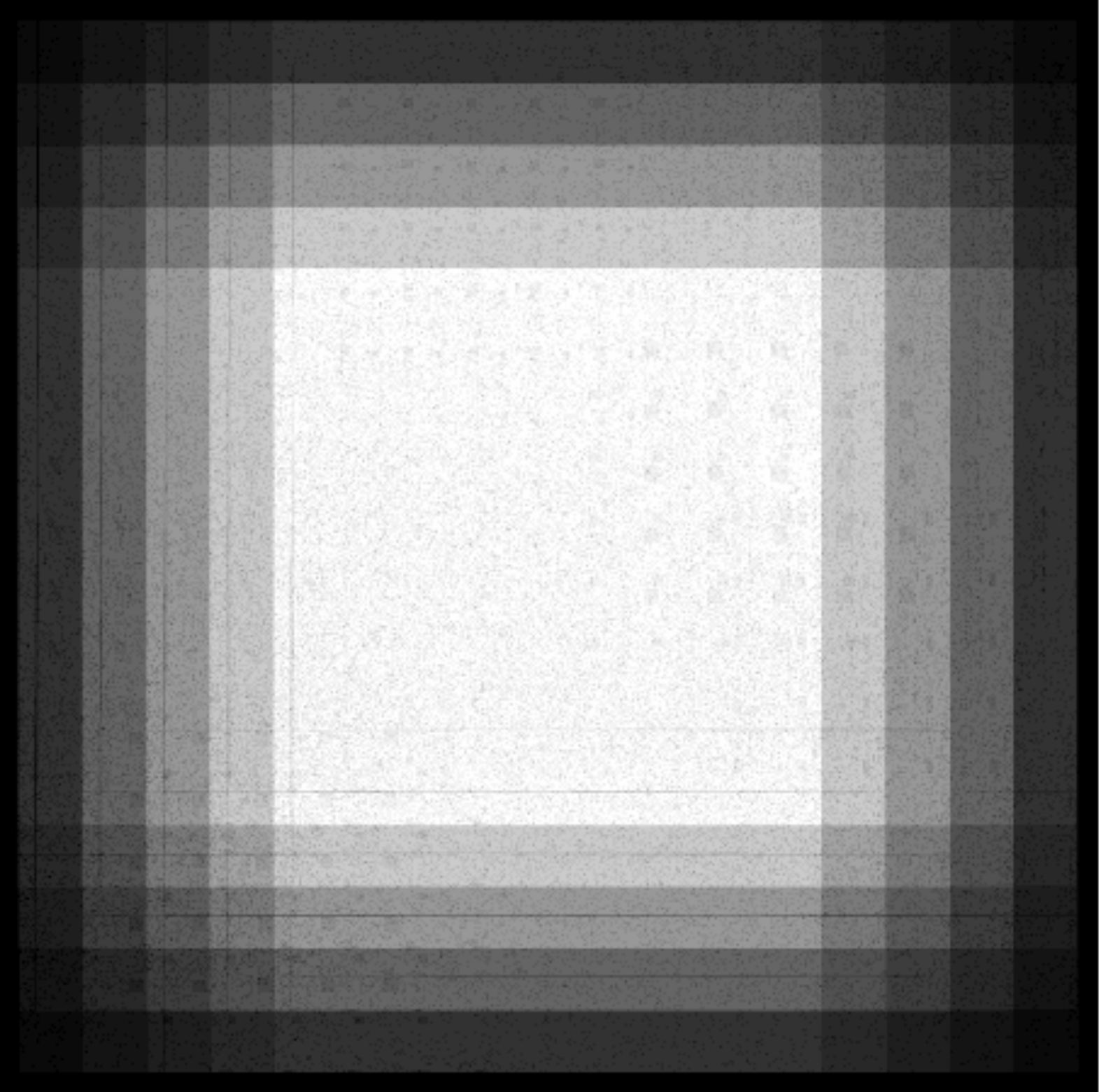}{0.3\textwidth}{(c)}}
\caption{Exposure maps of typical WIYN+WHIRC stacked observation sequences.  Our program employs dither patterns of: (a) 3x3 grid with 30\arcsec\ spacing; (b) 4x4 grid with 20\arcsec\ spacing; and (c) 5x5 grid with 15\arcsec\ spacing.
The displayed steps from black to white in each image above are (0, 1, 2, 3); (0, 1, 2, 3, 4); and (0, 1, 2, 3, 4, 5) exposures.
Exposure times at each individual pointing were 60~s for supernova targets and 10~s for Persson standard stars.  A combination of dither patterns were used to achieve different total exposure times.  E.g., for a 41-minute effective exposure time we would observe a 4x4 dither sequence followed by a 5x5 dither sequence.  The small scale structure in the pattern is from the bad pixel mask.}
\label{fig:weightmap}
\end{figure}

The major difference between optical CCD versus NIR observations is the significant background from both the sky and detector+telescope system.
To separate the effects of spatially and temporally variable sky background, dead and hot pixels, and other instrument and detector artifacts we obtained multiple images of the same field while moving the telescope a small amount between each exposure.
This dithering results in imaging the target at several locations on the array.\footnote{\url{http://www.noao.edu/kpno/manuals/whirc/WHIRC\_Datared\_090824.pdf}}
We employed three main dither patterns: a 3x3 grid pattern with 30\arcsec\ spacing, a 4x4 grid pattern with 20\arcsec\ spacing and a 5x5 grid pattern with 15\arcsec\ spacing. Each dither pattern travels over a 1\arcmin\ square centered about the target coordinates of equal exposure time.
Thus our co-added image stack for an observation typically covers 4\arcmin~$\times$~4\arcmin\ from a set of raw 3.3\arcmin~$\times$~3.3\arcmin\ ``raw'' dithers.
Figure~\ref{fig:weightmap} shows the total exposure time maps characteristic of each dither sequence.
Dither patterns were repeated or different dither patterns were combined to increase the total exposure time.

\clearpage  

We also obtain dome flat calibration images on each night in ($J$, $H$, $K_s$) consisting of eleven 5-s exposures with the ``high'' flat lamp on and eleven with the lamp off.
The lamp settings evolved slightly during the DR1 observing semesters.
In 2011B--2012A, we used settings of (600, 415, 380) respectively for ($J$, $H$, and $K_s$), while in 2012B, 2013A, and most of 2013B we used settings of (650, 450, 380).  These numbers are proportional to the voltage applied to the lamps.
The WHIRC users manual updated the flat lamp information in 2013 October based on the re-coating done in the summer 2013 and we switched to the newly recommend values of (600, 425, 380) on 2013-11-16.
Specifically, these lamp settings are the recommended flat-field lamp settings for the Fowler-1 readout mode (see Table~4 of the WHIRC user manual\footnote{\url{http://www.noao.edu/kpno/manuals/whirc/whirc.user.html}}) and yield 10,000--12,000 counts in the central region of each raw dome flat.
In $J$ and $H$, the lamp-off images have just tens of counts in this region.  In the $K_s$, this increased to $\sim2000$ counts.  The difference between the lamp-on and lamp-off images consistently yielded 10,000 counts for all filters, which at the standard WIYN detector gain of 3.4 e$^{-}$/ADU corresponded to 34,000 photoelectrons.

For a typical \snia\ target observation, we execute a dither pattern with a 60-s exposure at each pointing. Our observations are almost always executed in $JH$ order,
but with priority given to $H$ if there is only time for one filter.  For nearby targets ($z<0.02$) we also observe in $K_s$.
Beginning in the middle of the 2012B semester, we started observing photometric standard stars from \citet{Persson98} in $JHK_s$.  We list the Persson standards we observed in Table~\ref{table:persson_standards}.
For these Persson standard observations, we executed a 3x3x30\arcsec\ dither pattern with a 10-s exposure at each pointing in $JHK_s$.

\section{Processing of the Image Data}\label{sec:processing}

\subsection{Instrument Signature Removal and Co-addition}

The WHIRC images were processed using a
Pyraf\footnote{\url{http://www.stsci.edu/institute/software\_hardware/pyraf/}; v2.0 revision 1795}-based
 analysis program written by Ralf Kotulla.\footnote{\url{http://pubsvn.galev.org}}  
 
In 2012 we forked at revision 180 and made modifications to activate different modules and procedures for object masking and WCS registration.  

For each of the observed nights  in this data release we carried out the following steps:

\begin{enumerate}
\item
{\bf Data Transfer:}
To ensure repeatability and consistency of our analysis, we downloaded the raw images from the NOAO Science Archive server by retrieving images matching (PI, Program, telescope+instrument) = (Wood-Vasey, 2012B-0500, WIYN+WHIRC) for the desired observing date.
We downloaded the data using \code{lftp}\footnote{\url{http://archive.noao.edu/app_ext/contrib/ftp/lftp.readme.txt}} which allowed for parallel transfers of up to 16 simultaneous streams.  Typical download times for a night's 2.1~GB of compressed data were $\sim$10 minutes (3.5~MB/s).

\item
{\bf Uncompression:}
The raw images are stored on and transferred from the NOAO Science Archive as Rice-tile-compressed\footnote{\url{http://fits.gsfc.nasa.gov/registry/tilecompression.html}} FITS\footnote{\url{http://fits.gsfc.nasa.gov/}} files \citep{Wells81,Hanisch01,Pence10}.  We use \code{funpack}\footnote{\url{http://heasarc.nasa.gov/fitsio/fpack/}; Version 1.1.0 (August 2008)} to uncompress these files.  We re-name the files to their original filenames as when observed based on the \code{DTACQNAM} header keyword; this re-naming makes it easier to compare against the observer's log files from the night and to do some name-based filtering.

We also correct \code{IMGTYPE} headers that due to observer error often get written as ``acquire'' or otherwise not as ``object'' for science observations.
In addition there are occasional errors in the data archive headers for \code{DTACQNAM} and \code{DTTITLE} that we rewrite to preserve standard FITS compliance.

To ensure repeatability all corrections that change header information or mark certain types or names of files as bad are stored as version-controlled pre-analysis scripts.

\item
{\bf Nonlinearity Correction:}
We start using the functions in \code{reduce.py} with this step.  We first apply the non-linearity correction for the WHIRC detector using \code{reduce.nonlinearity}.  Applying non-linearity first is appropriate because the non-linearity applies at the raw count values.  The non-linearity correction applied is a 3rd-order polynomial in the original data values (in our Fowler-1 readout mode):
\begin{equation}
{\rm data}_{\rm corrected} =
  a \times {\rm data}_{\rm raw}   +
  b \times {\rm data}_{\rm raw}^2 +
  c \times {\rm data}_{\rm raw}^3
\label{eq:nonlinearity}
\end{equation}
using the correction coefficients listed in the WHIRC data reduction manual $(a, b, c)=(1, 1.29\times 10^{-7}, 2.506\times 10^{-11}$).

\item
{\bf Dome-flat construction:}
We prepare flats based on the dome-lamp observations and a known template for the WIYN+WHIRC pupil\footnote{\url{http://www.noao.edu/kpno/manuals/whirc/whircpupil.fits.gz}} using \code{reduce.makeflats}.
Each afternoon a set of 5-s flats are taken separately in (J,H,Ks) with the WIYN ``high'' lamp on the setting described in Section~\ref{sec:observations}.
A matching set of 5-s flats with the lamps off is taken in the same several-minute sequence.
By subtracting the median co-addition of the lamp-off flats from the median co-addition of the lamp-on flats, we obtained combined images that represent only the variation in effective sensitivity of the system from the flat-field screen through to the detector, and removed the effects of thermal radiation from the optics or telescope.
We take this master dome flat and then remove the effect of the pupil ghost based on the pupil template to create a FITS file that contains the response function for each filter \JHKs.  As part of the flat-generation step, we also generate a bad-pixel mask based on a stored \file{bad\_pixels.reg} region file with additions from the detected hot and cold pixels in the flat, defined as those pixels which differ by 5$\sigma$ from the local median value in a $5\times5$ square kernel.

Cleaning and correcting an image for the sky background is a two-step process.  We first estimate a rough sky image to be able to remove enough sky background from the image to identify astrophysical sources.  We then apply this master sky flat to each individual raw image.  After masking out the astrophysical sources, we re-estimate a master sky image and use this cleaner sky image to perform the final sky correction for the individual images.

\item
{\bf Sky-flat construction:}

The raw image headers are used to generate a \file{marksky.log} file that lists the raw image name, object, and if the file was a sky frame, object frame, or both.  This list is also where we later manually mark bad individual images.

The routine \code{reduce.makeskies} reads this list and sorts the images into lists for each field+filter.
An initial estimate of the sky in each raw frame is made based on the median of the 3-sigma-clipped image.  The raw frame is then divided by this sky background to normalize each image to a fiducial sky value of $1$.
These images are then median-coadded using Pyraf's \code{images.immatch.imcombine} for each field+filter for a night to generate a master sky image for that field+filter.  This co-addition is done in detector space; thus the dithered images median out the astrophysical sources and leave the (presumed constant) sky background.

\item
{\bf Sky correction to all individual frames:}
For each raw frame, the master sky image for each field+filter is applied to each image from the non-linear correction stage.
The master sky image is multiplied by the estimated sky background from the previous step for the given raw frame, and then subtracted from that frame to yield a sky-subtracted image.
This sky-subtracted image is then divided by the response file calculated during the flat fielding step above.
A bad-pixel mask for the particular image is generated based on the template bad-pixel mask.
This processing is done using \code{reduce.subtractsky\_flatfield}

We do not attempt to remove the Newton's ring pattern from Paschen-$\beta$ night sky lines.  Both the intensity and phase of this pattern varies due to the varying sky conditions.  While visible to the eye in some of the data when displayed at a \code{zscale} stretch, the effect on the photometry is minimal ($<0.001$~mag) as the contribution of the ring pattern is small in relation to the overall sky variation.

\item
{\bf Refine sky subtraction:}
The flattened images are processed using
Source Extractor~\citep{Bertin96} in \code{SEGMENTATION} mode to identify objects and to generate a mask file with every pixel that is ascribed to an object masked.
The segmentation mask is grown around the marked bad images updated by applying a 1-pixel Gaussian kernel to the segment mask (where masked pixels have a value of 1) and masking all pixels in the convolved mask that have a value greater than 0.005.
For an isolated bad pixel, this masks the 21 pixels in a 5-pixel diameter discretized circle around the bad pixel.
Using \code{reduce.refine\_skysubtraction}, the background is re-estimated from this masked image, and a new set of normalized sky frames are created.
This set is median combined into a new sky frame for each field+filter set.

The final flat-fielded raw image is generated using this new sky frame in the same sky subtraction and flat-fielding process as above.

\item
{\bf Apply geometric distortion correction:}
The known geometric distortion of WIYN+WHIRC is corrected using \code{reduce.geotran} to run \code{iraf.images.immatch.geotran} with a bicubic spline interpolation to generate undistorted individual images.
While not the theoretically optimal solution for maintaining information and minimizing covariance when coadding images, it is convenient to apply this transformation before shifting and combining the images for stacking.

The distortion solutions were based on the WHRIC online data reduction manual.\footnote{\url{http://www.noao.edu/kpno/manuals/whirc/datared.html}}
 We used the files prepared by NOAO\footnote{\url{http://www.noao.edu/kpno/manuals/whirc/whirc.distort.j.db}, {\url{http://www.noao.edu/kpno/manuals/whirc/whirc.distort.h.db}}, \url{http://www.noao.edu/kpno/manuals/whirc/whirc.distort.k.db}}
 as of 2009-03-05.

\item
{\bf Alignment of raw images to a common reference:}
The relative alignment of the set of images from a given observing sequence are now aligned to the middle image of the dither sequence.
The dither patterns used in this program are constructed so that the image in the middle of the sequence is also the image in the middle of the spatial pattern.
\code{Source Extractor} is run on each raw image and the alignment is calculated by matching the $(x,y)$ positions between the catalog for each raw image and the catalog of the central image.
This offset is recorded by updating the WCS information in each raw image (stored as a \code{CD} matrix) to be that of the central raw image plus the calculated transformation.
This is not necessarily a correct absolute WCS, merely a way of storing the transformation to the reference image.
The images are assumed to have no change in rotation or plate scale between them.
The specific $(x,y)$ offsets and uncertainties (in pixels) are also stored in keyword headers \code{ALIDX}, \code{ALIDXERR}, \code{ALIDY}, and \code{ALIDYERR}. The number of stars used to align the raw to the central reference raw is stored in \code{ALINSTAR}.
The images are written out to disk with header values noting the relative shifts, but they have not been actually moved.

\item
{\bf Stacking: combination of raw images into coadded stack:}
First the images are graded on the alignment offset uncertainties (ALIDXERR, ALIDYERR) from the previous step.  Any image with an uncertainty in the $x$ or $y$ offset of more than 10~pixels is rejected from being incorporated in the stack.  This quality cut rejected 11\% (1,504 out of 13,339) of the raw images.

The actual stacking begins with using Pyraf's \code{iraf.images.immatch.wregister} to shift the raw images, with no interpolation, to a common pixel grid based on the results of the previous step that calculated the alignment.

The shifted images are then median-combined using Pyraf's \code{iraf.images.immatch.imcombine} and an exposure map is generated to record the different effective exposure times of each pixel.  These exposure maps provide the information to properly track the pixel-by-pixel variance of the co-added stacks.

\item
{\bf WCS and preliminary photometric calibration:}
To determine the WCS calibration and estimated photometric calibration for these co-added stacks, \code{reduce.calibrate} runs \code{Source Extractor} on the co-added stacks.  The resulting catalog is compared against the 2MASS catalog to generate the astrometric calibration and to provide an estimate of the photometric calibration.  While the photometric calibration from this step is not used in the final calibration (see Section~\ref{sec:calibration}), a failure to generate a reasonable rough photometric calibration in this step is a used as an indicator that future photometric calibration will not be successful.

\end{enumerate}

In \citet{Weyant14} we used scripted but manually-run IRAF-based analysis following the steps outlined in the WHIRC Data Reduction manual.  These two approaches deal with the same basic issues of image reduction, but differ in some of the specifics.  The image reductions we present here supersede those of \citet{Weyant14} as do the photometric catalogs and lightcurves described in the following sections.

\subsection{Photometry}

We use aperture photometry on the stacked images to measure the instrumental magnitudes
of the \snia\ and stars.
This photometry relies on an accurate WCS solution to locate the 2MASS stars and the \snia\ in the field.
While all successfully generated stacks are provided as part of this data release,
we applied the following additional quality cuts to those used in photometric lightcurves.
First, a visual inspection was performed to determine if the stacked images were combined properly and to determine if the WCS solution is reasonable.
The inspection of each stacked image is detailed in Table~\ref{table:stackinspection}.

The orientation and scale of the detector are well known.  Thus, for the 66 images with visually good stacks but which failed the automatic WCS determination, manually-generated WCS solutions were generated by identification of a reference star x, y and RA, Dec.  Through this process we noted that the two images of SN~2012gm from 2011-11-22 are offset 4\arcmin\ from the intended pointing and thus do not include the supernova.
Any stack with a \code{ZP\_ERROR} outside the range of $(0,1]$~mag  
was rejected from further photometric processing.

These co-added images with bad astrometric or photometric solutions are included in this data release for completeness, but no corresponding catalogs or supernova lightcurve points are generated from these problematic co-additions.

We perform aperture photometry on stars in the field which are listed in the 2MASS point-source catalog
as downloaded from the VizieR catalog service\footnote{\url{http://vizier.u-strasbg.fr/}}.
We require that the 2MASS catalog entries for these stars
have a SNR$>5$ in each of $JHK_s$ and that the 2MASS catalog flag values indicate that the star is not blended or affected by known artifacts, e.g., contamination from a bright nearby star or extended source, diffraction spike, etc.
This is our operational definition of 2MASS ``star'' for the purposes of this present section.

Our median effective FWHM in our WIYN images is 0.7\arcsec\ and we are able to resolve some 2MASS ``stars'' as being extended or multiple objects (the 2MASS typical FWHM was 2.5--3\arcsec).  We perform an additional level of inspection of these objects when defining the standard stars that define our WIYN+WHIRC calibrators for each field --- see Section~\ref{sec:refstars} for details.
However, we provide aperture photometry for the full set of 2MASS ``stars'' in our fields.

We observed 6 Persson standard stars as part of our observations (Table~\ref{table:persson_standards}).  We updated the positions listed in the Persson catalog for these stars to the (RA, Dec) listed in the 2MASS catalog.

We updated the positions of the 2MASS stars (including the Persson standards) using the Goddard Space Flight Center (GSFC) IDL\footnote{IDL is a product of Harris Geospatial Solutions (formerly Exelis; ITT Visual Information Solutions; and Research Systems Inc.):  \url{http://www.harrisgeospatial.com/ProductsandSolutions/GeospatialProducts/IDL.aspx}}
Astronomy User's Library routine \code{gcntrd}.\footnote{W. Landsman, 2004--2009.  GSFC IDL Astron library.  \url{http://idlastro.gsfc.nasa.gov/}}
This re-centering procedure accounts for
 (a) uncertainties in the 2MASS positions that are better determined in our WHIRC images;
 (b) astrometric imprecision in our WHIRC image WCS solutions to the 2MASS frame (this imprecision comes from both uncertainties in position, and from the limited number of stars in the smaller WHIRC field of view in the sparser fields); 
 and 
 (c) proper motions of stars in the field.
Because we were using this re-centroiding to account for multiple factors, we adapted the \code{gcntrd} routine to allow for larger pixel shifts to accommodate the high-resolution, oversampled WHIRC images.
We did not use this re-centering for the SN positions as we did not want to be susceptible to variability in position as the SN fades.  The 2MASS stars in our WIYN+WHIRC images are all bright enough that they have plenty of S/N to spare for re-centering.

The location of the \snia\ was originally taken from the reported position of the \snia\ from its original ATEL or CBET (see Table~\ref{table:snsummary}).
This position was then translated to a region file and visually inspected in \code{ds9}.\footnote{\url{http://ds9.si.edu}}
If the position was discrepant from the location of the SN in our WIYN+WHIRC postage stamps, it was updated based on visual selection of the (RA, Dec) position of the \snia\ in the highest SNR image available (either in $J$ or $H$).
Many (48/\numSne) of the SN positions were shifted in this way by $<1$\arcsec.  We attribute these shifts to a combination of the uncertainties in real-time detection systems, and the often larger effective seeing of SN search programs (1--2\arcsec) particularly compared to the often excellent seeing at WIYN+WHIRC (median 0.7\arcsec).
The list of 2MASS (RA, Dec) and the (updated) position of the SN~Ia
were then used as the centers for aperture photometry.
It is these updated positions that we report as the SN in Table~\ref{table:snsummary}.

We measure the FWHM from the image by fitting a 2D Gaussian to each 2MASS star in the image using Eric Deutsch's IDL routine \code{starfit.pro}.\footnote{\url{http://www.astro.washington.edu/docs/idl/htmlhelp/library03.html\#STARFIT}}
 We then choose the median FWHM from this set of stars as the best estimate for the FWHM of the image.
We performed aperture photometry at an array of radii --- [2, 3, 4, 5, 6, 7, 8, 9, 10, 12, 15, 20, 25, 30, 40] pixels --- for later uses in computing aperture corrections and selecting the optimal S/N radius for the final quoted photometry.
The background was estimated from a sky annulus from 41--50 pixels (i.e., 0.1\arcsec--1.0\arcsec ~beyond the maximum adopted aperture size).

We used these aperture and sky annulus values to measure the detected counts of objects in the field using the GSFC IDL routine \code{aper}.

The stacks are computed in terms of the integration time of the raw dither in the sequence.
This exposure time is generally 60~s for SN images and 10~s for Persson standard star fields.
We record the instrumental magnitudes in standardized counts-per-second convention.

\begin{equation}
m_{{\rm inst},f}=-2.5\log_{10}{\frac{\rm ADU}{{\rm exptime}_{\rm raw}}}
\label{eq:instrumental_magnitude}
\end{equation}
\section{Calibration}\label{sec:calibration}

We here describe the analysis part of our processing which starts with the raw instrumental photometric catalogs derived in the previous section.  We have split these Sections~\ref{sec:processing} and \ref{sec:calibration} thusly, as all of the steps in this section can be reproduced from the photometric catalogs without need to refer to the imaging data.

\subsection{Definition of WHIRC Natural System}\label{sec:whirc_natural_system}

We base our WIYN+WHIRC natural system on the 2MASS system at a given set of characteristic colors.  We then calculate fit coefficients for color terms and airmass terms.  The color term is a description of the difference between the WIYN+WHIRC and 2MASS systems, while the airmass term is a property of the atmosphere above KPNO.

Specifically, we use the following standard transformation between the measured instrumental magnitudes and the 2MASS system as a function of the airmass, $X$, and 2MASS color, $\Delta_{m_1, m_2}^{\rm 2MASS}$, using the following equation
\begin{equation}
m^{\rm 2MASS}_f = m_{{\rm inst,}f}^{\rm WHIRC} + {\rm zpt}_f - k_f \left(X-1\right) + c_f \left(\Delta_{m_1,m_2}^{\rm 2MASS} - \Delta_{\rm ref}^{\rm 2MASS}\right)
\label{eq:transform}
\end{equation}
where $f$ designates the filter;
${\rm zpt}$ is the per-second zeropoint of the WIYN+WHIRC system in filter $f$;
$k_f$ is the airmass coefficient\footnote{Our chosen sign convention means that $k$ should be positive.  The opposite convention is also common in the literature, including in \citet{Weyant14}.} for filter $f$;
and $c_f$ is the color coefficient.

We define the reference point for translations between the WIYN+WHIRC natural system and the 2MASS system at these reference 2MASS colors: $\Delta_{\rm ref}^{\rm 2MASS} = m_J^{\rm 2MASS}-m_H^{\rm 2MASS}=0.5$~mag  for $J$ and $H$ and $\Delta_{\rm ref}^{\rm 2MASS} =m_H^{\rm 2MASS} - m_{K_s}^{\rm 2MASS} = 0.5$~mag for $K_s$.
These reference colors are selected as being characteristic of the stellar populations observed.  They also conveniently roughly describe the NIR colors of a \sneia\ after maximum light.
We obtain $m^{\rm 2MASS}_f$ and $\Delta_{m_1,m_2}^{\rm 2MASS}$ from the 2MASS Point-Source Catalog~\citep{Cutri11}.
We determine $X$ from the metadata for the observation, and measure $m_{{\rm inst,}f}^{\rm WHIRC}$ from our WIYN+WHIRC images (as described in Section~\ref{sec:processing}).
Our calibration process then becomes fitting for the calibration coefficients: ${\rm zpt}_f $, $k_f$, and $c_f$.

We base our absolute photometric calibration using 2MASS stars observed on photometric nights.  Photometric nights were defined as those in which we had no contamination from clouds as judged by the observer.  Table~\ref{table:nights_observed} lists those nights along with the number of Persson star observations each night.

We calibrate using 2MASS stars in the magnitude ranges of $10<H<15.5$~mag, $10<J<15.8$~mag, and $10<K_s<14.3$~mag.  The bright limit avoids stars that will likely be saturated in our WIYN+WHIRC images, while the faint limit loosely corresponds to requiring a SNR $>10$ for the 2MASS measurements of the stars.
To minimize sensitivity to variations in the PSF across different images, we calculate an aperture correction to standardize the effective aperture at 40 pixels (4\arcsec).

\subsection{WTTM Refurbishment}\label{sec:wttm_refurbishment}
In 2013 February, the WIYN engineering crew removed the WTTM module from the telescope.  The WTTM instrument and the 4 mirrors were inspected.  The surface of the tip/tilt mirror was found to be significantly degraded and the other three mirrors (M1, M2, and M3) were dusty.
All the WTTM mirrors were cleaned with CO$_2$ during this 2013 February servicing.
Motivated by this inspection, and because WHIRC is the only instrument now mounted on WTTM, NOAO decided to re-coat the WTTM mirror with gold for improved NIR performance (it previously had a silver-based coating more appropriate for its previous use with the Mini-Mosaic Imager\footnote{\url{https://www.noao.edu/kpno/manuals/minimo/minimo.html}}).
Unfortunately, the M2 mirror was destroyed in shipment and the M1 and M3 mirrors suffered minor damage.  The tip/tilt mirror was undamaged.  The M1 and M3 mirrors were repolished.  The M2 mirror was replaced with an engineering spare which was polished to match the figure of the original M2 mirror.  The mirrors were then re-coated.
The replaced mirror changed the effective figure of the WTTM system, making the focus a bit more sensitive.
A more extensive record of updates to WHIRC is available at the NOAO WHIRC website.\footnote{\url{http://www.noao.edu/kpno/manuals/whirc/hotnews.html}}

\subsection{Calibration of Data Before and After WTTM Refurbishment}\label{sec:calibration_old_new}
The cleaning and re-coating of the WTTM mirrors
led to a notable, apparently grey, improvement in system throughput (0.7~mag in $J$, 0.7~mag in $H$, and 0.5 mag in $K_s$; see below).
For the purposes of calibration, we therefore split the data into two samples: 2011B--2013A (before the mirrors were replaced, ``old'') and 2013B (after, ``new'').
Assuming the re-coating was grey in its increase in sensitivity, the color term should not change.
According to its physical meaning $k_f$ should not have changed (to first order) across the mirror replacement and re-coating; any change in $k_f$ from a change in the color response would be second order.  Given that we do not fit for a color$\times$airmass cross term originally, we similarly do not fit for a change in such a term.
We verified that the chisq surfaces are consistent between the ``old'' and ``new'' samples
for the values of $k_f$ and $c_f$.
We thus fit a constant $k_f$, $c_f$ across 2011B--2013A and 2013B, but separate ${\rm zpt}_f^{\rm old}$ and ${\rm zpt}_f^{\rm new}$ coefficients.

To provide a clean measurement of $k_f$ from bright stars, we use our repeated observations of Persson standard stars: P161-D, P212-C, P330-E, P525-E, S791-C, and S840-F.  These have the advantage of being very high signal-to-noise measurements ($100<{\rm SNR}<1000$, median 500) and confirmed stars.
 Our observations of these standards span an airmass range of 1.001--1.930.  The Persson standards we observed cover a color range from $0.15 < J-H < 0.70$ mag.

\begin{figure}
\plottwo{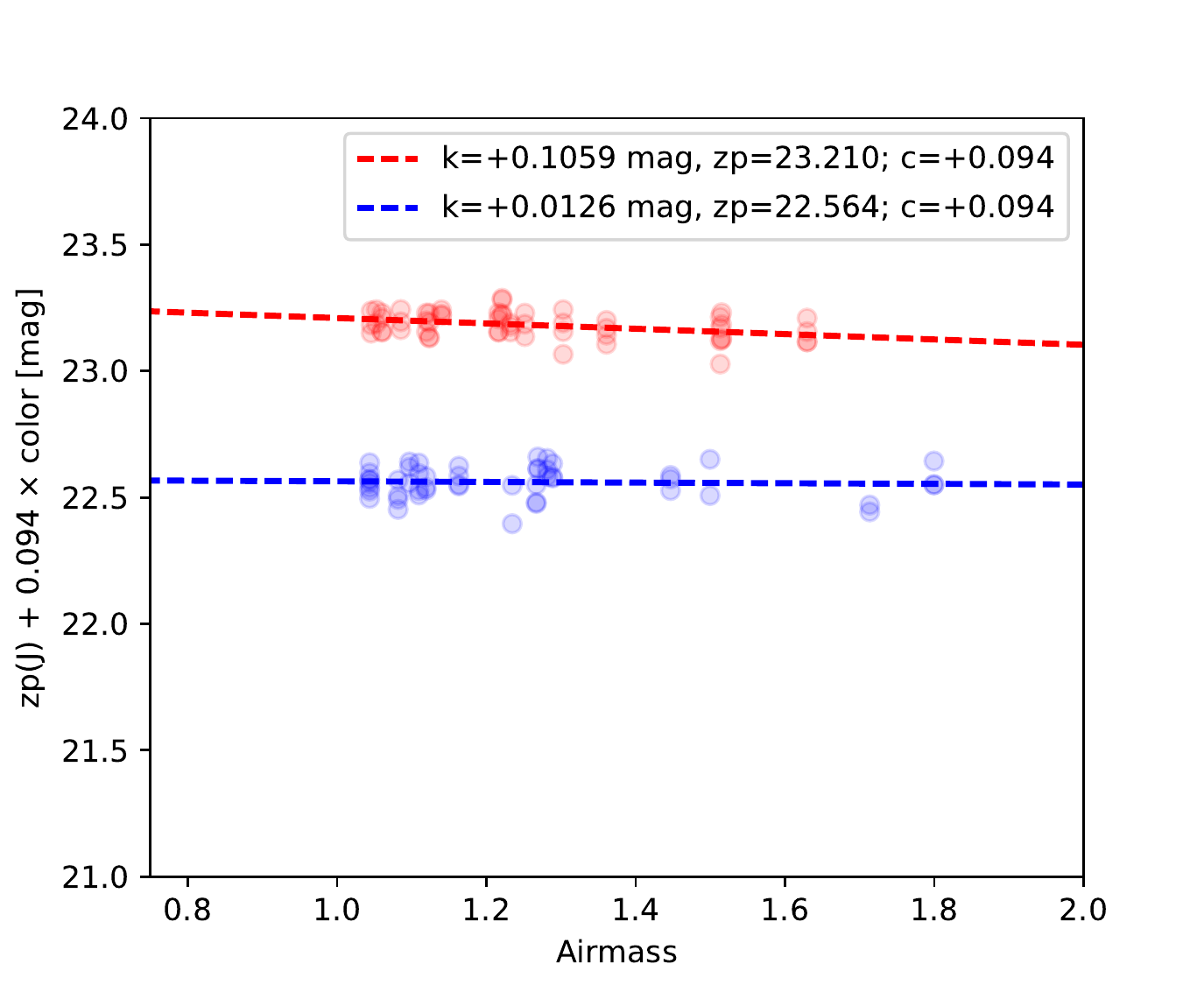}{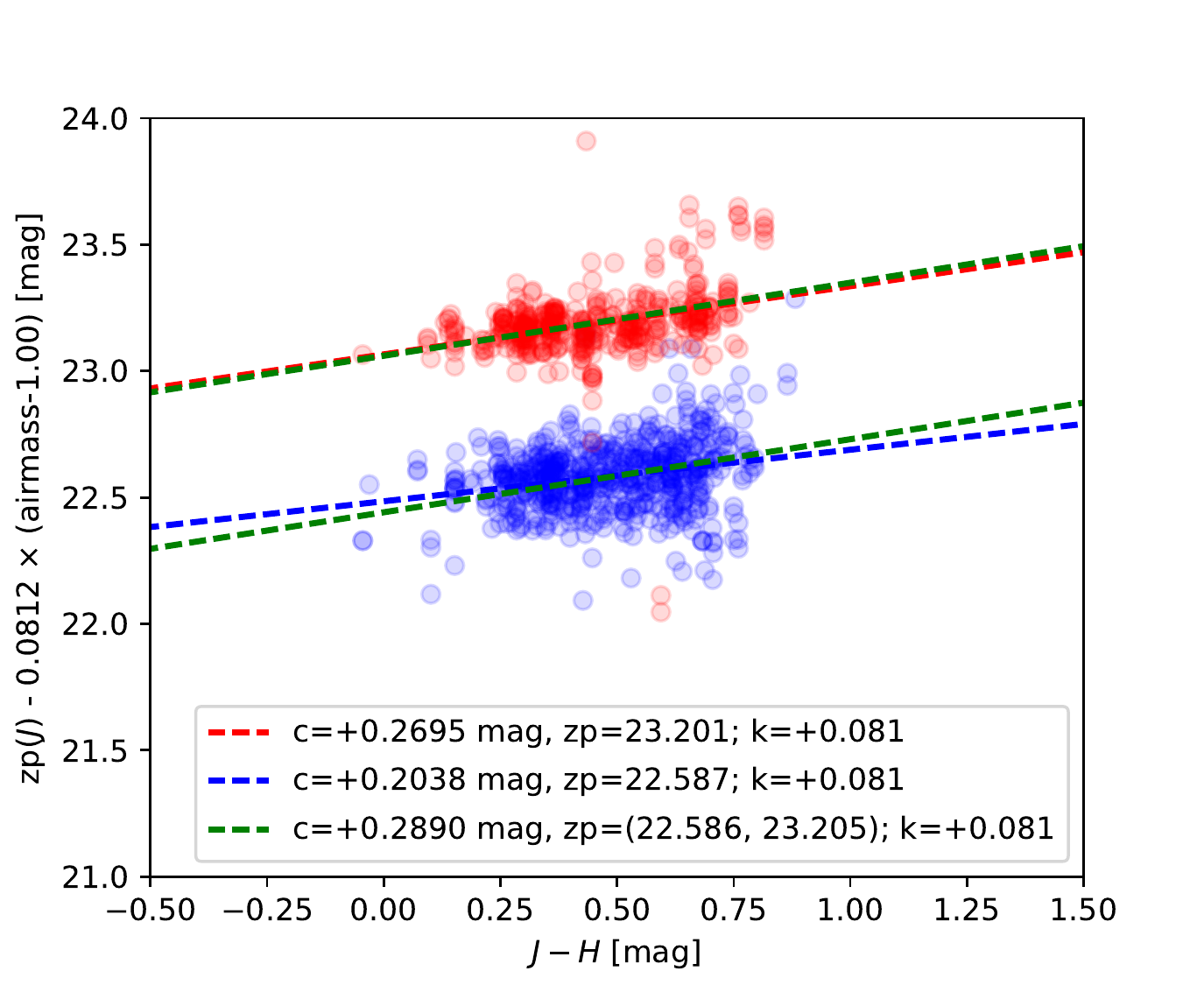}
\plottwo{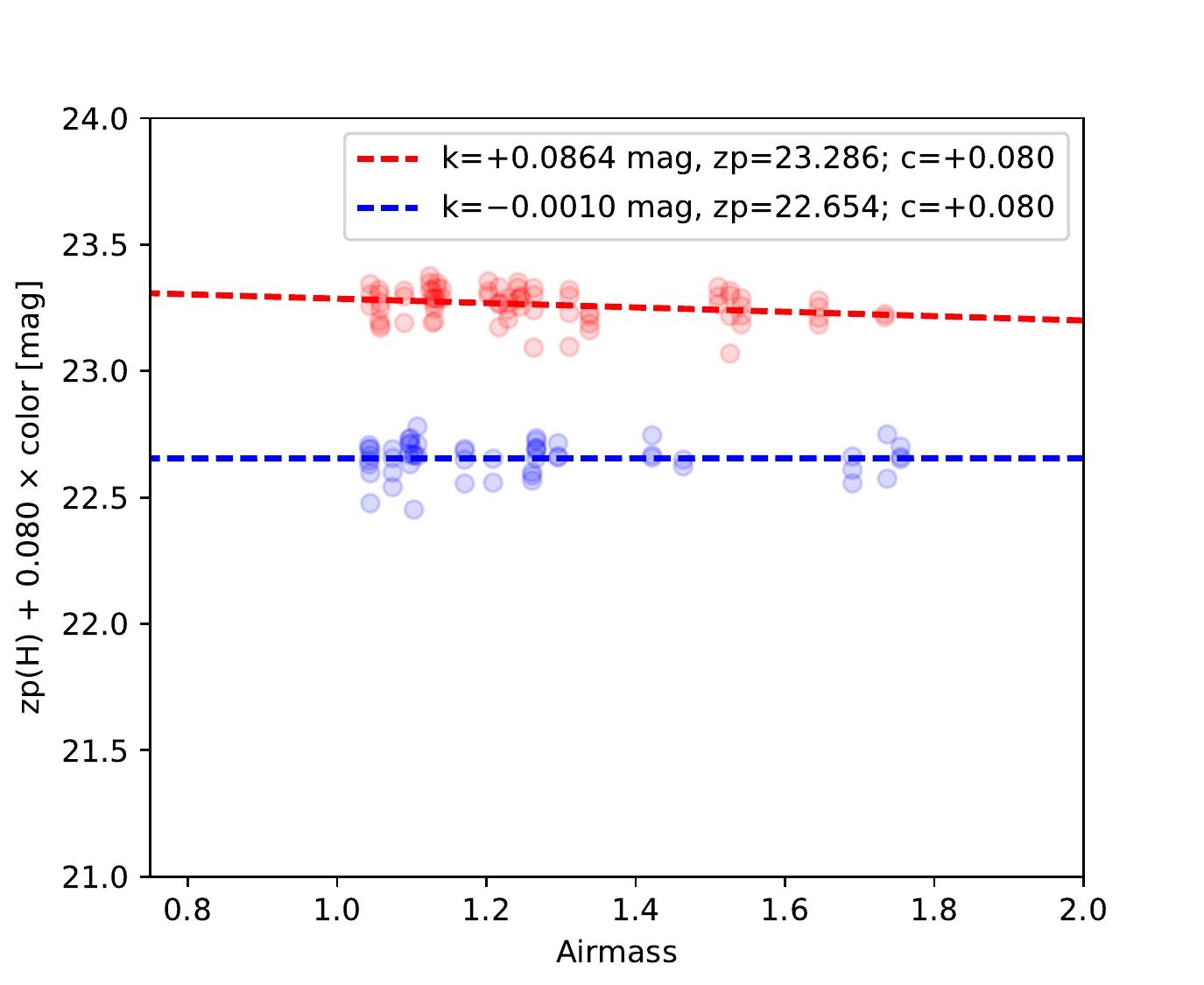}{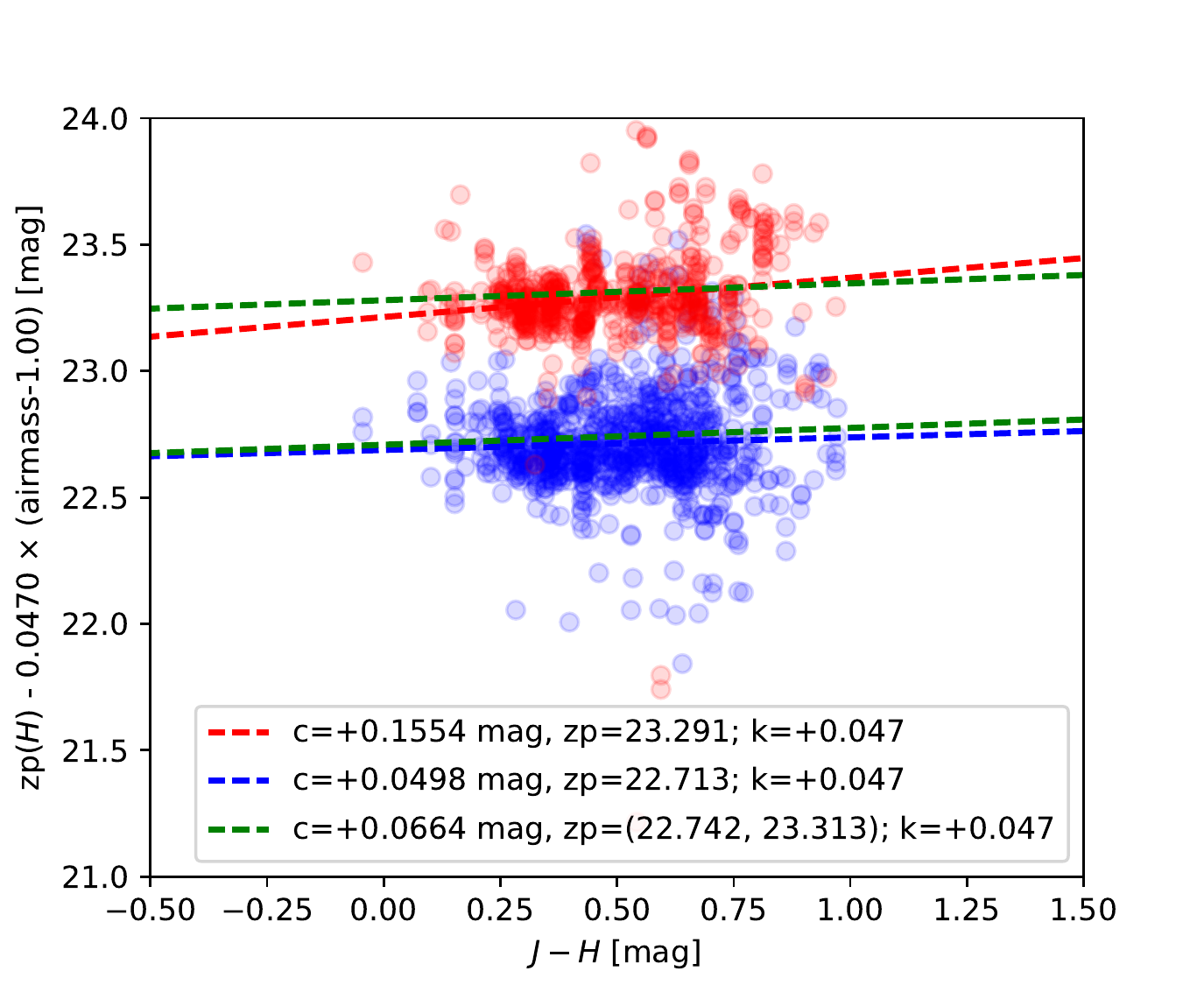}
\plottwo{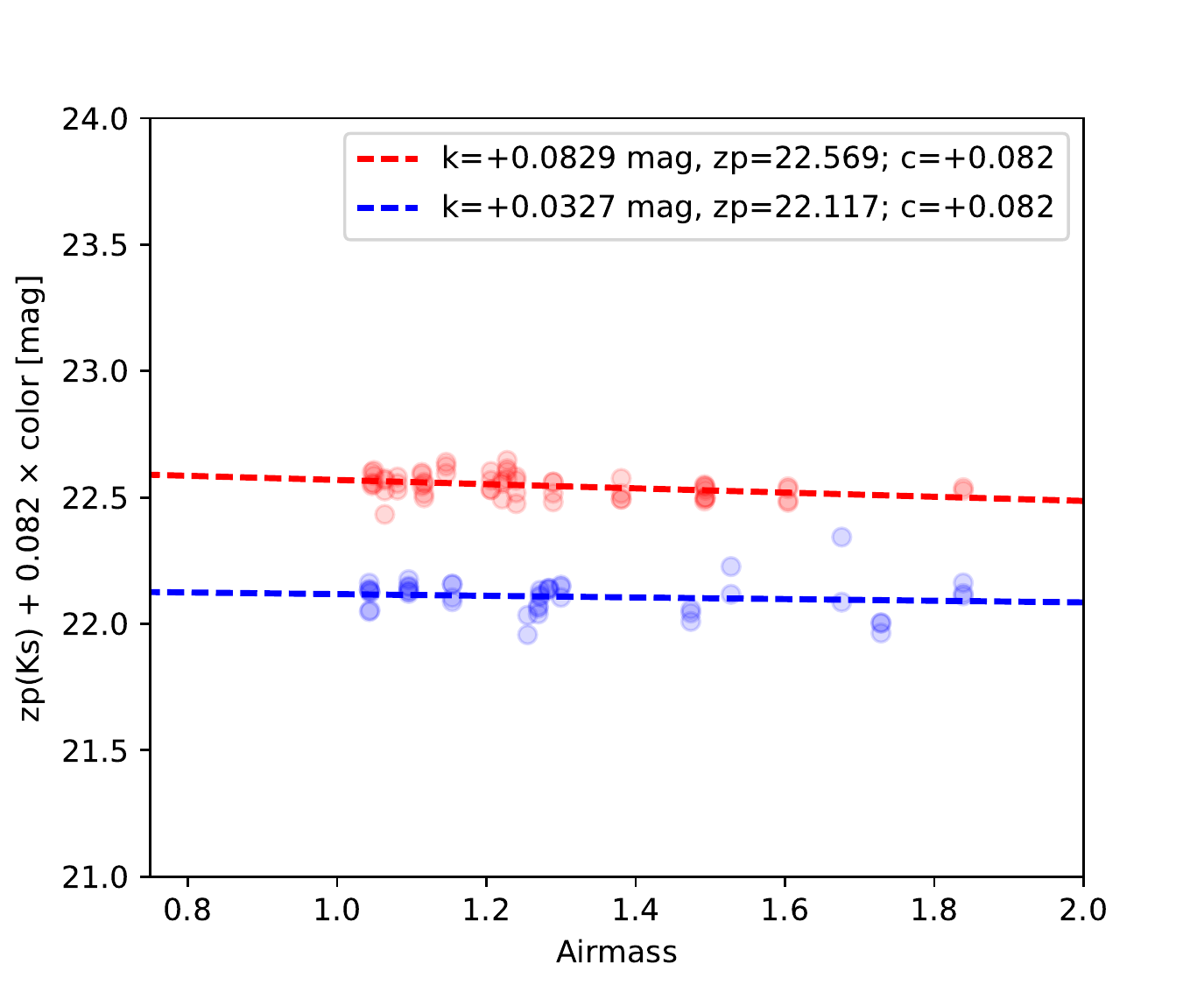}{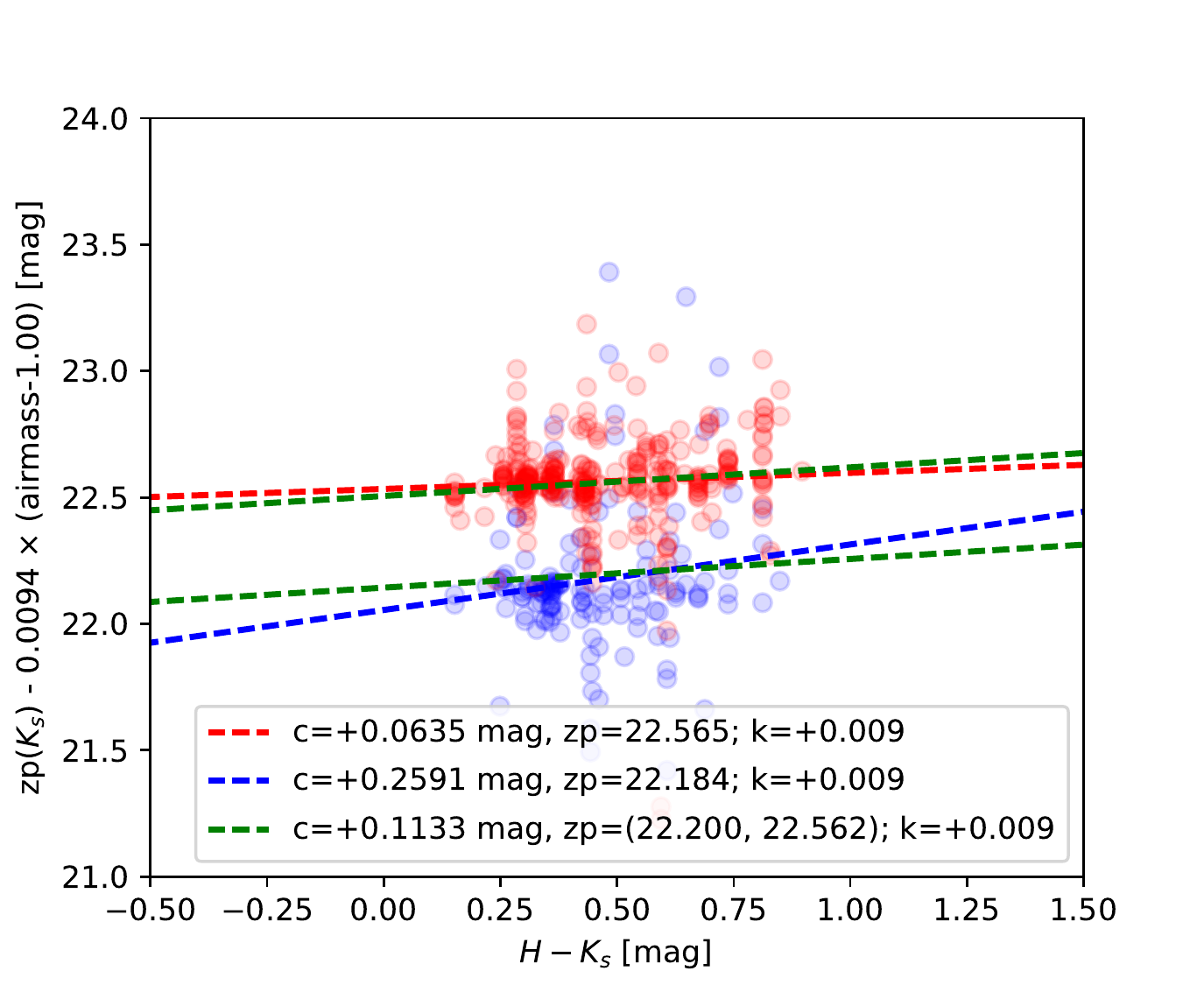}
\caption{The difference in 2MASS magnitude and WHIRC instrumental magnitude
for the Persson standard stars for airmass corrected for color (left column),
and for the 2MASS stars for color and zeropoint using the fixed Persson-derived airmass coefficient (right column).
The rows are for $J$, $H$, and $K_s$.
Results prior to the mirror re-coating and replacement (2011B-2013A) are in blue, results from after (i.e., 2013B) are in red, and results from a joint fit to $k_f$, $c_f$, ${\rm zpt}_f^{\rm old}$, and ${\rm zpt}_f^{\rm new}$ are in green.
For illustration, the green dashed lines are the values for the old (lower) and new (higher) zeropoints vs. airmass coefficient when the color coefficient is fixed at $c_f=0$.  
We use the joint-fit values in our photometric calibration.
}
\label{fig:calibration_coefficients}
\end{figure}

\begin{figure}
\plotone{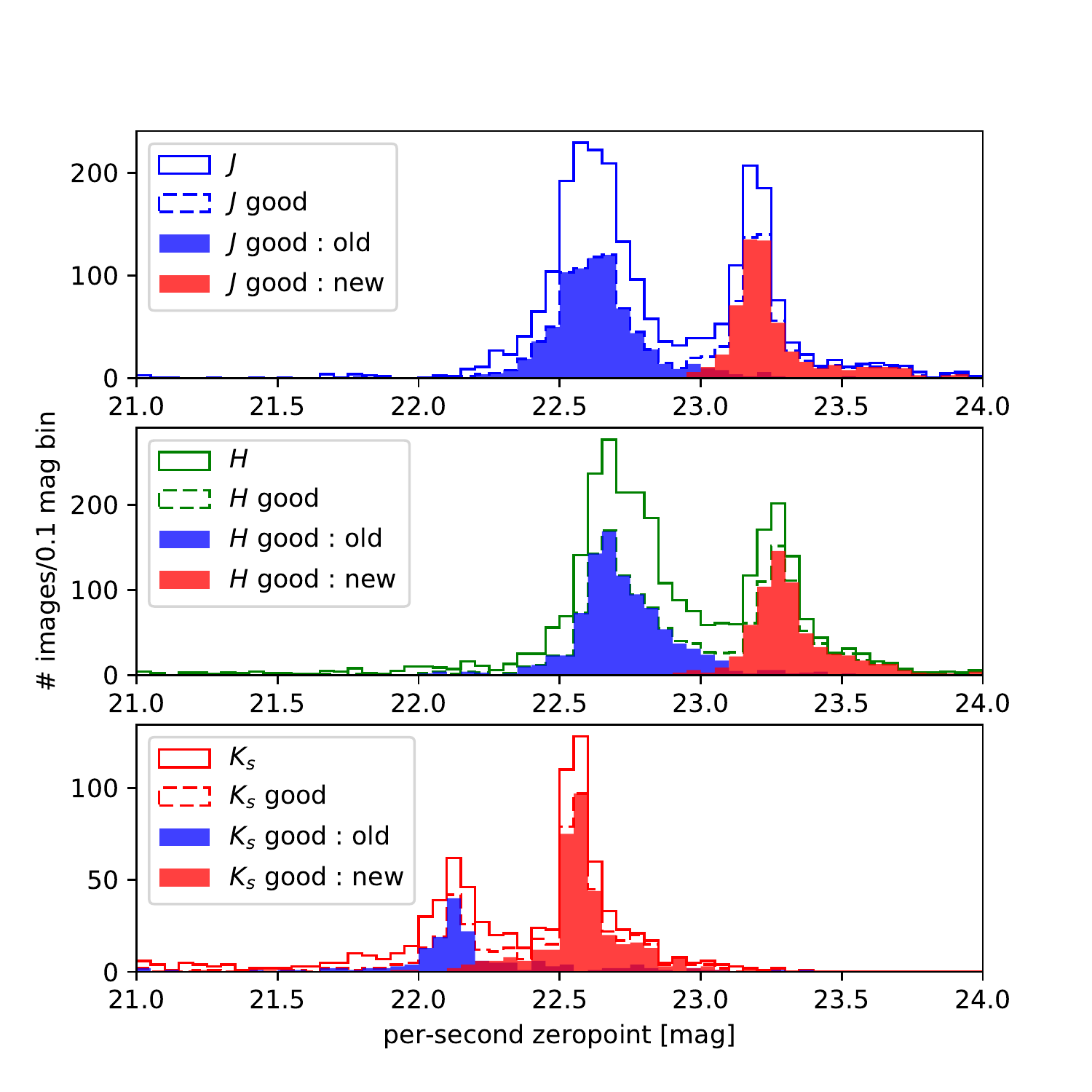}
\caption{Comparison of the WIYN+WHIRC per-second zeropoint from (2011B--2013A; blue) and (2013B; red) for ($J$, $H$, $K_s$) (top, center, bottom).
Solid lines show the distribution for all data combining both eras.
Dashed lines show the distribution for all photometric (``good'') data combining both eras.
Color-coding of these solid and dashed lines follows the filters
(blue, green, red) for ($J$, $H$, $K_s$).
Solid filled histograms show the results for ``good'' data from each era.
Color-coding of these histograms displays the era (filled blue: old, filled red: new).
Re-coating the optics resulted in a clear improvement in throughput of
approximately (0.7, 0.7, 0.5)~mag in ($J$, $H$, $K_s$).
}
\label{fig:zpt}
\end{figure}

Holding $k_f$ constant, we use the comparison of WHIRC instrumental magnitudes and 2MASS catalog values for 2MASS stars from the photometric nights to jointly solve for ${\rm zpt}_f^{\rm old}$, ${\rm zpt}_f^{\rm new}$, and $c_f$.
These derived photometry coefficients are given in Table~\ref{table:transform}.
Figure~\ref{fig:calibration_coefficients} shows the Persson stars used to derive the airmass coefficient and the 2MASS stars used to derive the color terms and per-second zeropoint of the system.
The resulting per-image ${\rm zpt}$ values are showing in Figure~\ref{fig:zpt}, where the distinction between the ``old'' and ``new'' set is quite evident.

\begin{figure}
\plotone{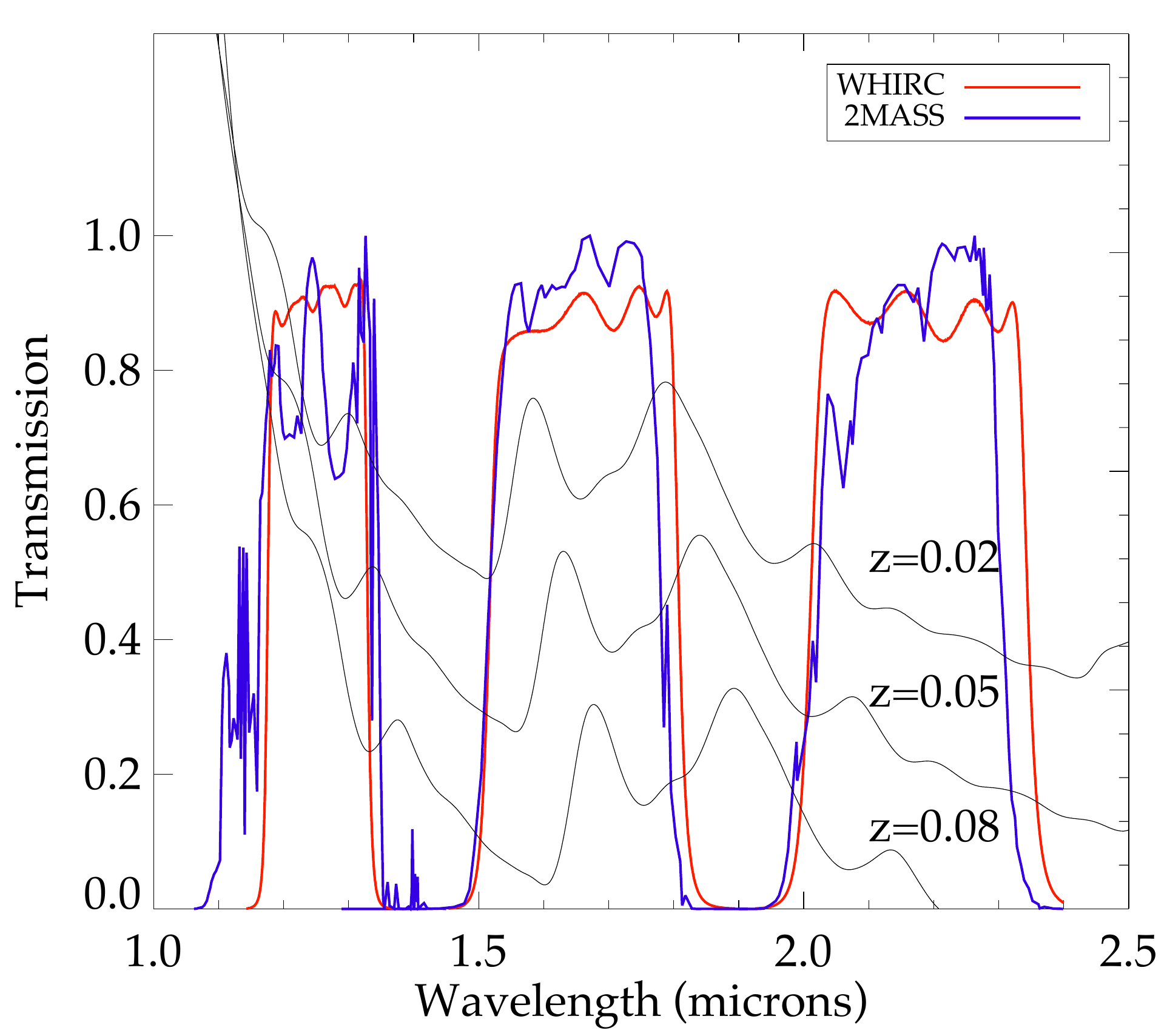}
\caption{Filter transmission for WHIRC compared with 2MASS system response function.
Over-plotted is a synthetic spectrum for a Type Ia which is 30 days old from \citet{Hsiao07} at three different redshifts (with arbitrary ordinate offset for clarity).
Note the significant bumps in the NIR spectrum that move out of the $J$- and $H$-band filters from $0<z<0.1$.
A $J$- and $H$-band only version of this figure was shown in \citet{Weyant14}.
These filter transmissions function curves were not used in determination of the WIYN+WHIRC natural system that we present here, which is based purely on color terms from broad-band colors.
Note that because the WHIRC transmission curves do not include the optics, and more importantly the atmosphere, this figure isn't quantitatively correct, but it illustrates the key point of the different wavelength cutoffs (filter edges) in the system response.
Despite this inconsistency, this figure still illustrates that expected the color terms to convert form the WHIRC system to the 2MASS system should be positive.
That is, for the same total counts in the WHIRC system, a redder star will be fainter in the 2MASS system than in the WHIRC system.
}
\label{fig:filterresponse}
\end{figure}

Figure~\ref{fig:filterresponse} illustrates the WHIRC filter functions\footnote{\url{http://www.noao.edu/kpno/manuals/whirc/filters.html}} vs. the 2MASS system (from the SNooPy reference file; which is originally from the 2MASS Explanatory Supplement\footnote{\url{http://www.ipac.caltech.edu/2mass/releases/allsky/doc/sec6\_4a.html\#rsr}}).  While we did not quantitatively use the WHIRC filter functions in calibrating our system, the positive color terms we find with respect to the 2MASS system are consistent with the redder extents of the WHIRC filters.  

Section~\ref{sec:refstars} details how the photometric coefficients for the WIYN+WHIRC system are used to derive the WIYN natural system.

\subsection{Reference star for each field}
\label{sec:refstars}

The zeropoint of the WIYN+WHIRC system was calibrated
to that of 2MASS because of the all-sky coverage of 2MASS and the availability of 2MASS objects in each of our SN fields.
A reference star is selected in each field that was observed on a photometric night (see Table~\ref{table:nights_observed}).
We calculate the difference between the WHIRC instrumental magnitude and the 2MASS catalog magnitude in $H$ (i.e., the ``zeropoint'' uncorrected for airmass or color).  To make a robust selection of the comparison star, we select a star whose ``zeropoint'' is consistent with the median of this ``zeropoint'' distribution.
We also require that this star is within 2\arcmin\ of the SN~Ia, and has an $H$-band magnitude between 10 and 15.5~mag.

The increased resolution of WIYN+WHIRC ($H$-band median FWHM 0.87\arcsec) over 2MASS ($H$-band median FWHM 2.8\arcsec) allows for the resolution of a number of 2MASS point sources into resolved galaxies or multiple point sources.
While we do not explicitly present a star-galaxy catalog as part of this work, the catalogs we present here do provide measured FWHM values for each object and characteristic FWHM values for each image.
For the standard stars as originally selected below, we inspected postage stamps for all of our WIYN+WHIRC images of each star visually and identified that 8 of these 2MASS point sources were either extended objects or consisted of multiple objects: 7 were resolved into galaxies and 1 into a ``visual'' binary.
Alternate 2MASS stars that were still point sources in the WIYN+WHIRC images, still otherwise satisfying the above criteria, were then selected by hand for these 8 fields.

The appropriate zeropoint-per-second, $k_f$, and $c_f$ coefficients from Table~\ref{table:transform}
were applied to the observed WHIRC instrumental magnitude of the star and its observed airmass to generate the defining WHIRC reference magnitude for each field's reference star.
\begin{equation}
m^{\rm WHIRC}_{ {\rm cal}, f} = m_{{\rm inst,}f}^{\rm WHIRC} + {\rm zpt}_f - k_f \left(X-1\right)
\label{eq:whirc_reference_mag}
\end{equation}

Eight supernovae: SN~2011io, PTF11qzq, PTF12ikt, SN~2012go, SNhunt175, SN~2013bq, CSS130218:092354+385837, and SN~2013cb were not observed on a photometric night.
For these objects, the first night that the object was observed on was chosen as the reference night.

The full list of reference stars is given in Table~\ref{table:calibrationstars} and provided in \file{WHIRC\_standard\_star\_catalog.fits}.
The calibrated standard stars in the WHIRC natural system were generated using Equation~\ref{eq:transform} and the results from Table~\ref{table:transform}.

The WHIRC standard star for each field is used to find the zeropoint for each stacked image as follows
\begin{equation}
{\rm zpt}_{f,i} = m_{{\rm cal},f}^{\rm WHIRC} - m_{{\rm inst},f,i}^{\rm WHIRC}\label{eq:zpt}
\end{equation}
where the $i$ subscript indexes over the stacked images and $m^{\rm WHIRC}_{\rm cal}$ is the single WHIRC standard star for that field.
The calibrated supernova magnitude is the WHIRC instrumental magnitude plus the zeropoint as described in Equation~\ref{eq:zpt}.  The \snia\ lightcurves in the WHIRC natural system are presented in Table~\ref{table:lightcurves}.  We estimate that we have calibrated our ensemble average of WIYN+WHIRC observations to the 2MASS system to within 0.02~mag, with an additional field-to-field variation of 0.03 mag due to the limited number of 2MASS stars available in many of the fields.  In future work we will refine this calibration by using all detected stars in our images to propagate a photometric solution across our fields, instead of relying on only the 2MASS catalog stars in each field.

\section{Data}\label{sec:data}

With this data release we publish processed stacked images, photometric catalogs for each stacked image, lightcurves for each 2MASS star in each field, and lightcurves of supernovae which have little enough host-galaxy background light that reliable lightcurves could be generated without subtraction of reference images.

\begin{figure}
\plotone{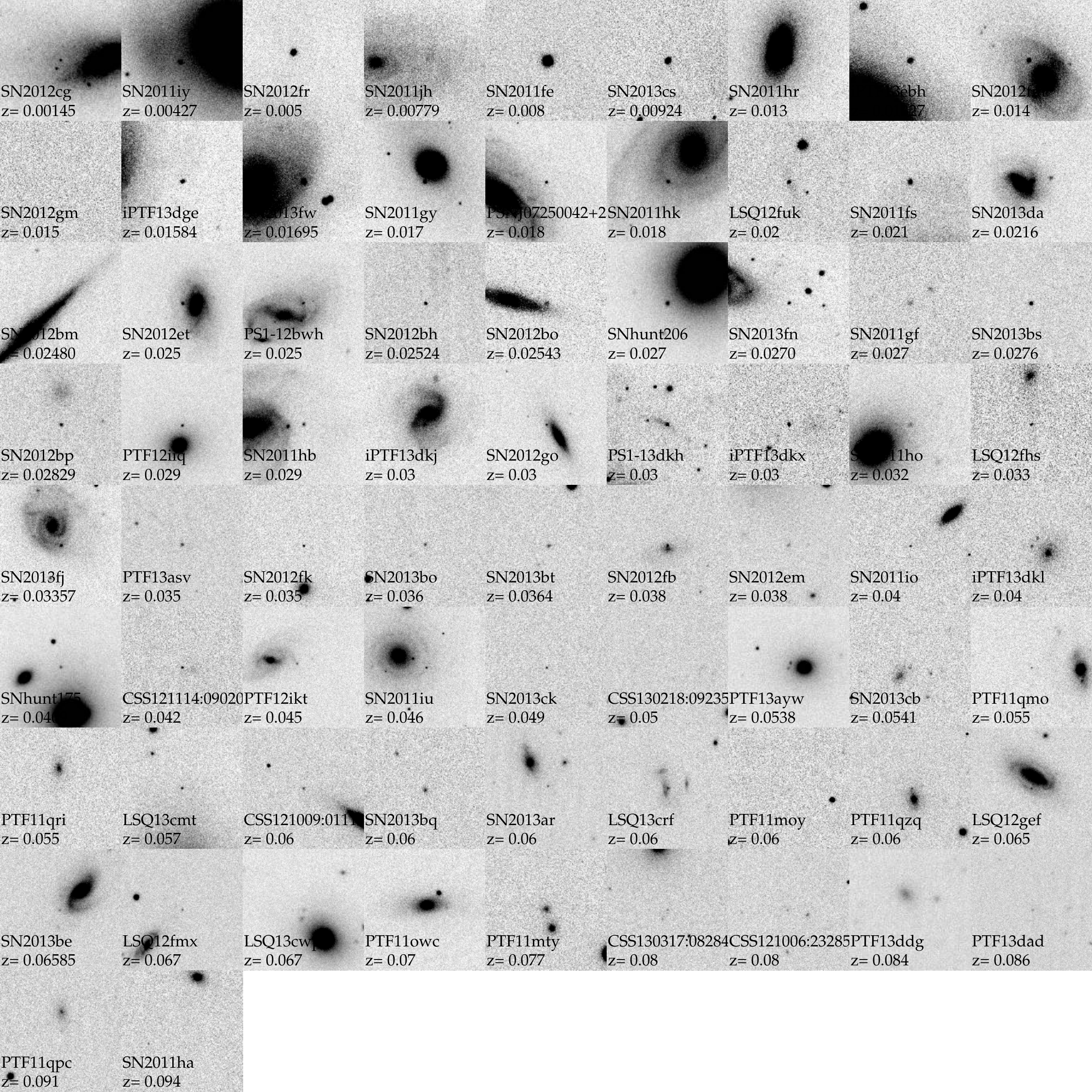}
\caption{$H$-band images of the 74 SNe~Ia presented in this data release from our WIYN+WHIRC stacked images.  The stamps are oriented North up, East left.  The SN is at the center of each postage stamp.  The SNe are presented in order of increasing redshift.  Each image is 40\arcsec$\times$40\arcsec\ square.
Note that the image for SN~2011bh is from the host-galaxy template.  The ``live'' supernovae observations for SN~2011bh were not successfully processed by our current pipeline.
Please see online version of paper for a full-resolution version of this figure.
}

\label{fig:postage_stamps}
\end{figure}

Table~\ref{table:dr1} summarizes the observations and resulting supernova lightcurves after the processing described in Section~\ref{sec:processing}.
Table~\ref{table:imageinfo} lists each of the stacked images along with their filter, date, and for SN fields, the approximate phase of the observation.  These are the original estimates of the phase -- refer to the actual lightcurve files for more accurate phases.
Figure~\ref{fig:postage_stamps} shows postage stamp images of the 74 supernovae whose images are included in this release.

\subsection{Lightcurves of 2MASS stars}\label{sec:2masslcs}
We present $JHK_s$ lightcurves for \numTwomassLc\ 2MASS stars plus the 6 Persson standards (denoted by the Persson catalog name rather than their 2MASS names, see Table~\ref{table:persson_standards}).
These ``2MASS stars'' are based on the 2MASS classification as a star (point-like object).
We have {\em not} updated these star/galaxy classification based on our higher-resolution WIYN+WHIRC data.
Thus some of the 2MASS ``stars'' for which we report lightcurves are likely galaxies or blended stars.

\subsection{Lightcurves of Supernovae}\label{sec:snlcs}
Table~\ref{table:lightcurves} presents $JHK_s$ lightcurves for the \numLc\ \sneia\ that were in sufficiently low surface-brightness regions of their host galaxy to lead to accurate lightcurves.

We used $r$-band measurements from SDSS DR9~\citep{SDSSDR9} and Pan-STARRS1 DR1~\citep{PS1,PS1DR1}
images at the location of the SNe to quantitatively estimate the host-galaxy background contamination at the location of each SN.
We judged all SNe that had an estimated $r$-band host surface brightness / sq.~arcsec within 5 mag/sq.~arcsec of the estimated peak H-band magnitude of the supernova to be too contaminated: 0.6\arcsec\ was a typical aperture radius for our supernova photometry which is an area of $\sim$1~sq.~arcsec.  Our typical lightcurves are between 10--20 days after maximum light, when a SNe is within 0.5~mag of peak brightness in H.

For these SNe that we have selected using the above $r$-band criterion, we estimate the photometric uncertainty from not doing host-galaxy background subtraction as a systematic 3\% of the flux of the SN at maximum brightness.

\begin{figure}
\figsetstart
\figsetnum{1}
\figsettitle{Supernova Lightcurves}

\figsetgrpstart
\figsetgrpnum{1.1}
\figsetgrptitle{CSS121009:011101-172841}
\figsetplot{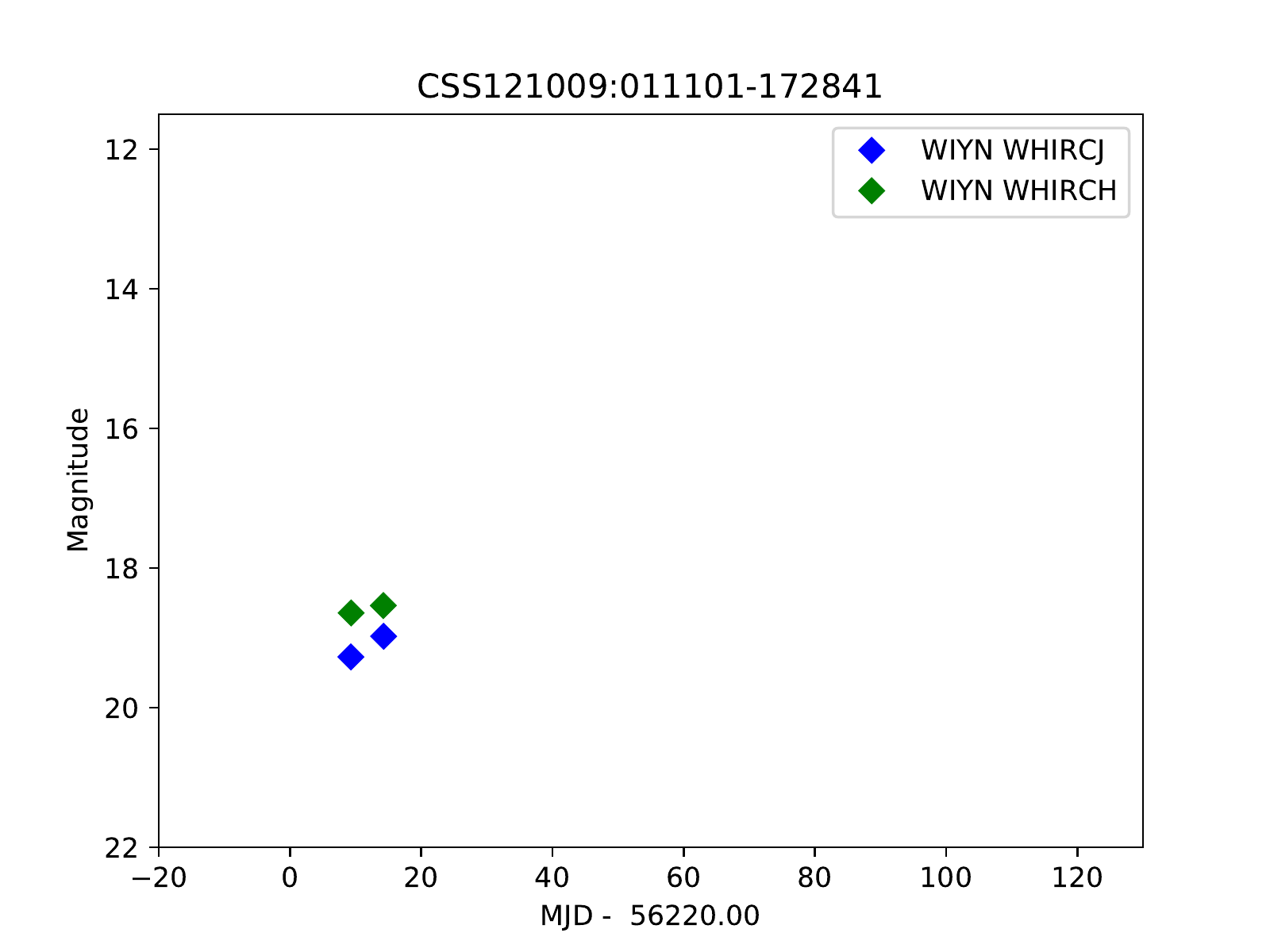}
\figsetgrpnote{CSS121114:090202+101800}
\figsetgrpend

\figsetgrpstart
\figsetgrpnum{1.2}
\figsetgrptitle{CSS121114:090202+101800}
\figsetplot{lightcurves/CSS121114:090202+101800.pdf}
\figsetgrpnote{CSS121114:090202+101800}
\figsetgrpend

\figsetgrpstart
\figsetgrpnum{1.3}
\figsetgrptitle{LSQ12fhs}
\figsetplot{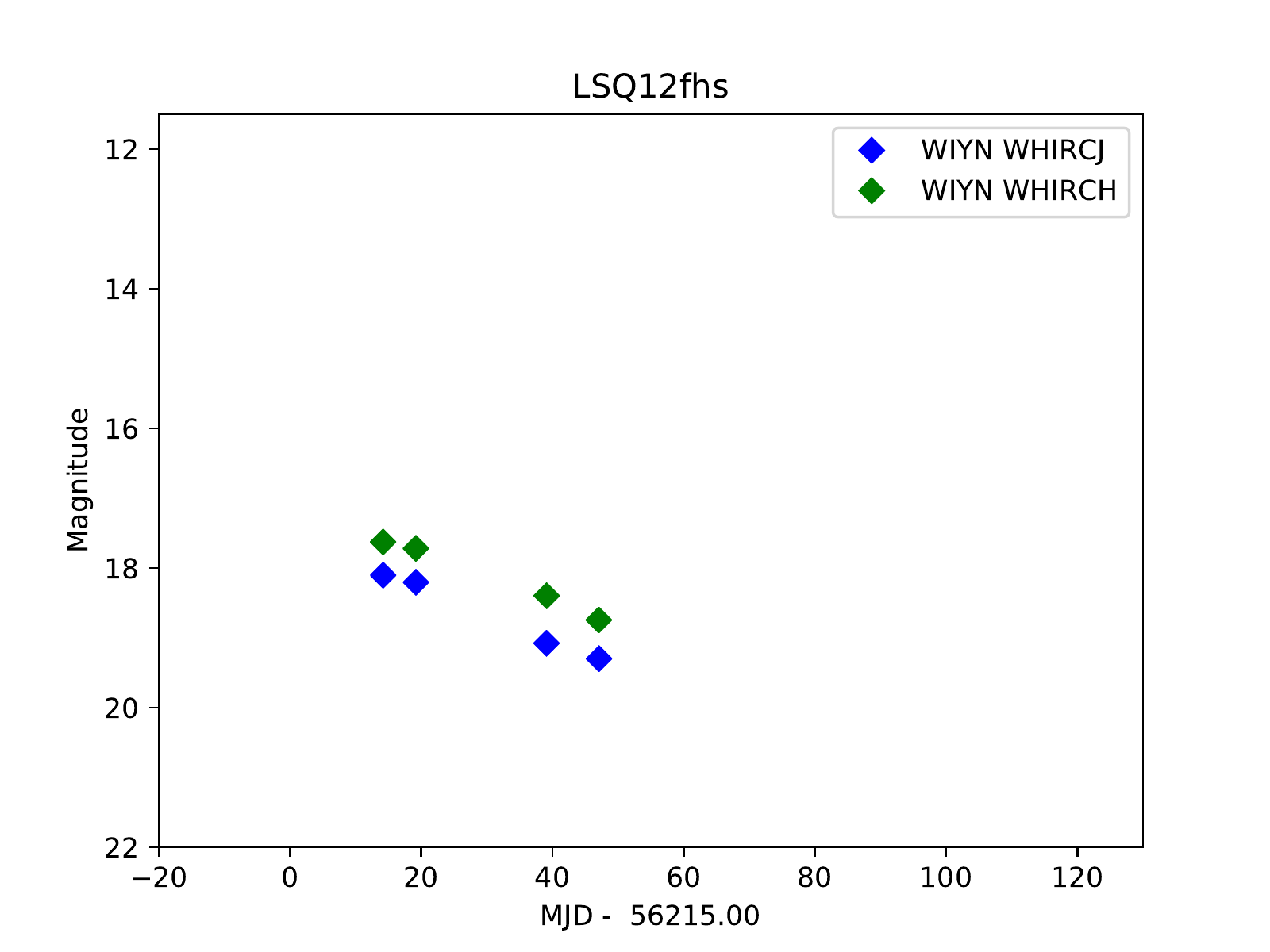}
\figsetgrpnote{LSQ12fhs}
\figsetgrpend

\figsetgrpstart
\figsetgrpnum{1.4}
\figsetgrptitle{LSQ12fmx}
\figsetplot{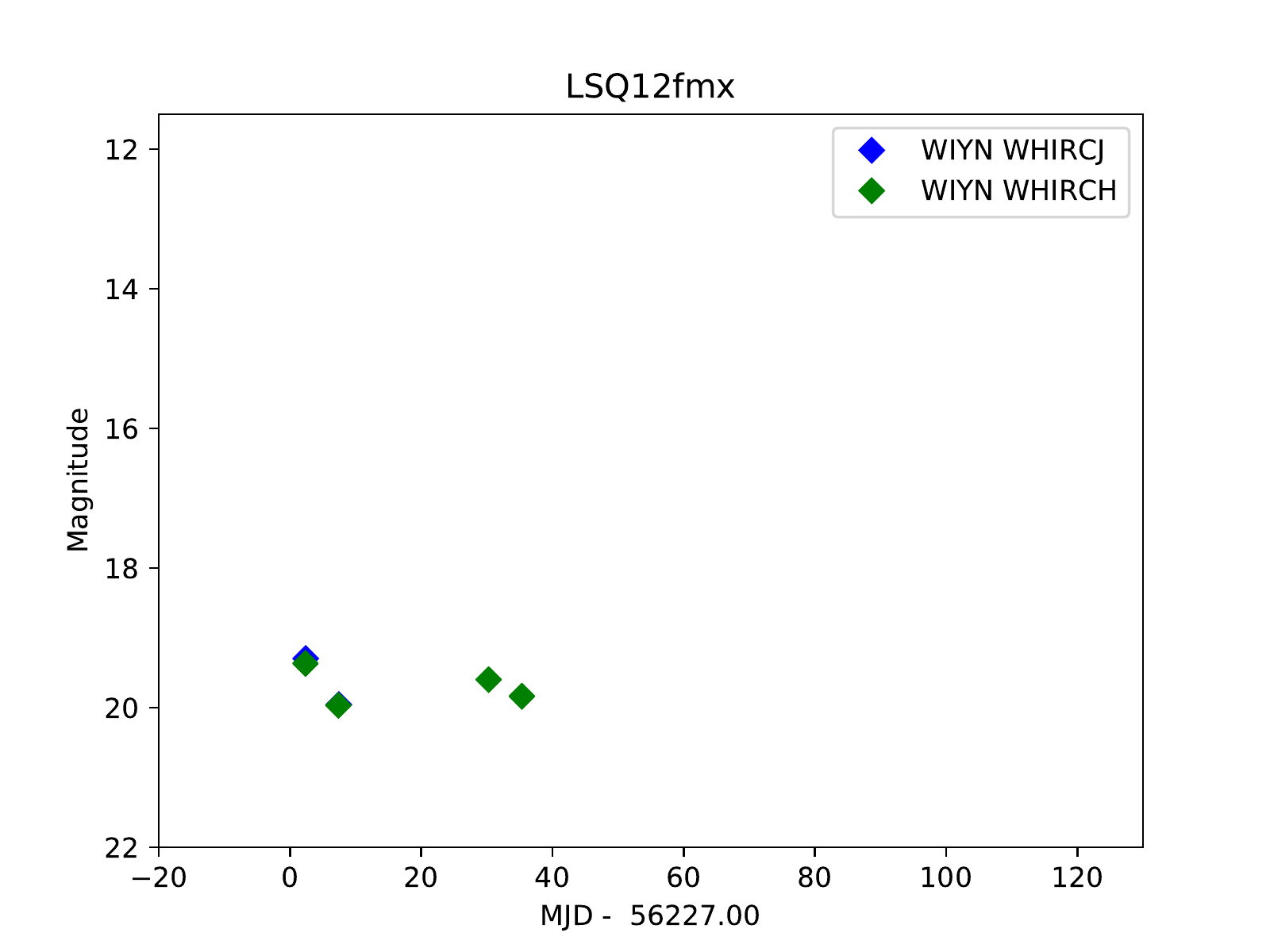}
\figsetgrpnote{LSQ12fmx}
\figsetgrpend

\figsetgrpstart
\figsetgrpnum{1.5}
\figsetgrptitle{LSQ13cmt}
\figsetplot{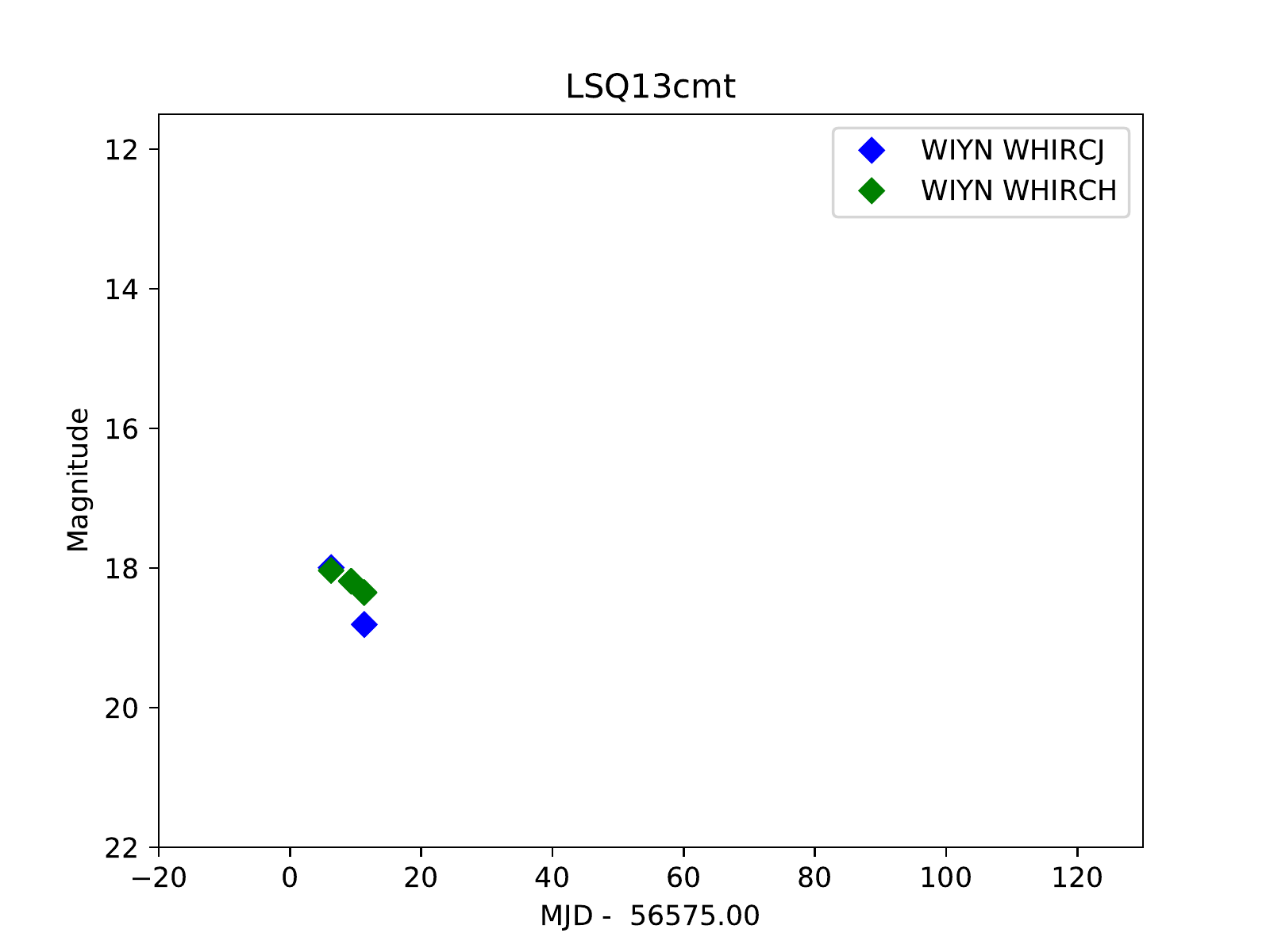}
\figsetgrpnote{LSQ13cmt}
\figsetgrpend

\figsetgrpstart
\figsetgrpnum{1.6}
\figsetgrptitle{LSQ13cwp}
\figsetplot{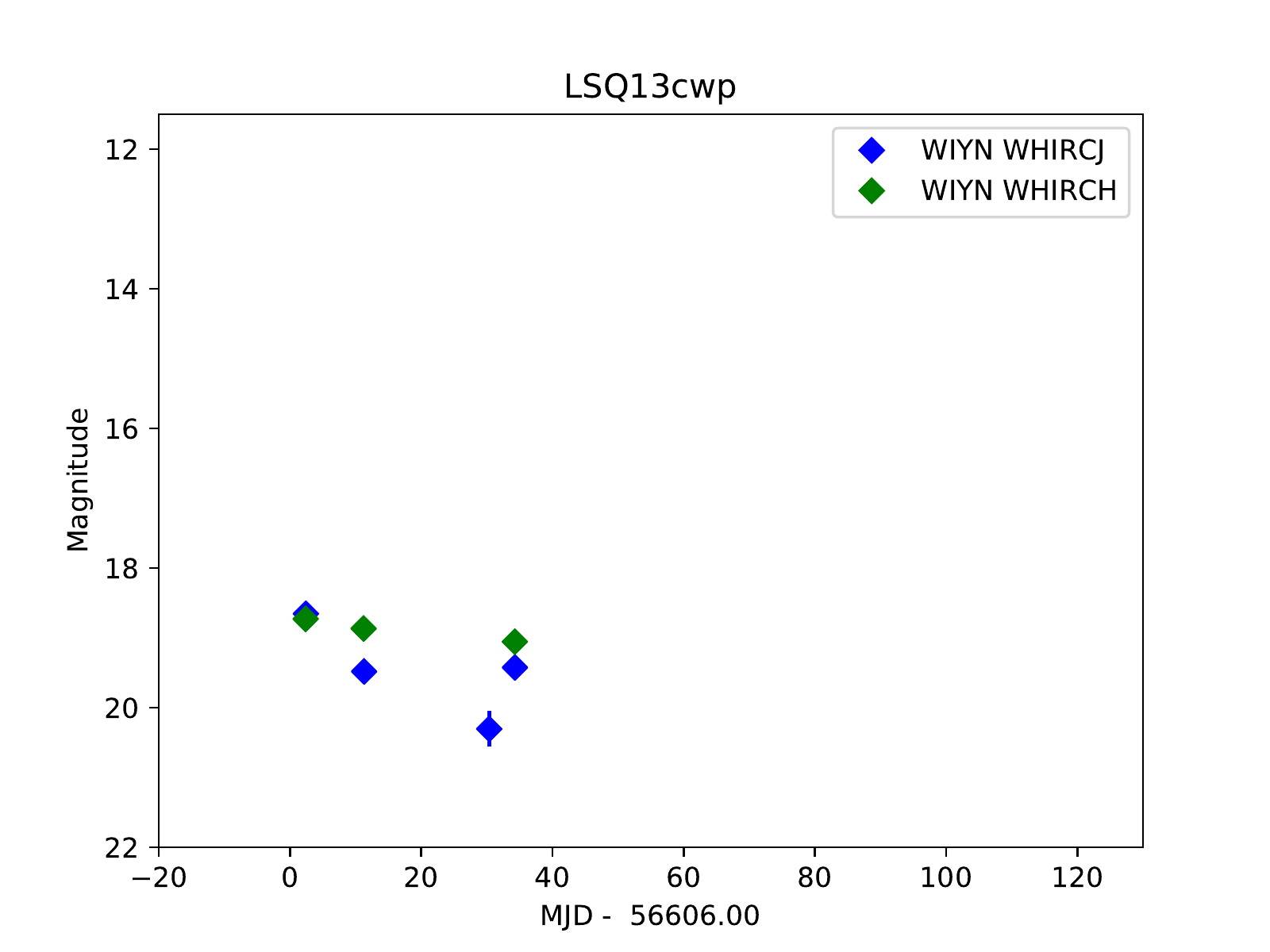}
\figsetgrpnote{LSQ13cwp}
\figsetgrpend

\figsetgrpstart
\figsetgrpnum{1.7}
\figsetgrptitle{PS1-13dkh}
\figsetplot{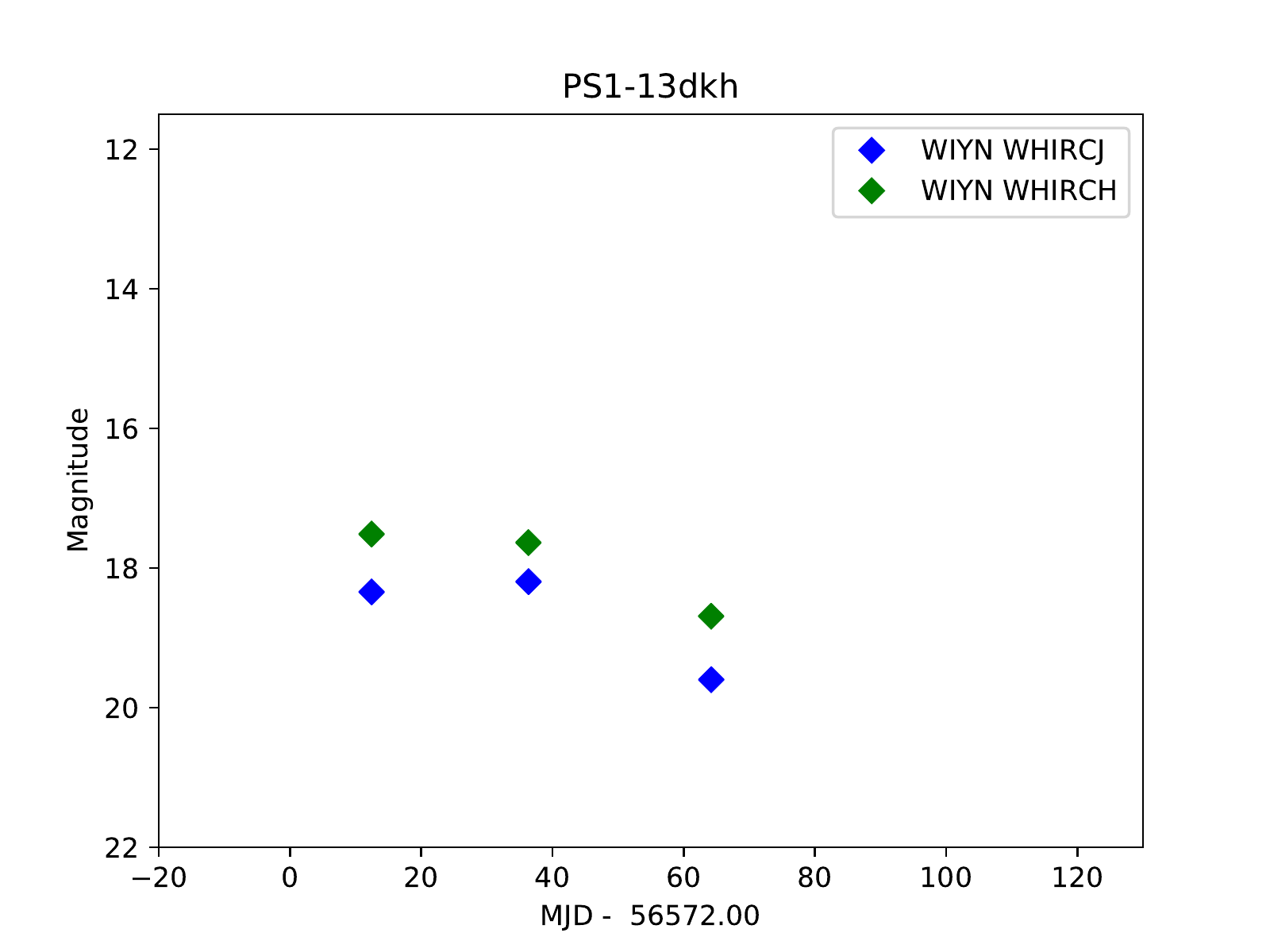}
\figsetgrpnote{PS1-13dkh}
\figsetgrpend

\figsetgrpstart
\figsetgrpnum{1.8}
\figsetgrptitle{PTF11moy}
\figsetplot{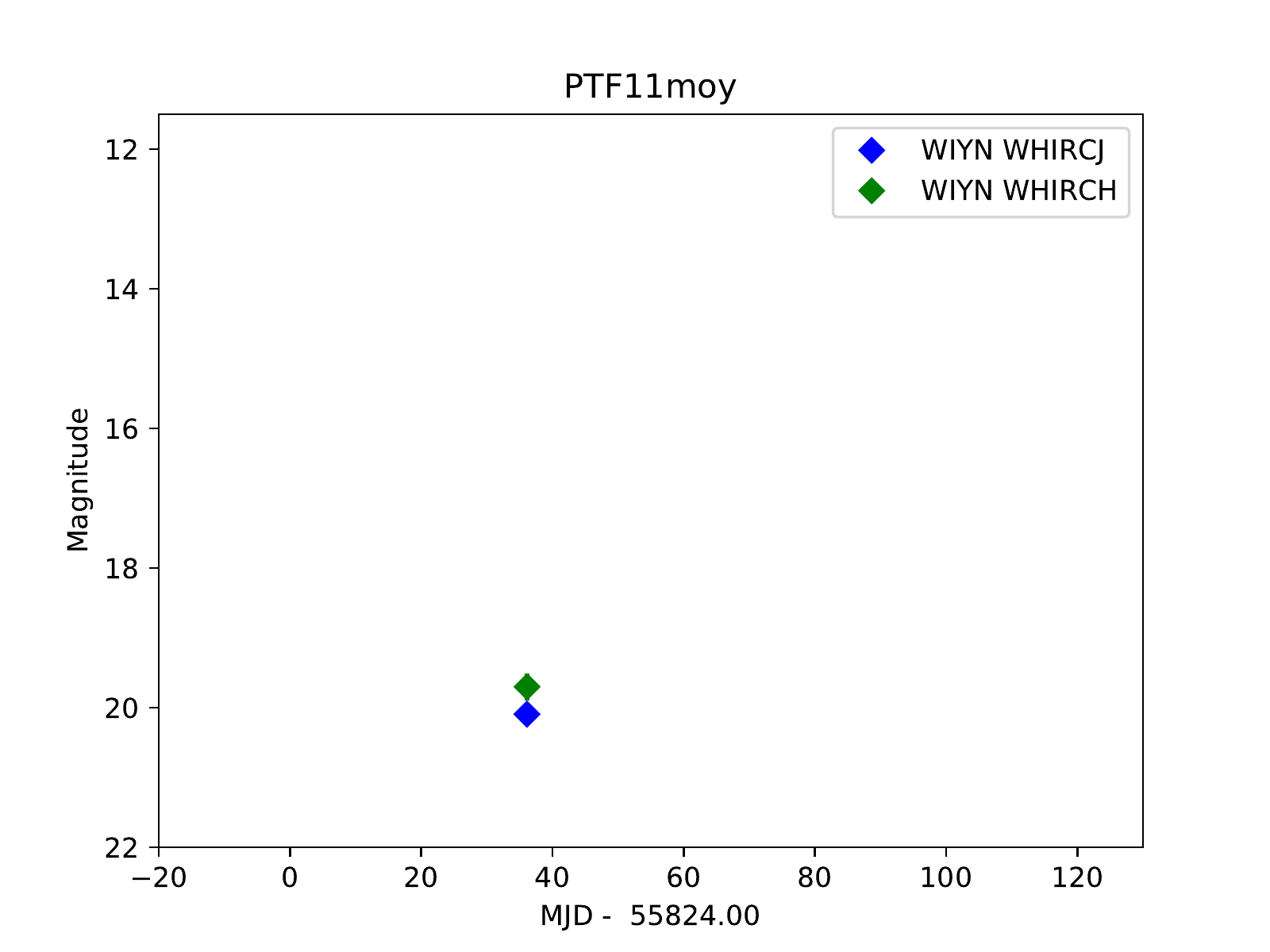}
\figsetgrpnote{PTF11moy}
\figsetgrpend

\figsetgrpstart
\figsetgrpnum{1.9}
\figsetgrptitle{PTF11qmo}
\figsetplot{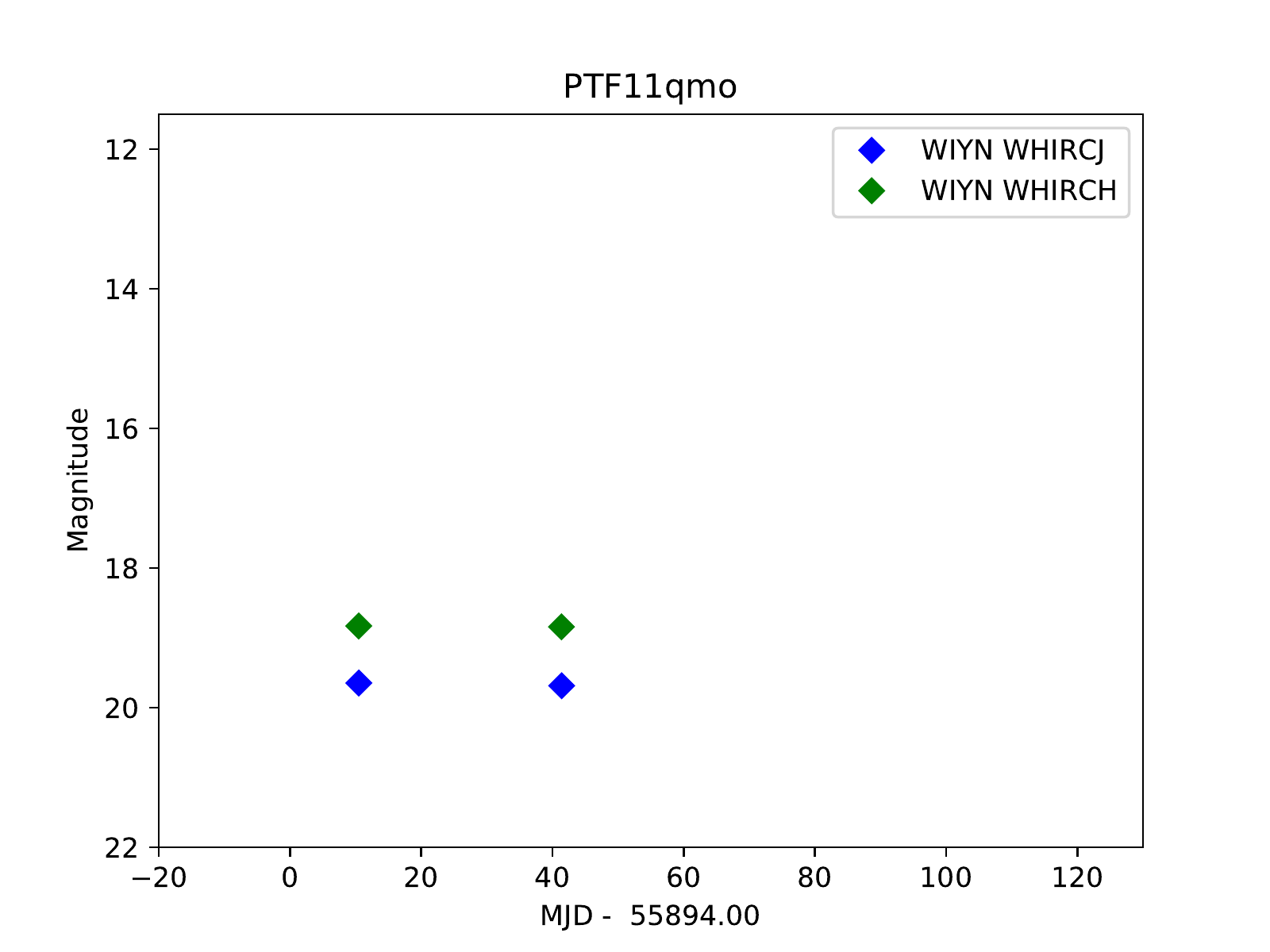}
\figsetgrpnote{PTF11qmo}
\figsetgrpend

\figsetgrpstart
\figsetgrpnum{1.10}
\figsetgrptitle{PTF13asv}
\figsetplot{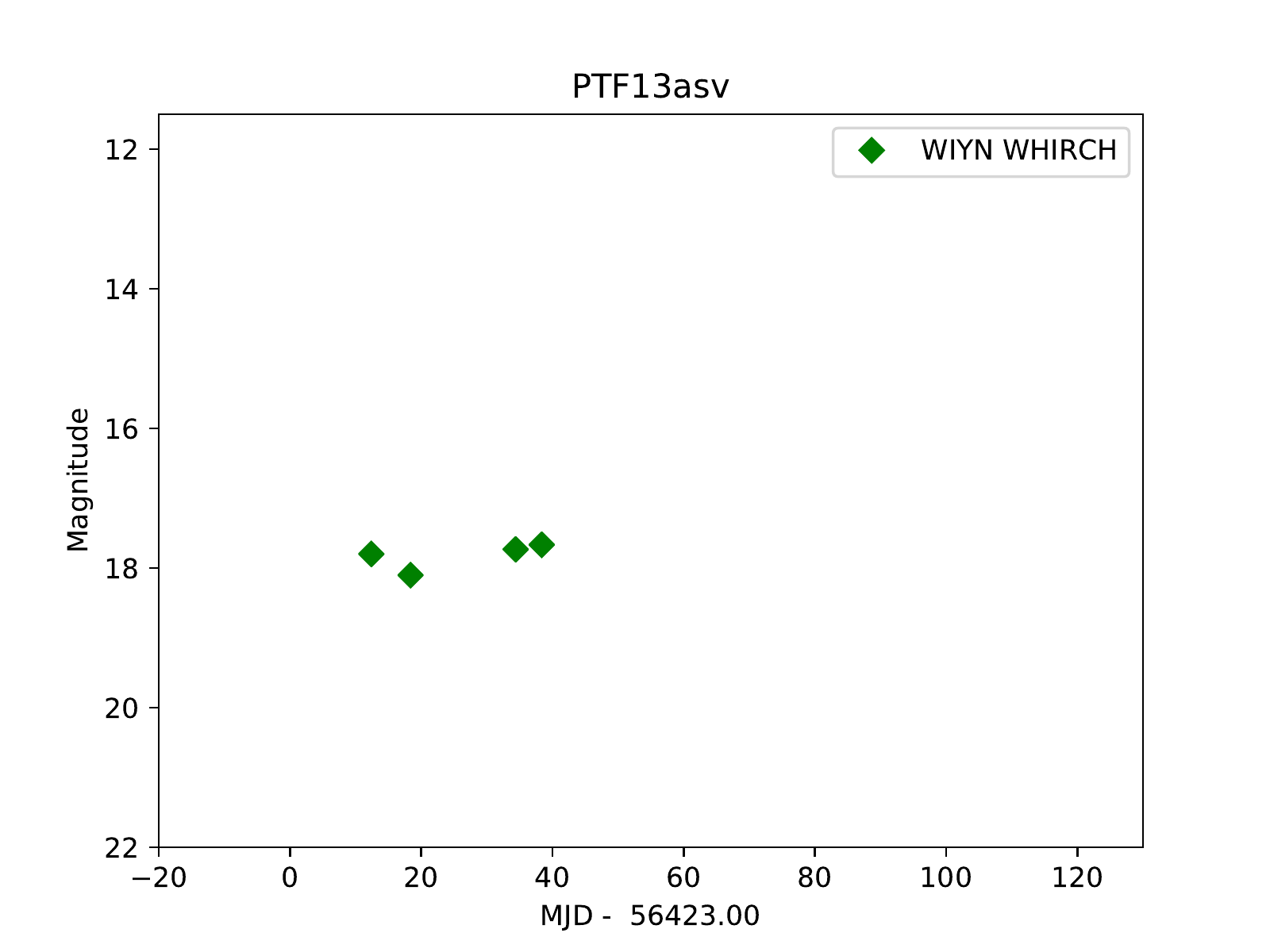}
\figsetgrpnote{PTF13asv}
\figsetgrpend

\figsetgrpstart
\figsetgrpnum{1.11}
\figsetgrptitle{PTF13dad}
\figsetplot{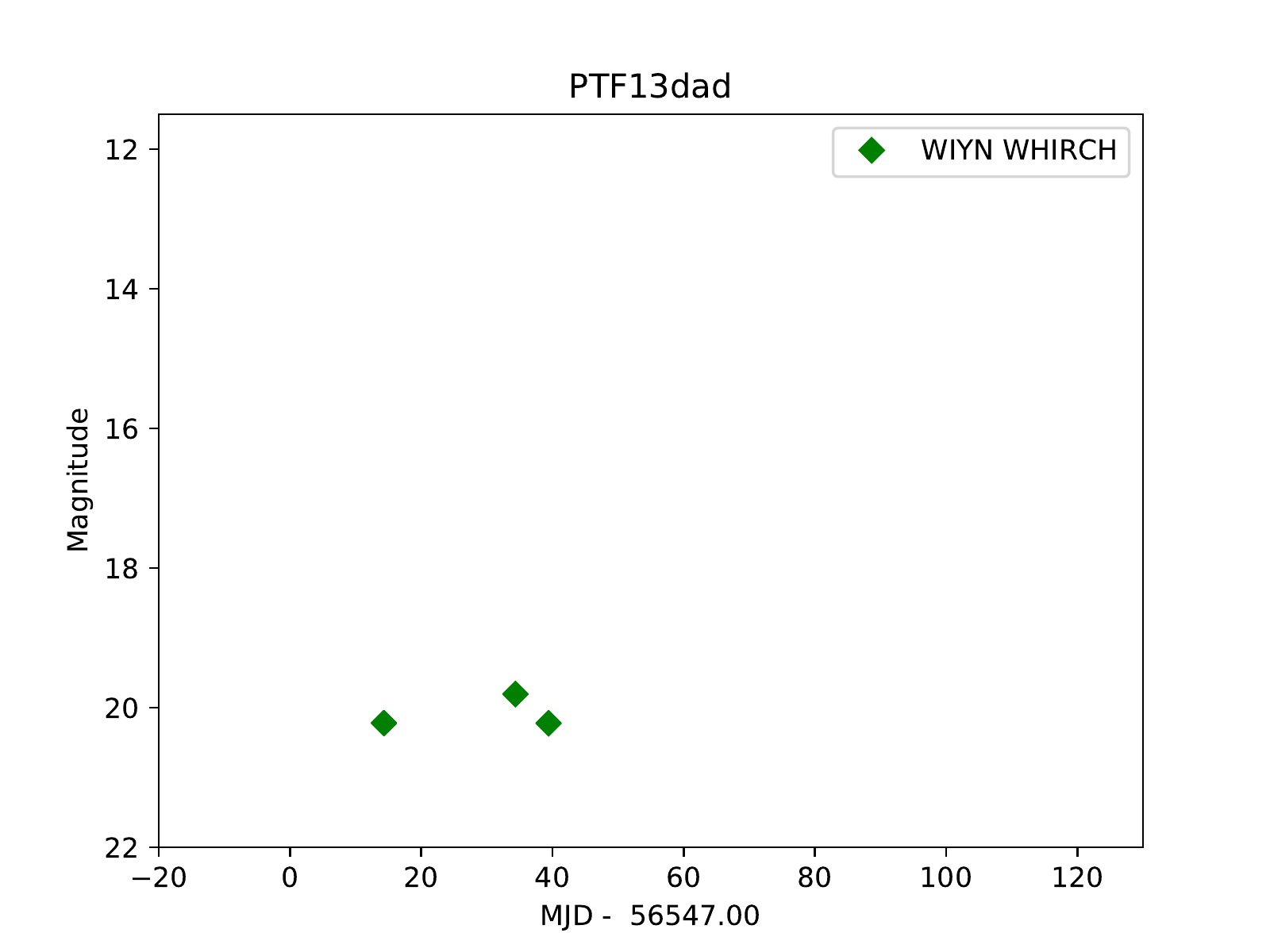}
\figsetgrpnote{PTF13dad}
\figsetgrpend

\figsetgrpstart
\figsetgrpnum{1.12}
\figsetgrptitle{PTF13ddg}
\figsetplot{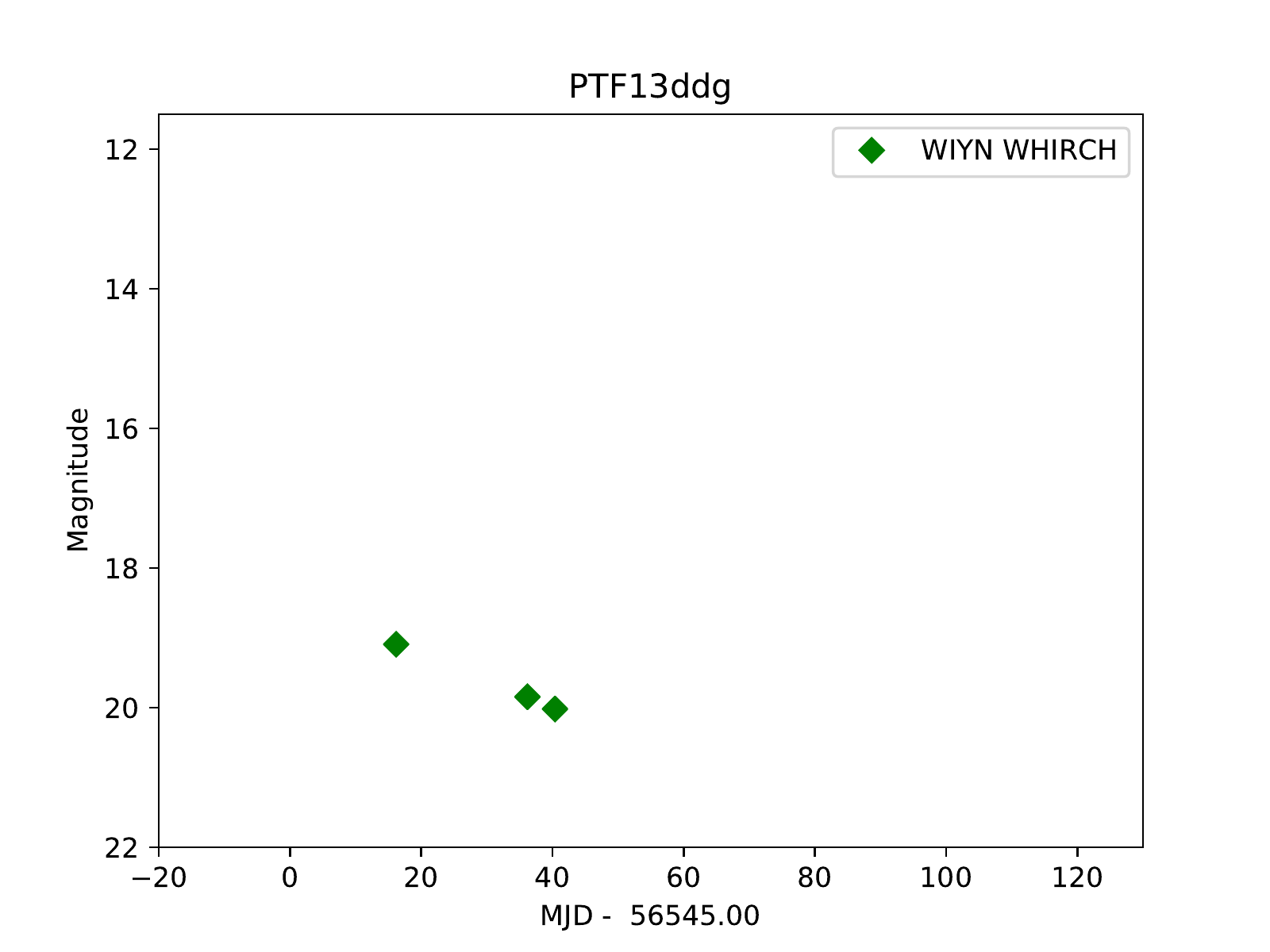}
\figsetgrpnote{PTF13ddg}
\figsetgrpend

\figsetgrpstart
\figsetgrpnum{1.13}
\figsetgrptitle{SN2011fe}
\figsetplot{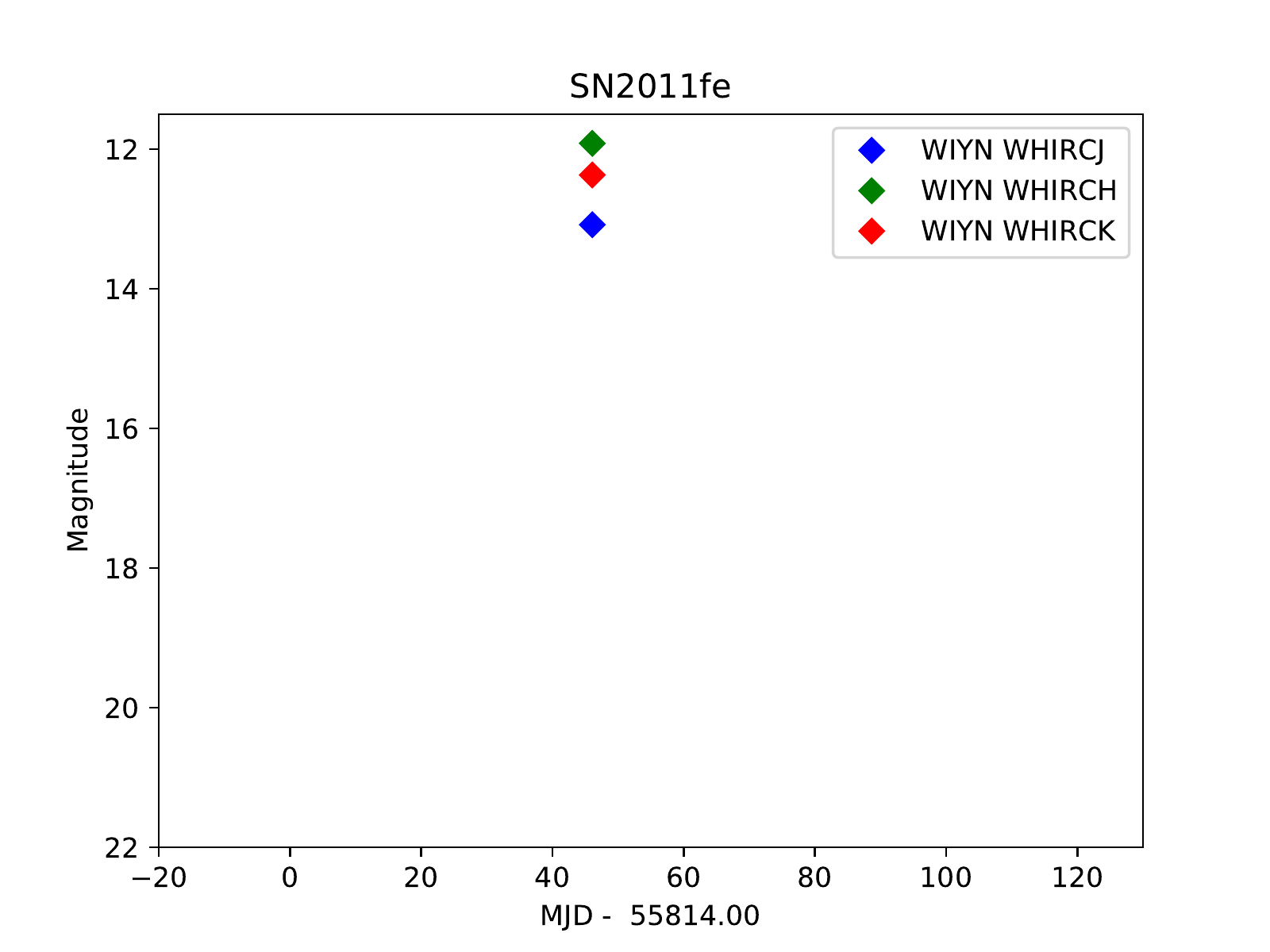}
\figsetgrpnote{SN2011fe}
\figsetgrpend

\figsetgrpstart
\figsetgrpnum{1.14}
\figsetgrptitle{SN2011fs}
\figsetplot{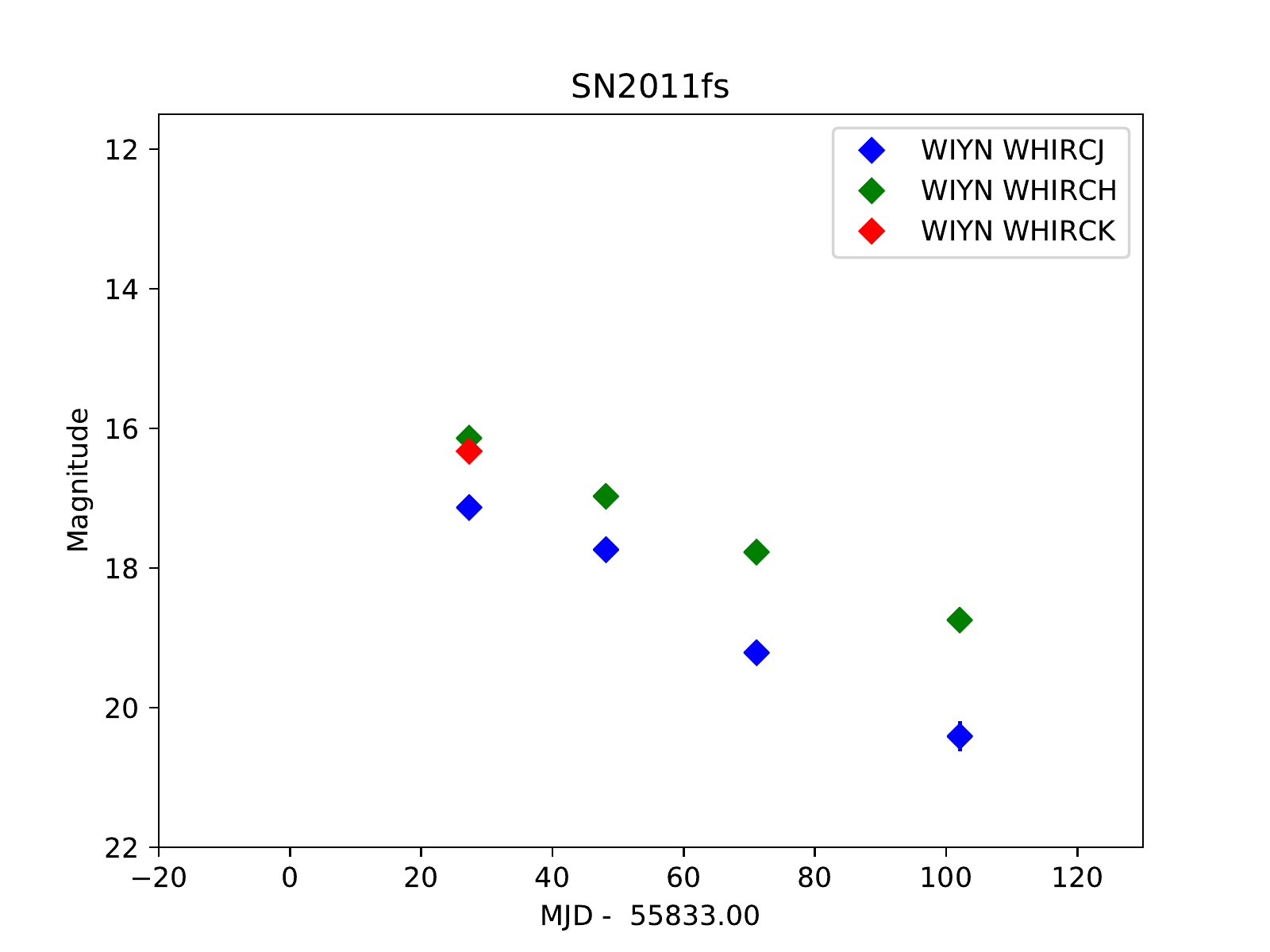}
\figsetgrpnote{SN2011fs}
\figsetgrpend

\figsetgrpstart
\figsetgrpnum{1.15}
\figsetgrptitle{SN2011gf}
\figsetplot{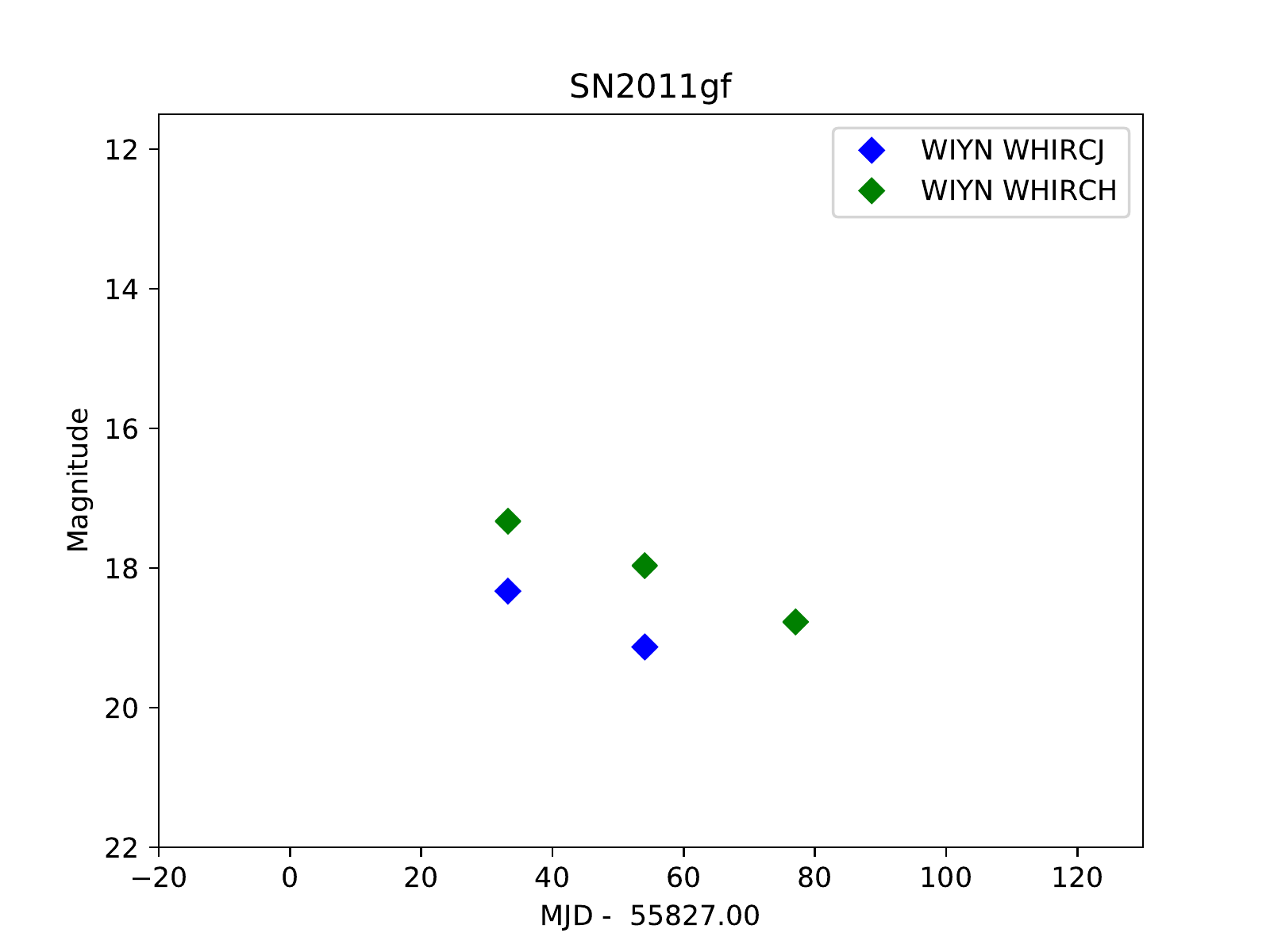}
\figsetgrpnote{SN2011gf}
\figsetgrpend

\figsetgrpstart
\figsetgrpnum{1.16}
\figsetgrptitle{SN2011io}
\figsetplot{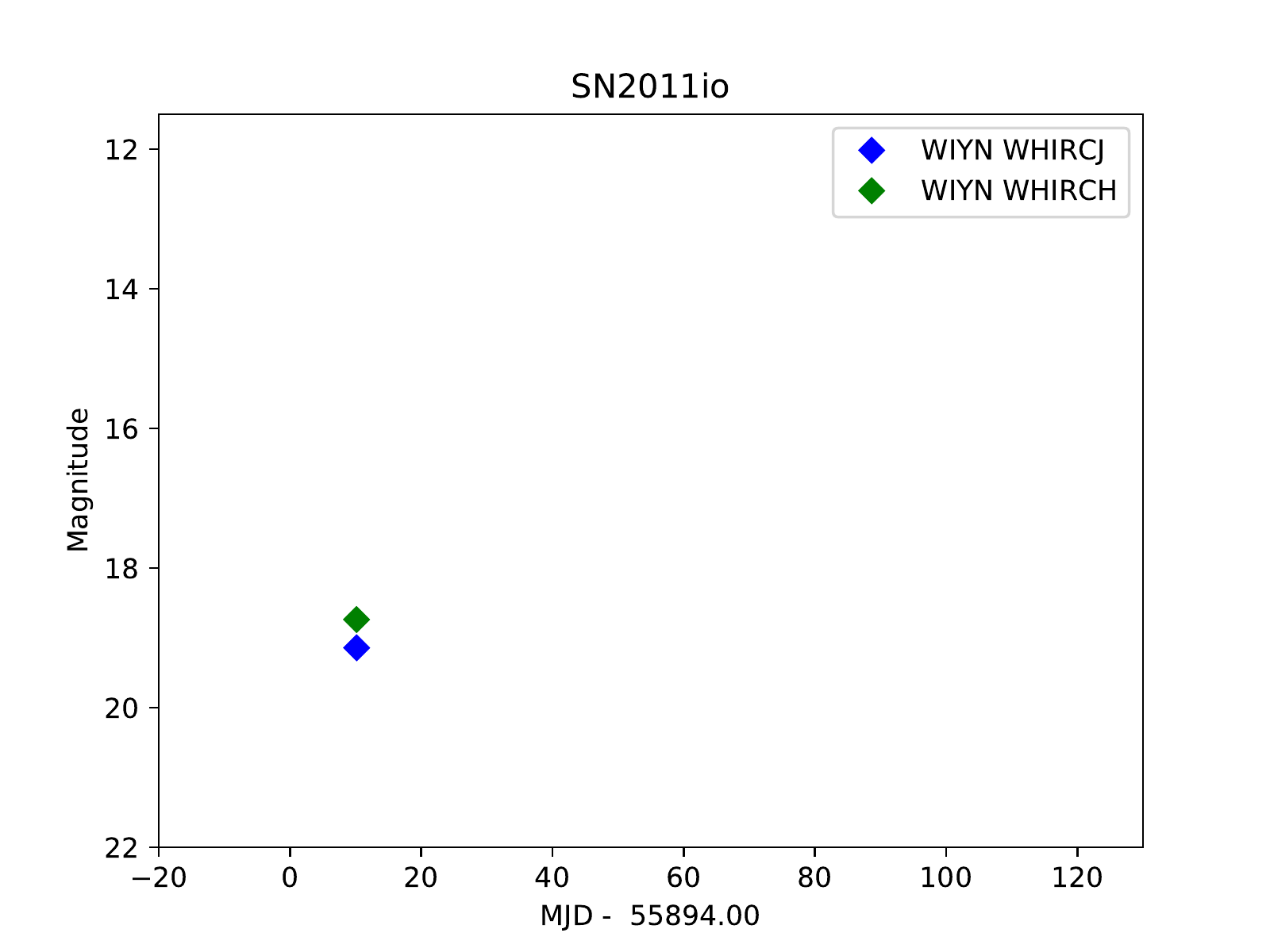}
\figsetgrpnote{SN2011io}
\figsetgrpend

\figsetgrpstart
\figsetgrpnum{1.17}
\figsetgrptitle{SN2011iy}
\figsetplot{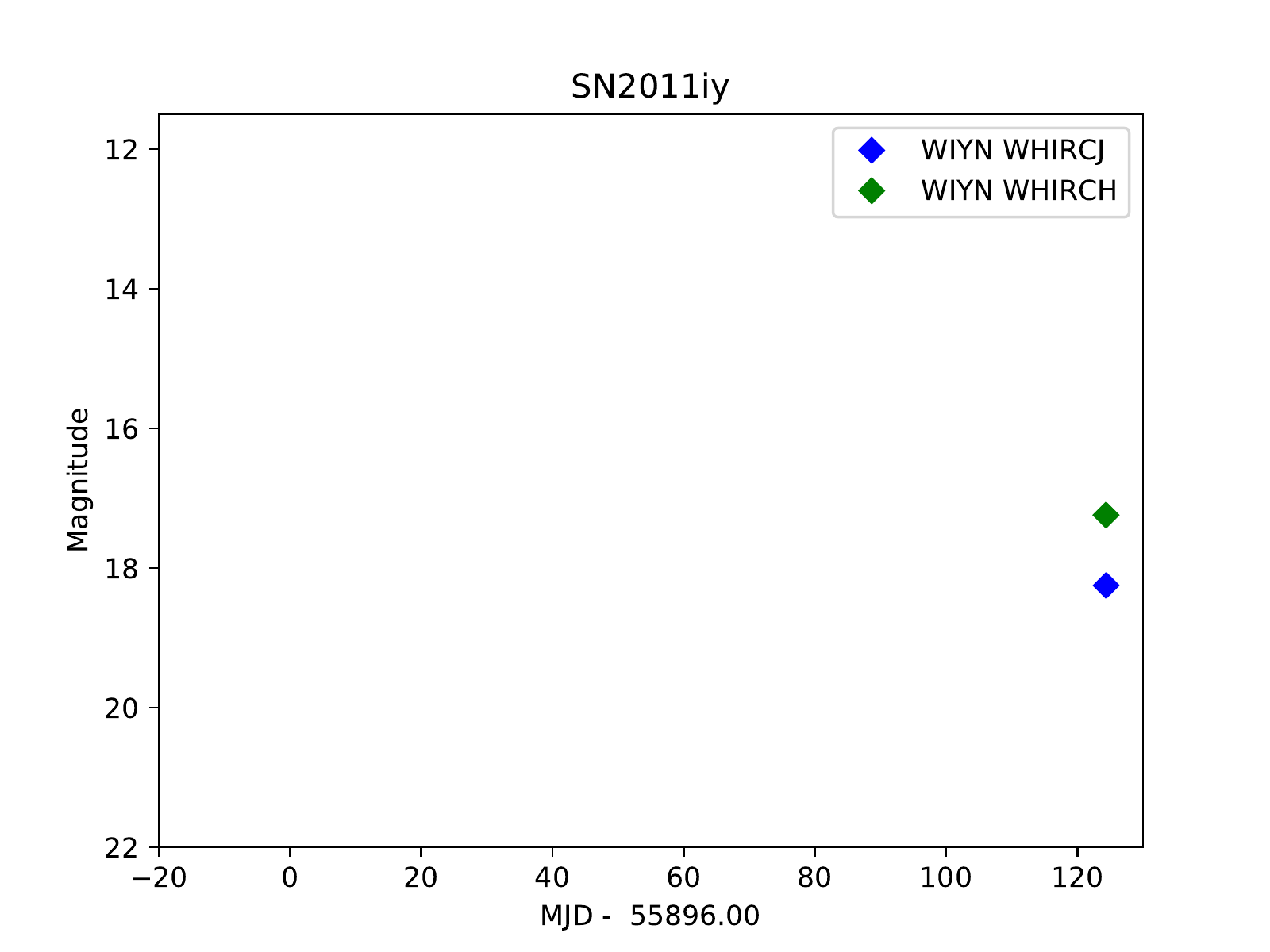}
\figsetgrpnote{SN2011iy}
\figsetgrpend

\figsetgrpstart
\figsetgrpnum{1.18}
\figsetgrptitle{SN2011jh}
\figsetplot{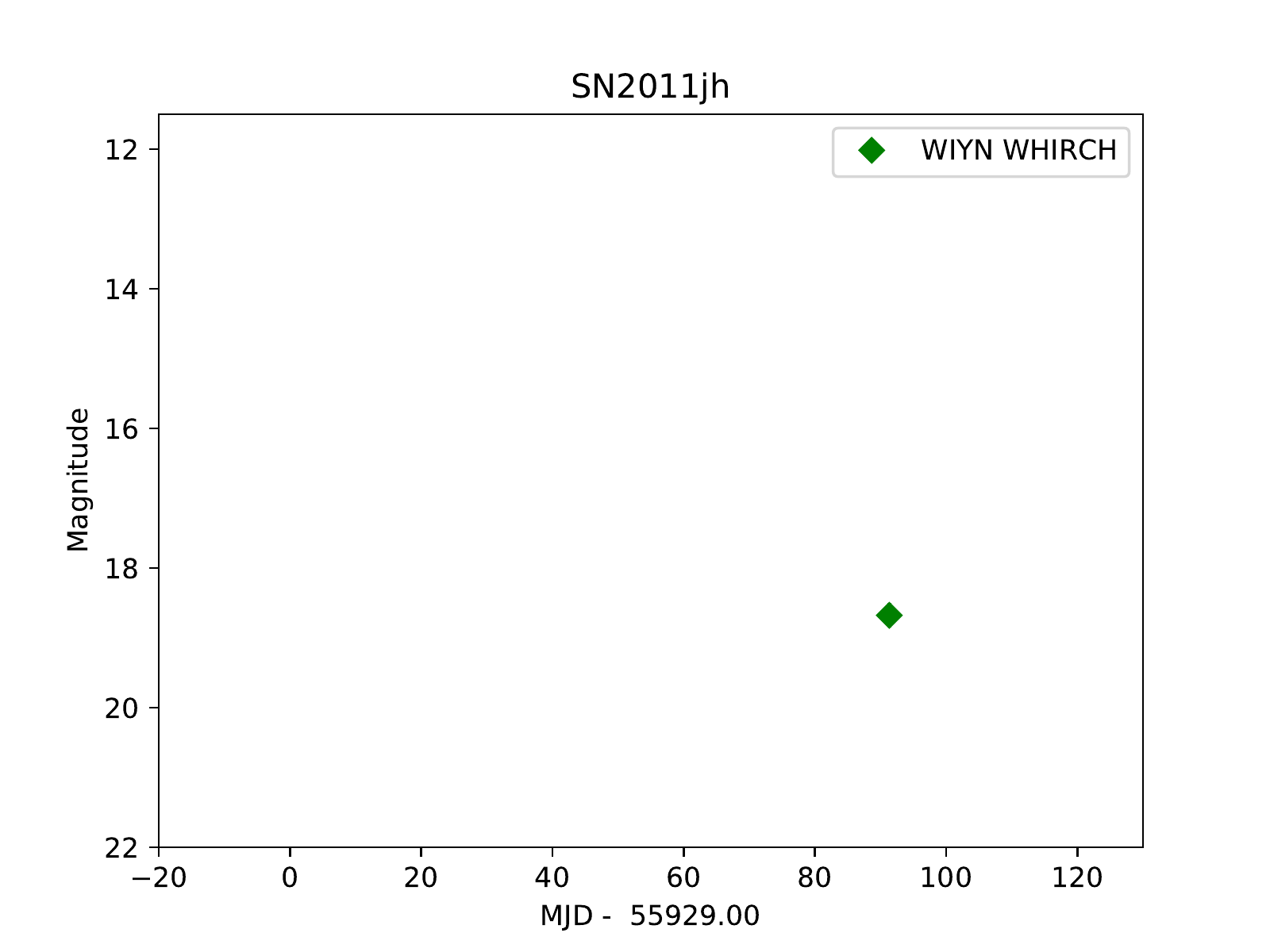}
\figsetgrpnote{SN2011jh}
\figsetgrpend

\figsetgrpstart
\figsetgrpnum{1.19}
\figsetgrptitle{SN2012bh}
\figsetplot{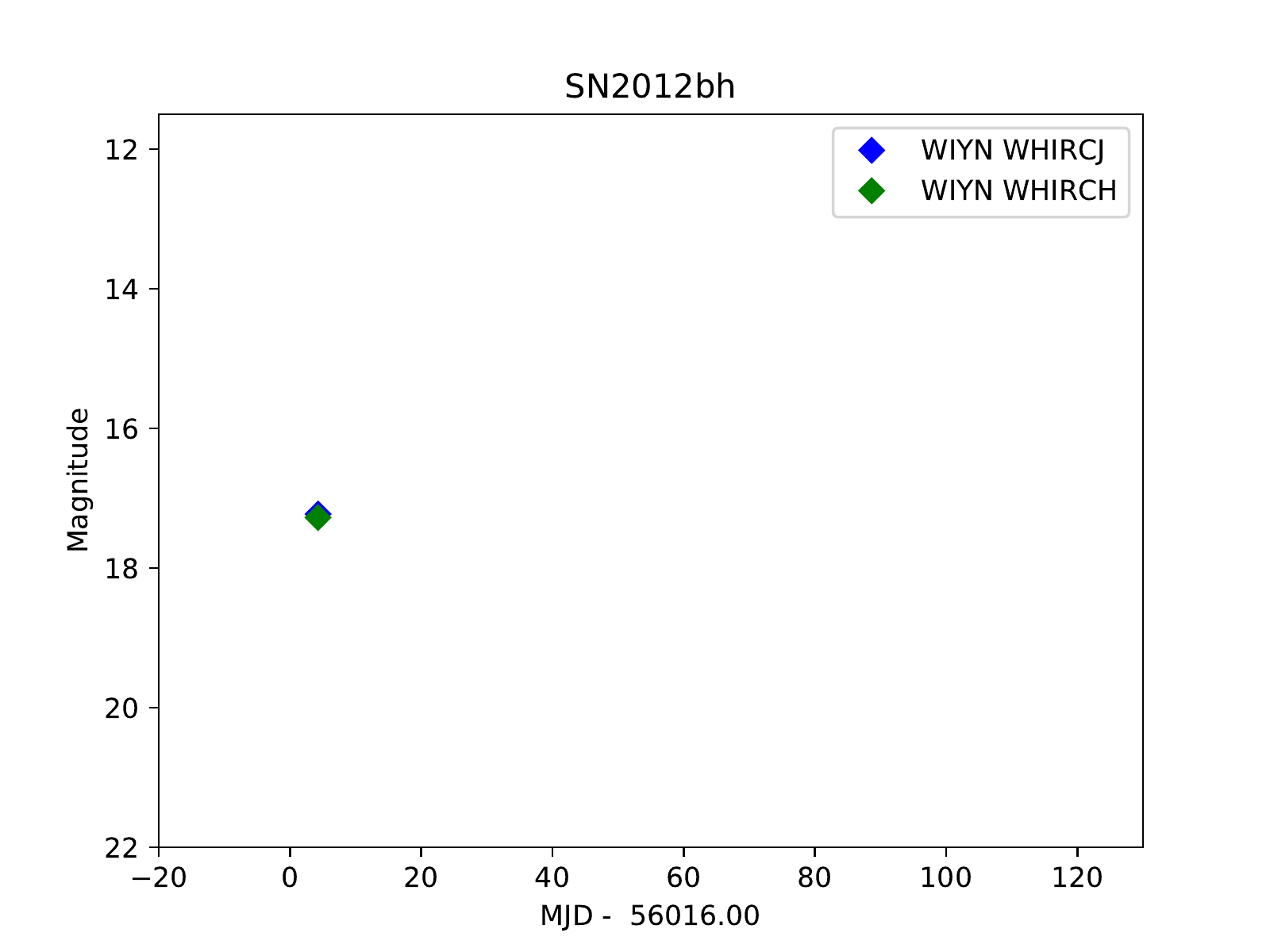}
\figsetgrpnote{SN2012bh}
\figsetgrpend

\figsetgrpstart
\figsetgrpnum{1.20}
\figsetgrptitle{SN2012bo}
\figsetplot{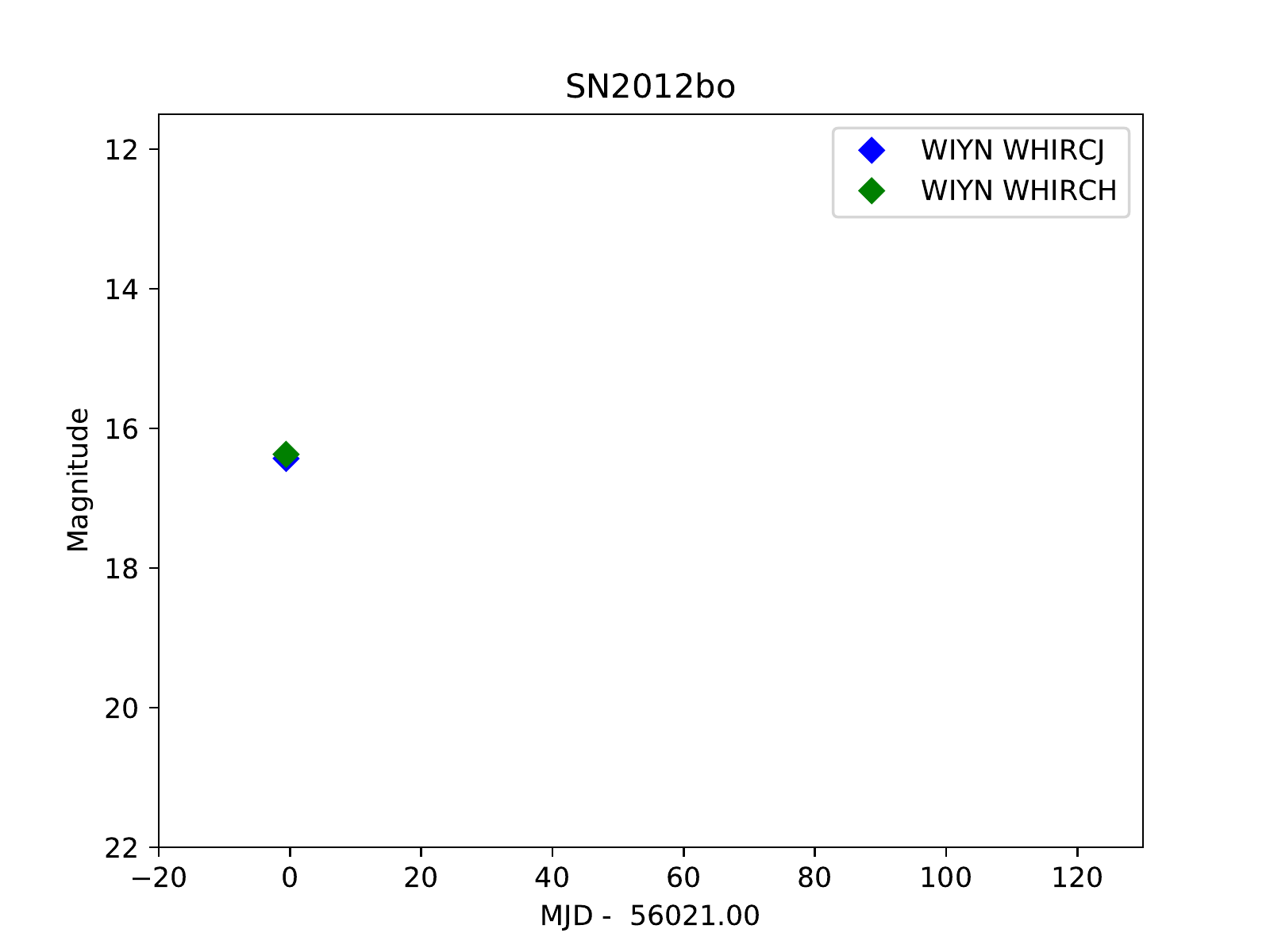}
\figsetgrpnote{SN2012bo}
\figsetgrpend

\figsetgrpstart
\figsetgrpnum{1.21}
\figsetgrptitle{SN2012bp}
\figsetplot{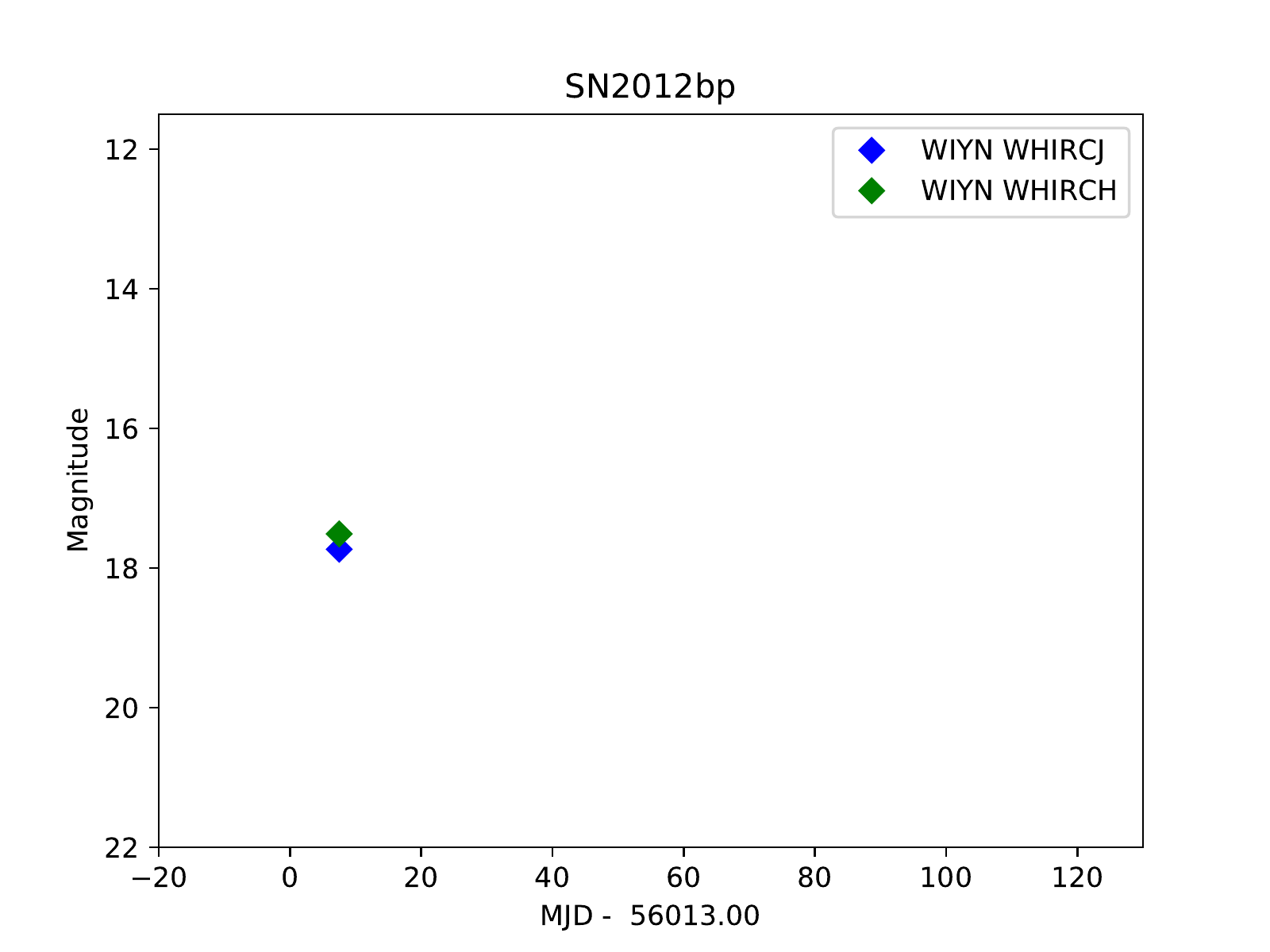}
\figsetgrpnote{SN2012bp}
\figsetgrpend

\figsetgrpstart
\figsetgrpnum{1.22}
\figsetgrptitle{SN2012em}
\figsetplot{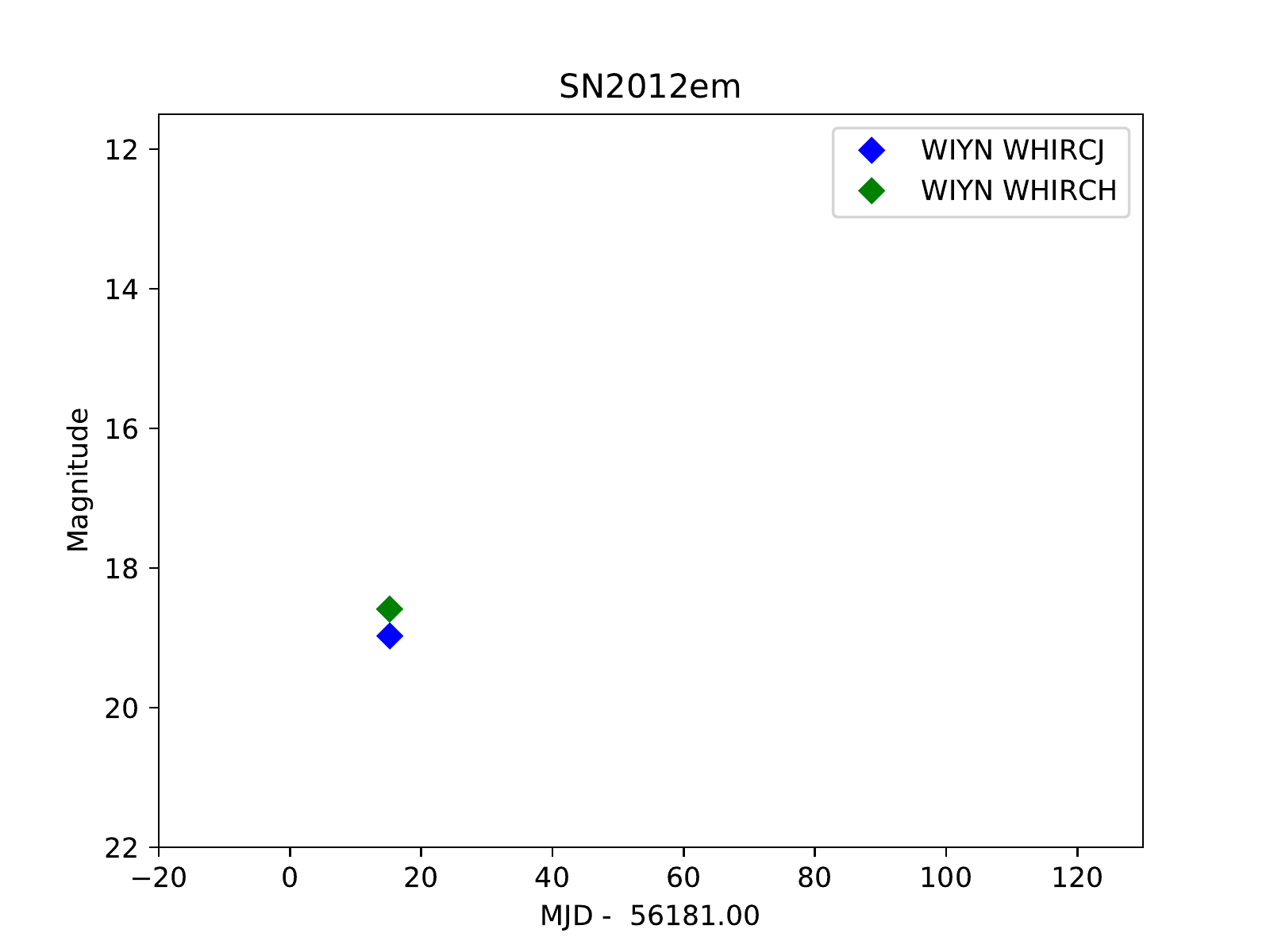}
\figsetgrpnote{SN2012em}
\figsetgrpend

\figsetgrpstart
\figsetgrpnum{1.23}
\figsetgrptitle{SN2012fk}
\figsetplot{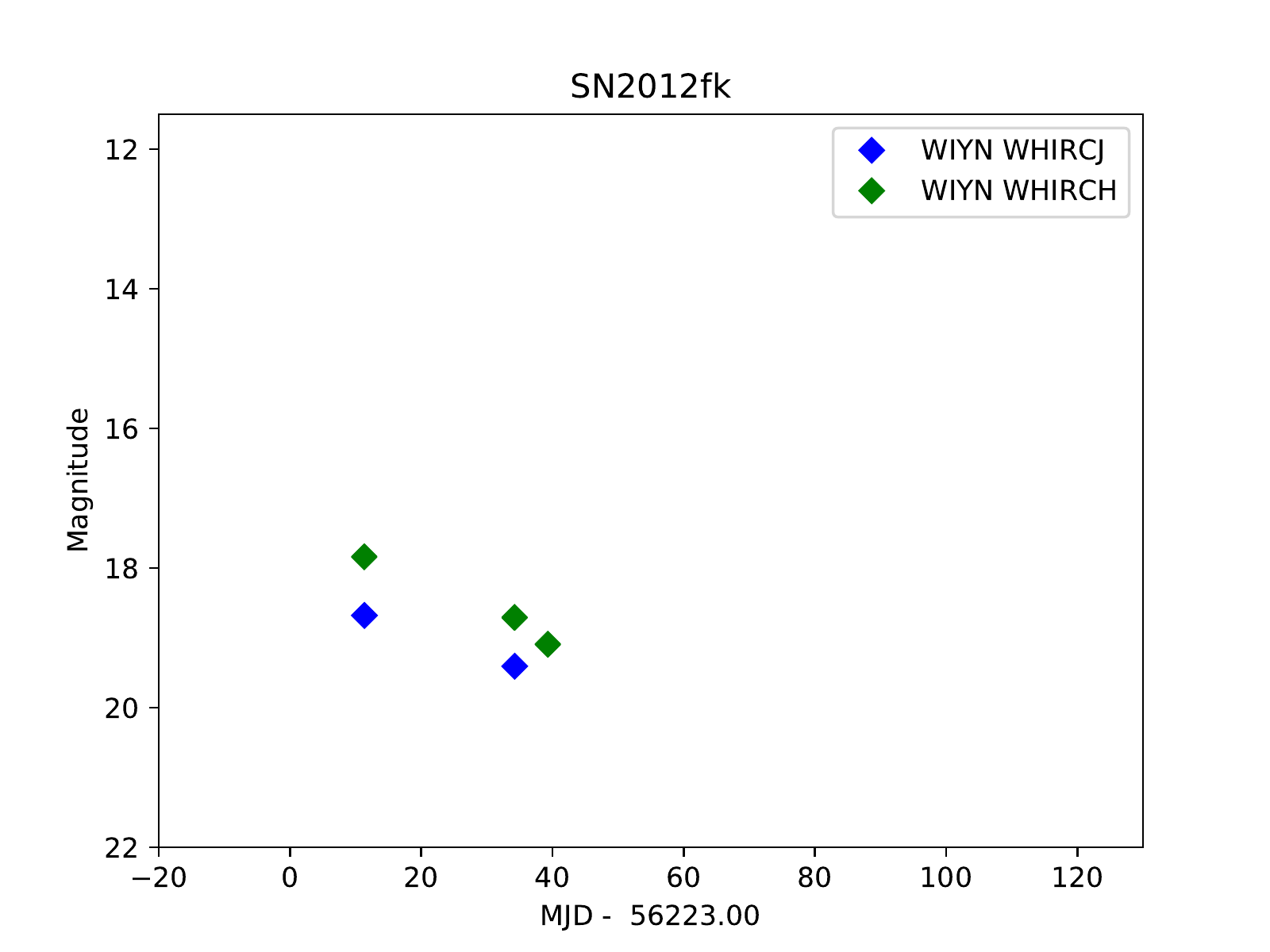}
\figsetgrpnote{SN2012fk}
\figsetgrpend

\figsetgrpstart
\figsetgrpnum{1.24}
\figsetgrptitle{SN2012fr}
\figsetplot{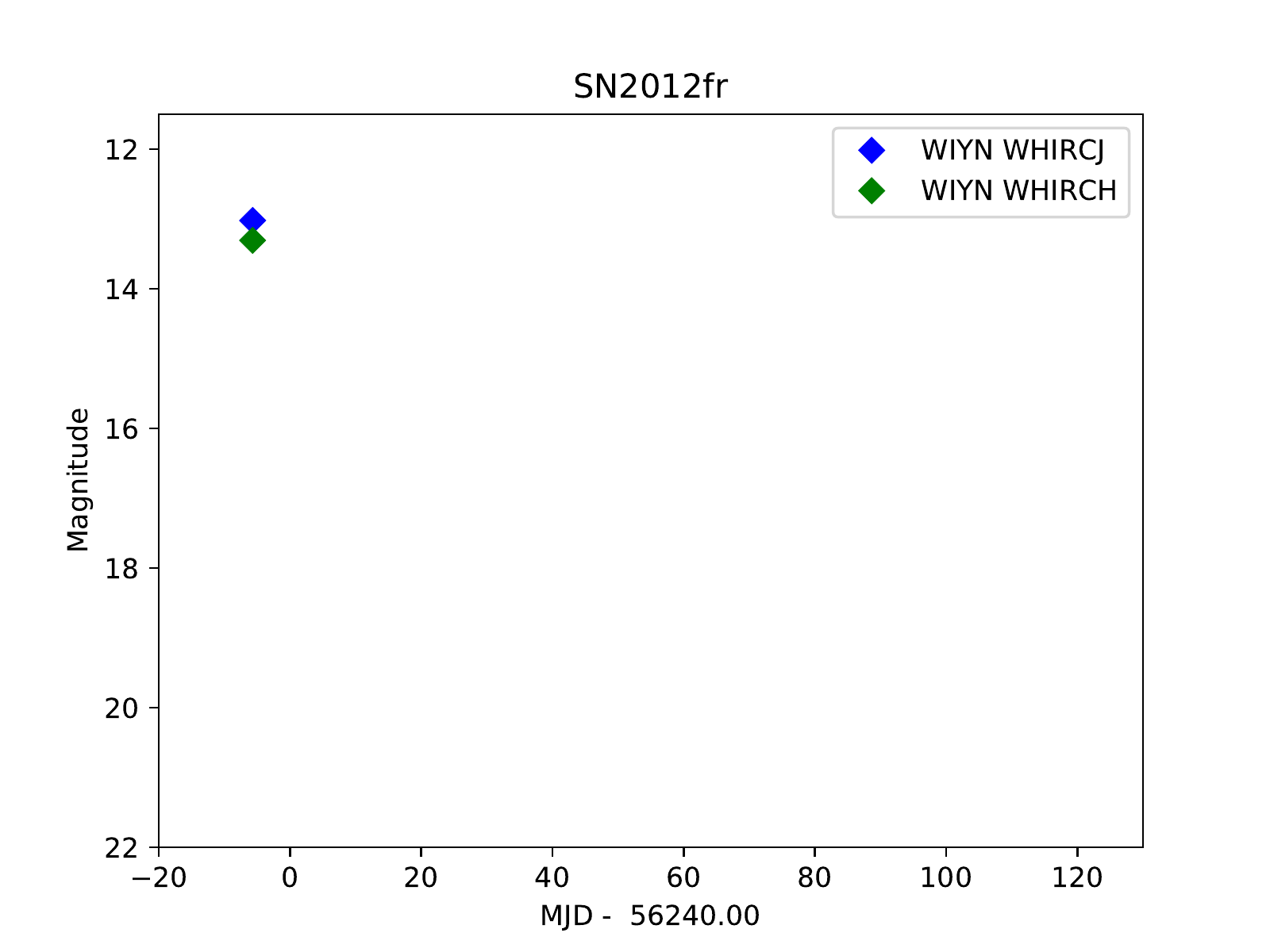}
\figsetgrpnote{SN2012fr}
\figsetgrpend

\figsetgrpstart
\figsetgrpnum{1.25}
\figsetgrptitle{SN2013ar}
\figsetplot{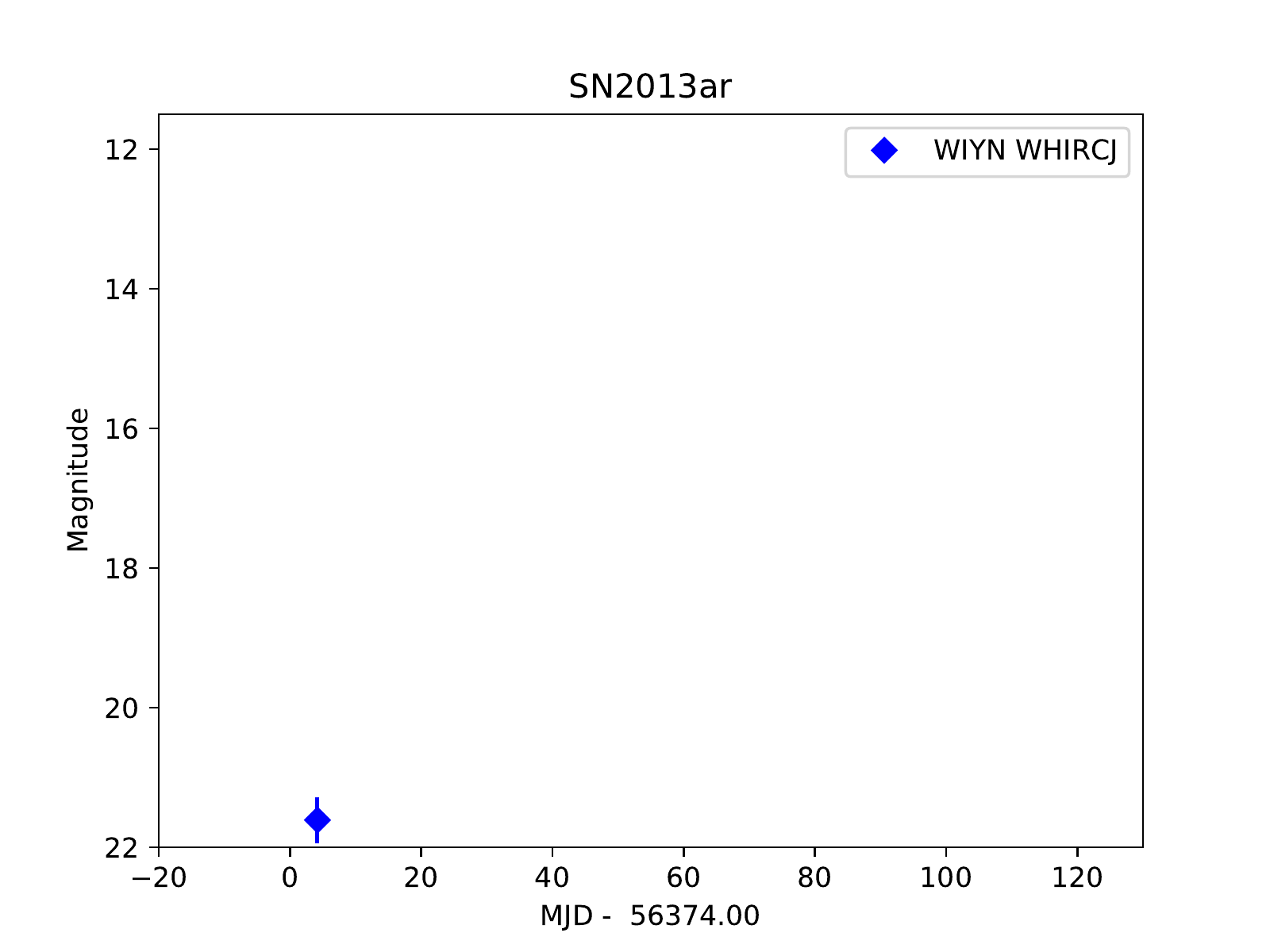}
\figsetgrpnote{SN2013ar}
\figsetgrpend

\figsetgrpstart
\figsetgrpnum{1.26}
\figsetgrptitle{SN2013bs}
\figsetplot{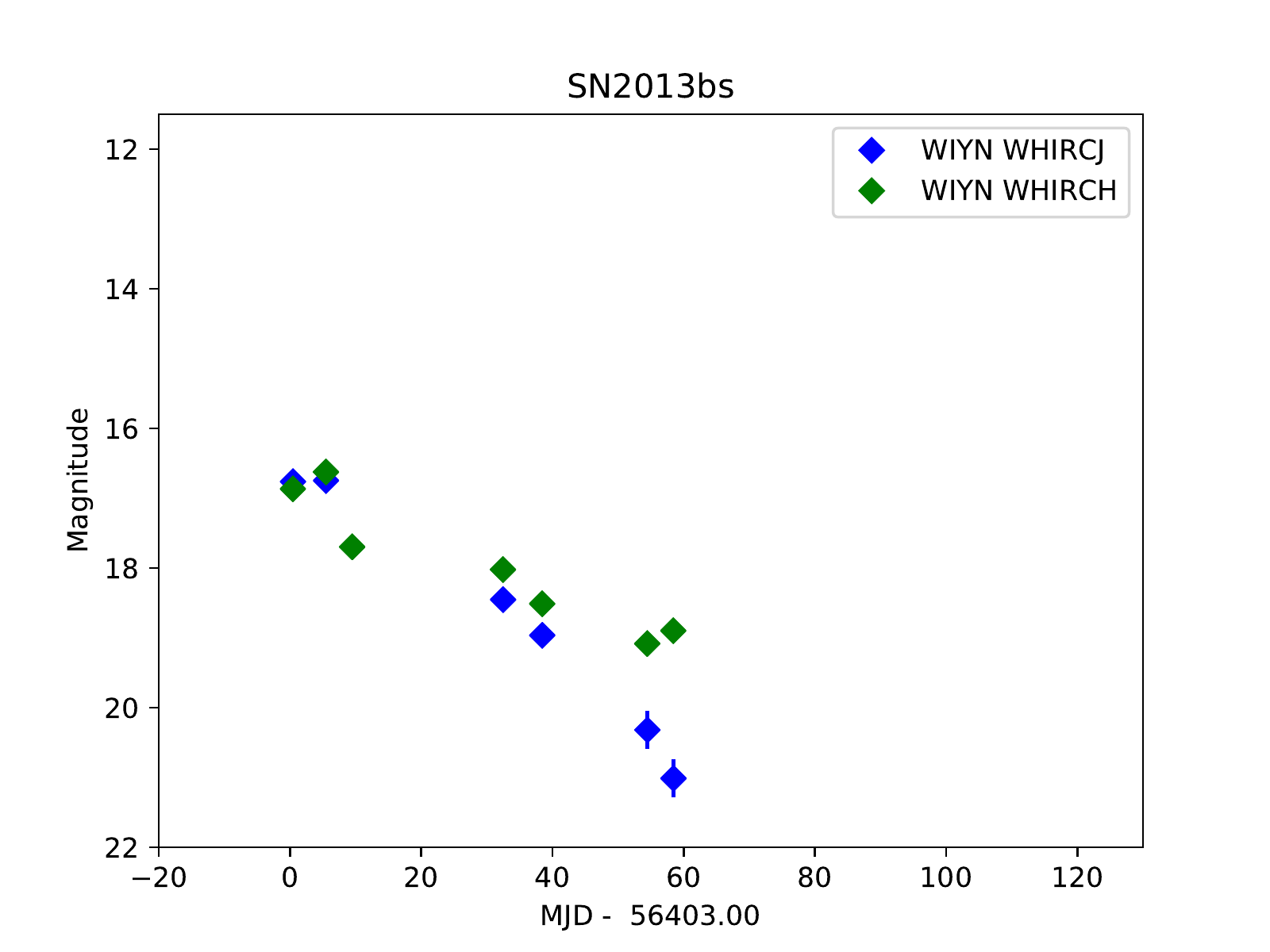}
\figsetgrpnote{SN2013bs}
\figsetgrpend

\figsetgrpstart
\figsetgrpnum{1.27}
\figsetgrptitle{SN2013bt}
\figsetplot{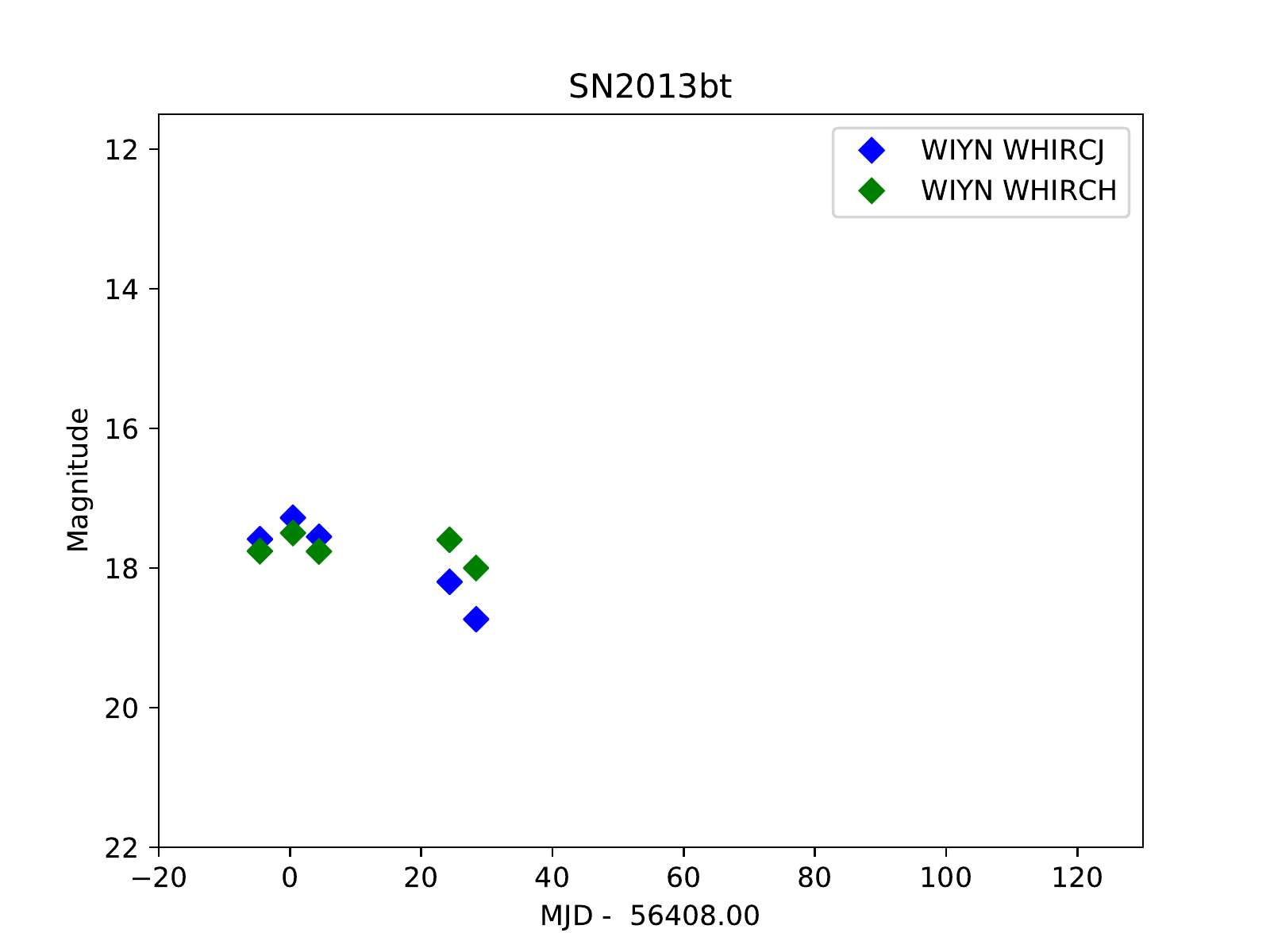}
\figsetgrpnote{SN2013bt}
\figsetgrpend

\figsetgrpstart
\figsetgrpnum{1.28}
\figsetgrptitle{SN2013cs}
\figsetplot{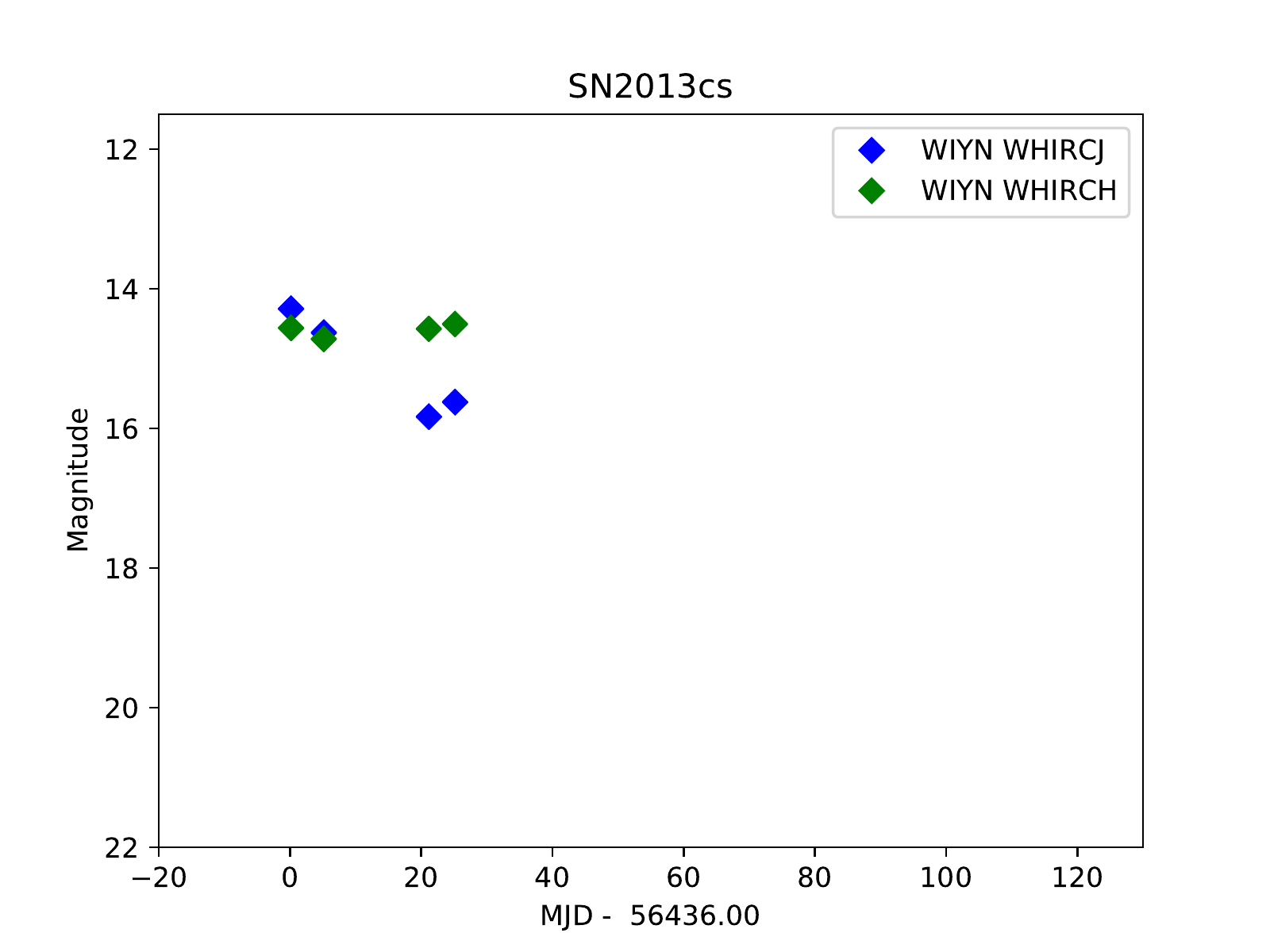}
\figsetgrpnote{SN2013cs}
\figsetgrpend

\figsetgrpstart
\figsetgrpnum{1.29}
\figsetgrptitle{SN2013fn}
\figsetplot{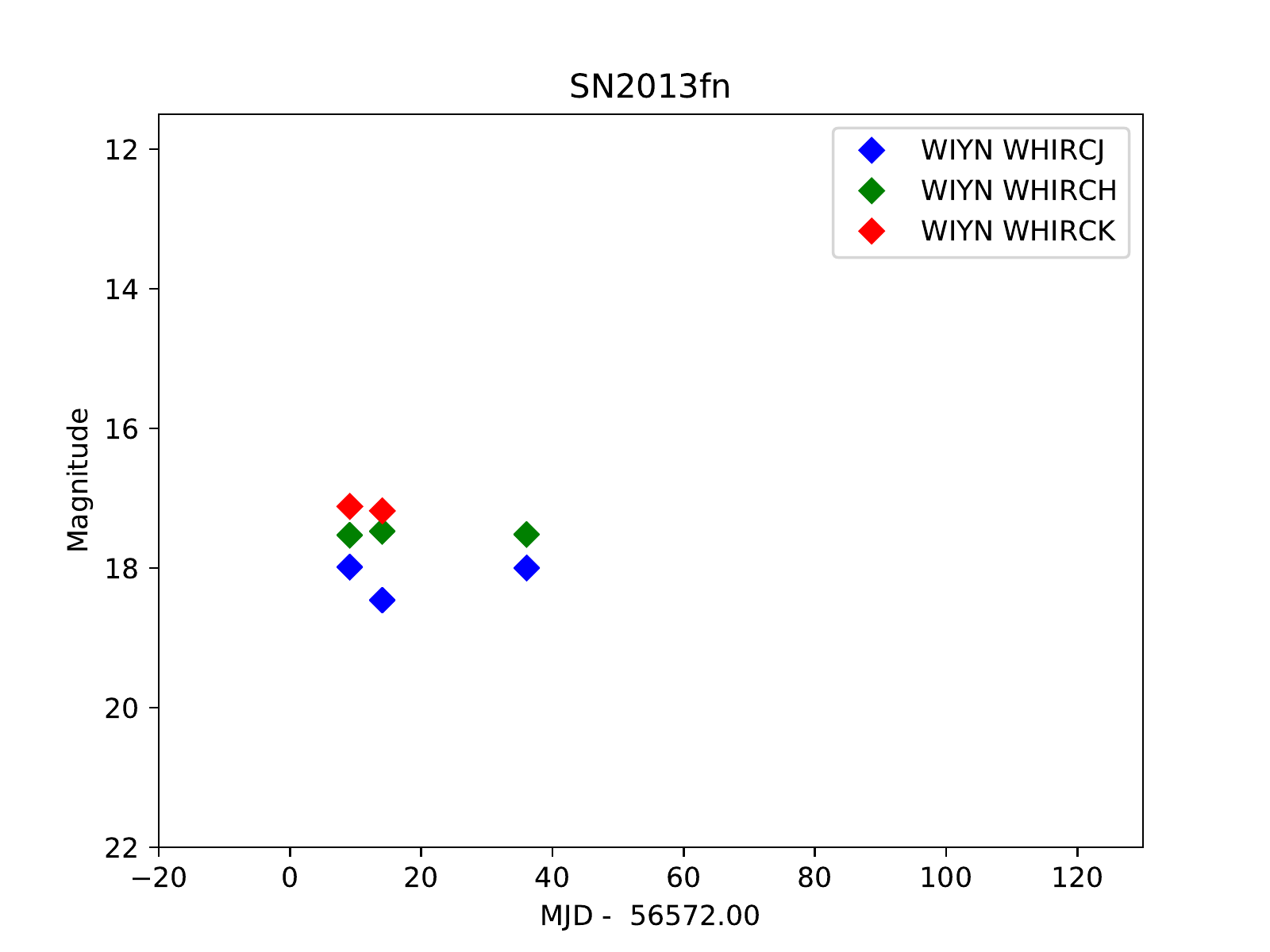}
\figsetgrpnote{SN2013fn}
\figsetgrpend

\figsetgrpstart
\figsetgrpnum{1.30}
\figsetgrptitle{iPTF13dge}
\figsetplot{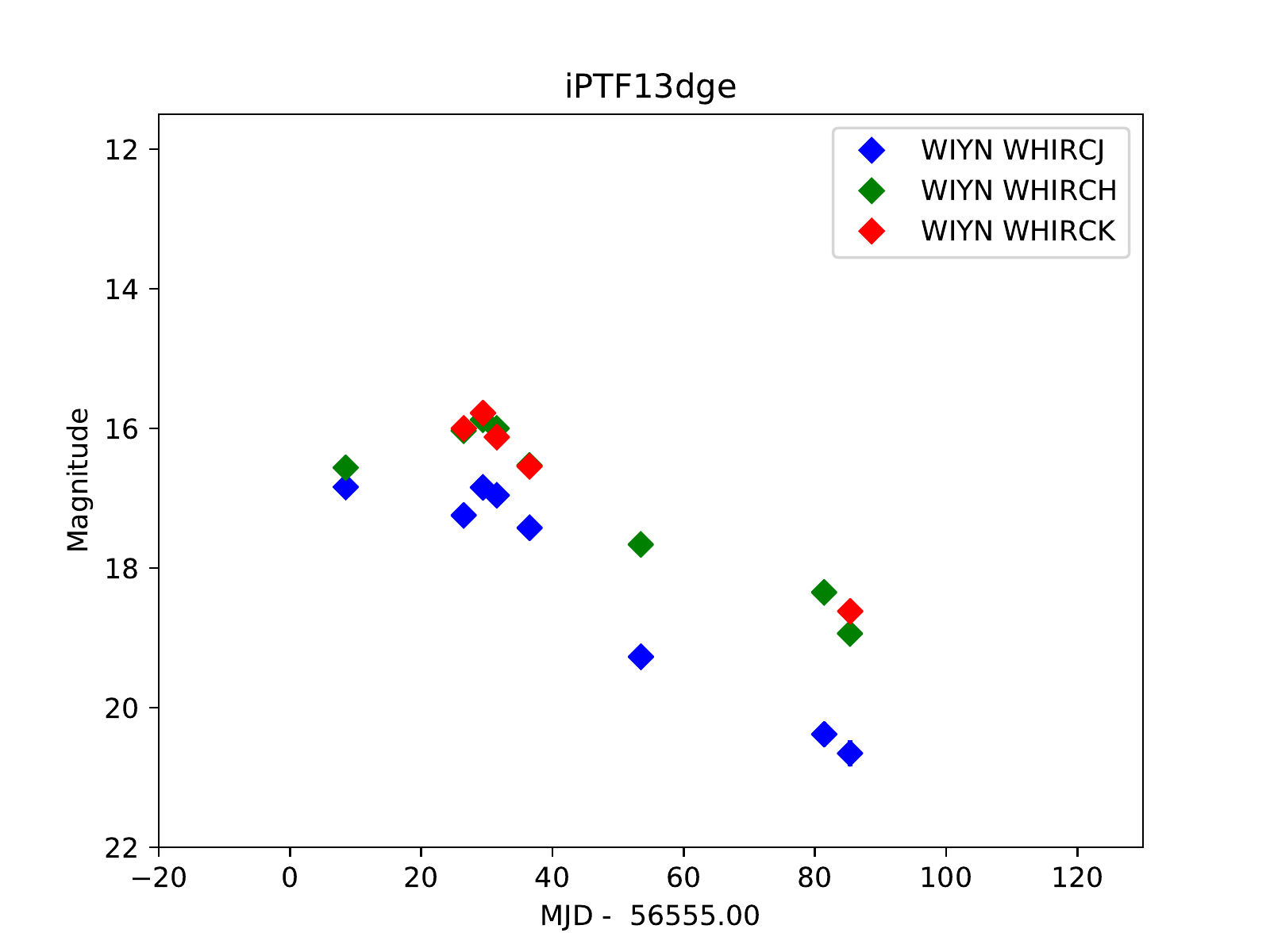}
\figsetgrpnote{iPTF13dge}
\figsetgrpend

\figsetgrpstart
\figsetgrpnum{1.31}
\figsetgrptitle{iPTF13dkl}
\figsetplot{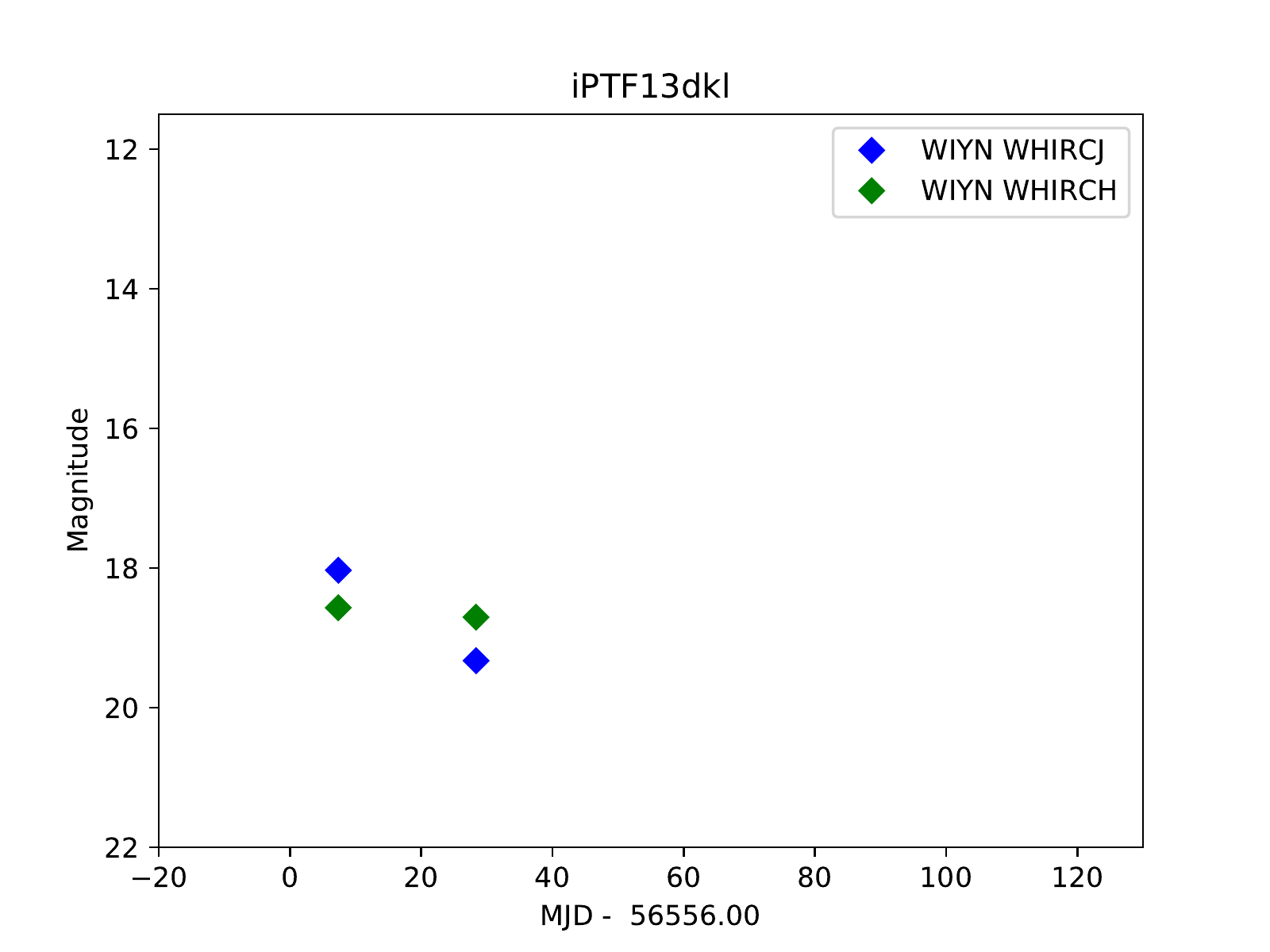}
\figsetgrpnote{iPTF13dkl}
\figsetgrpend

\figsetgrpstart
\figsetgrpnum{1.32}
\figsetgrptitle{iPTF13dkx}
\figsetplot{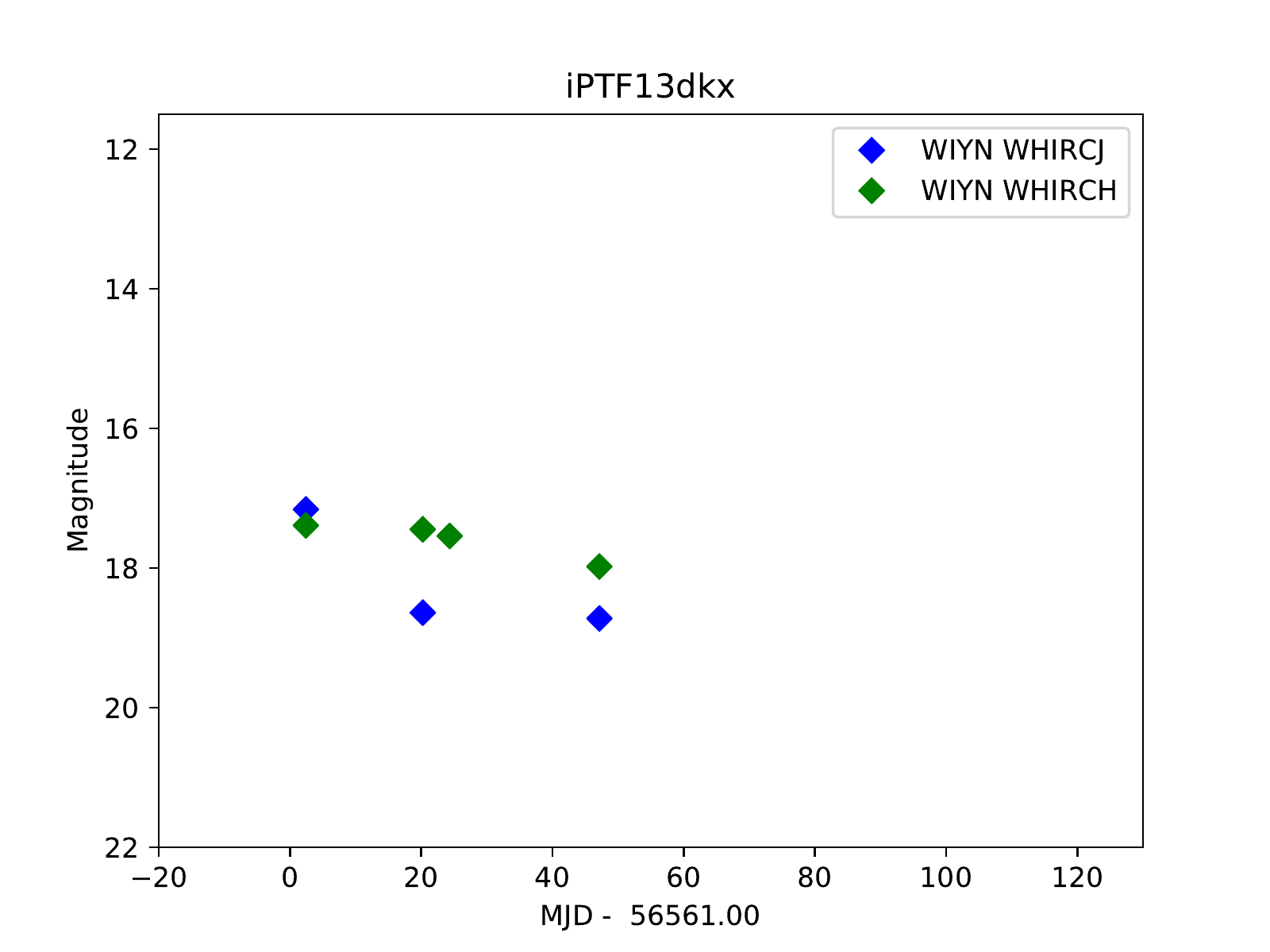}
\figsetgrpnote{iPTF13dkx}
\figsetgrpend

\figsetgrpstart
\figsetgrpnum{1.33}
\figsetgrptitle{iPTF13ebh}
\figsetplot{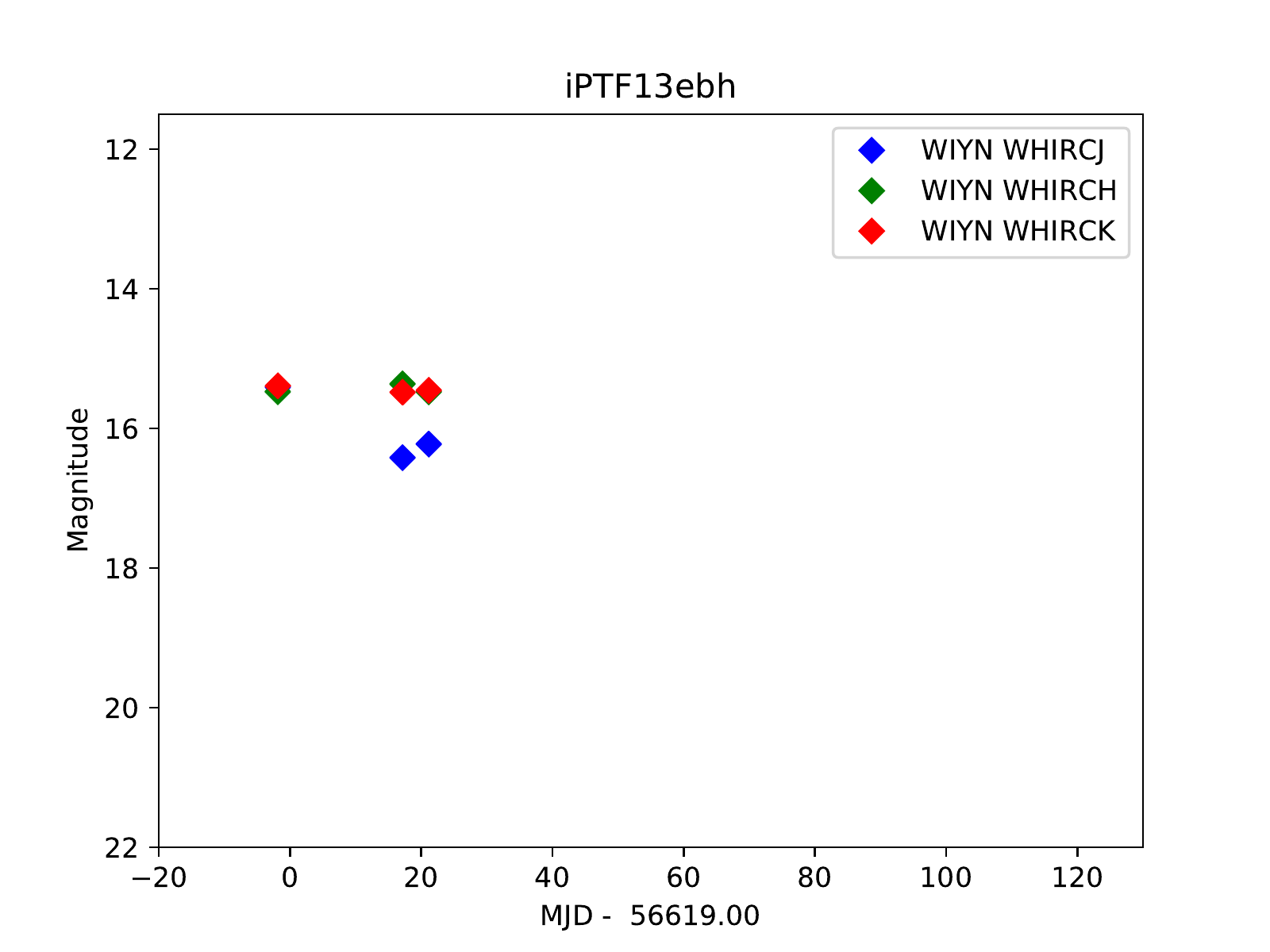}
\figsetgrpnote{iPTF13ebh}
\figsetgrpend

\figsetend
\gridline{\fig{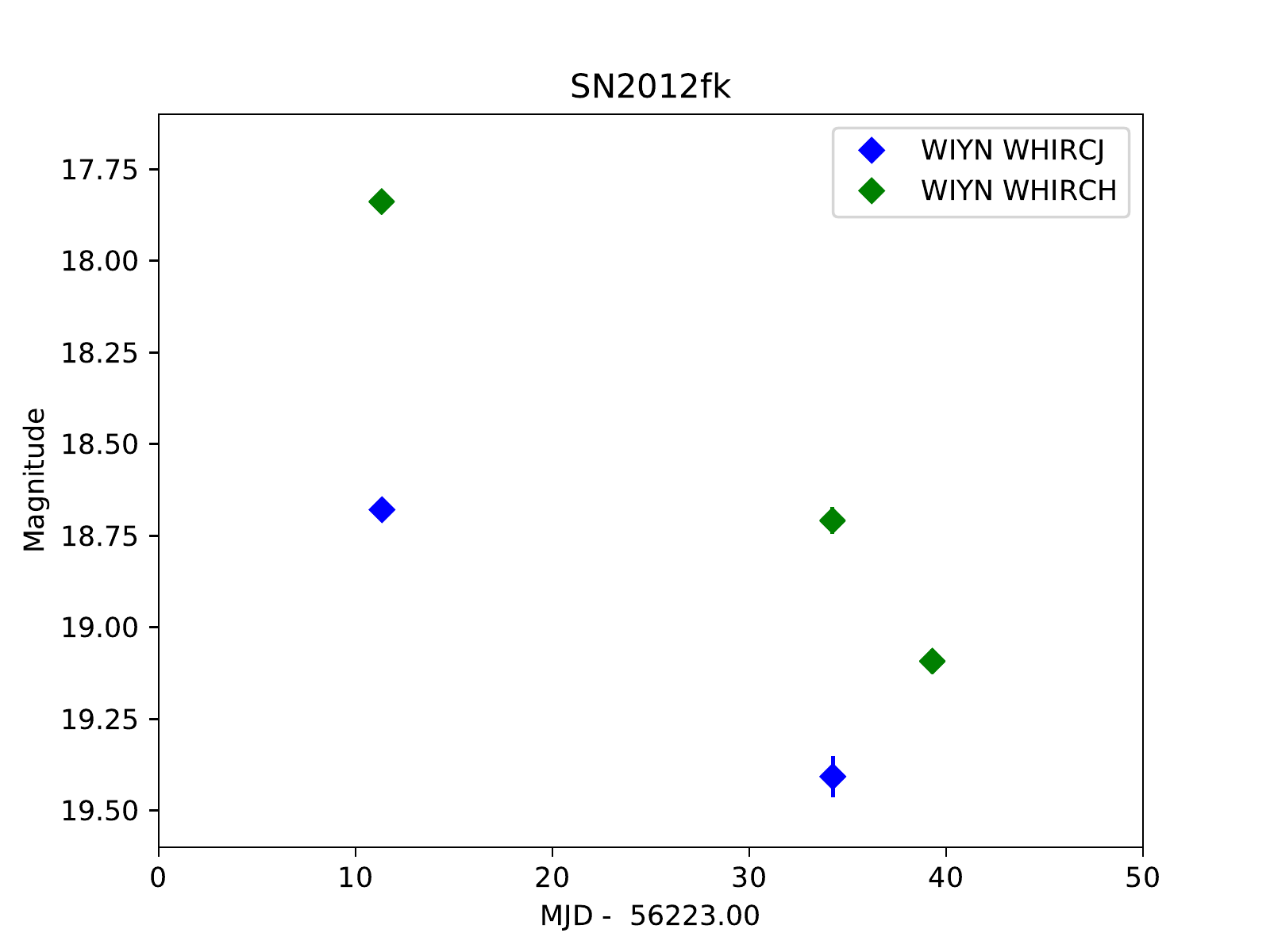}{0.3\textwidth}{}\fig{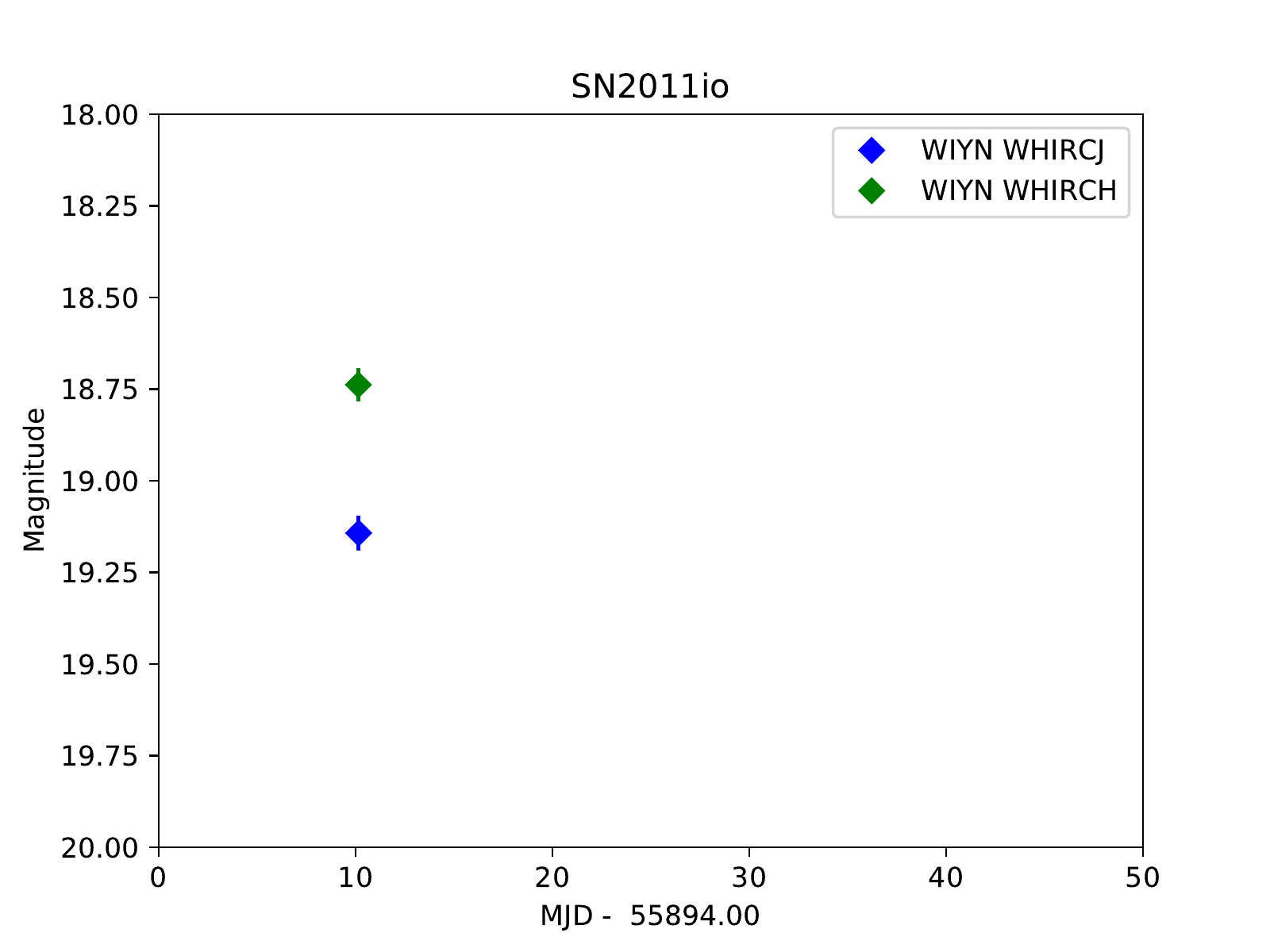}{0.3\textwidth}{}\fig{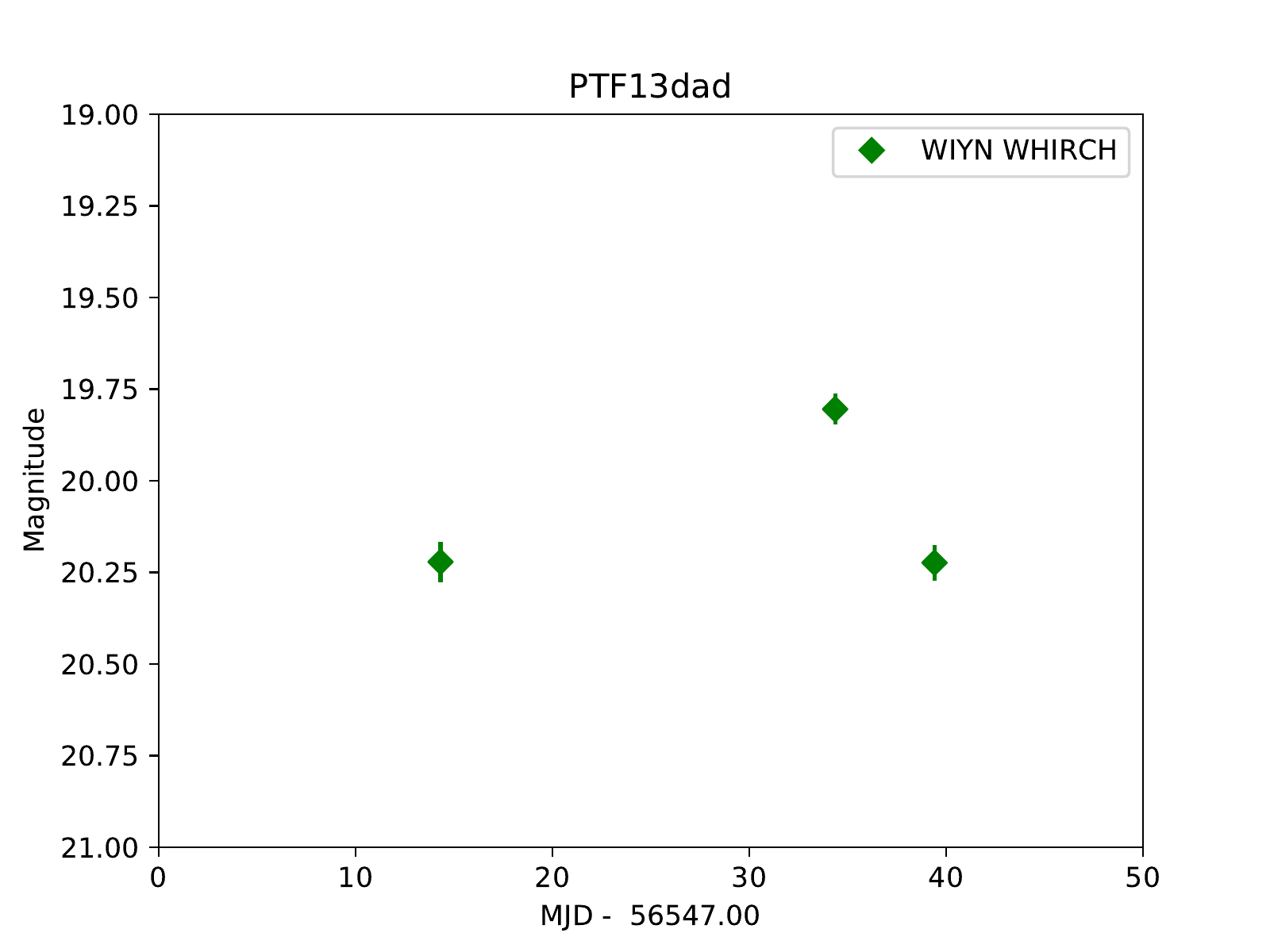}{0.3\textwidth}{}}
\caption{
Lightcurves for the \numLc\ \sneia\ with lightcurves presented in this paper.
The individual lightcurve plots have been roughly normalized with respect
to an estimated $B$-band maximum.
The time of $B$-band maximum light was taken as that reported from the spectral classification in the respective ATel or CBET.
We here show three representative lightcurves: SN~2012fk ($z=0.035$), SN~2011io ($z=0.04$), and PTF13dad ($z=0.086$).
The complete figure set of \numLc\ images is available in the online journal.
}
\label{fig:lightcurves}
\end{figure}

The lightcurves for these \numLc\ \sneia\ are presented in Figure~\ref{fig:lightcurves}.
We have obtained template images of the host galaxies for the remaining 41 \sneia\ over the 2013A--2017A semesters.  Lightcurves for these supernovae will be presented in a future data release.

\section{Access}\label{sec:access}

We here detail the processing and provide specific information about file names and structure that describe the data as released.  Raw data are available through the NOAO Science Archive.  Processed data -- stacks, catalogs, and lightcurves -- are available at \url{http://www.phyast.pitt.edu/\~wmwv/SweetSpot/DR1\_data/}.  Subdirectories listed below are with respect to this URL.

\begin{itemize}
	\item  {\bf raw data} The raw data, both science data and calibration frames, are available through the standard interface to the NOAO Science Archive.  All raw data presented in this data release have been now released as non-proprietary as the standard NOAO 18-month waiting period has elapsed.
	\item  {\bf stacks}  We provide coadded stacked images for each observation which succeeded in processing through to this stage.  This includes some stacks with poor WCS solutions -- see Table~\ref{table:stackinspection} for details.  \\
    These images are named as \file{FIELD\_SET\_FILTER\_YYYYMMDD.fits}.  E.g., \file{iPTF13dkx\_A\_H\_20130927.fits} is an observation of the field of the supernova iPTF13dkx, the first set of observations during that night (``A''), in filter $H$, taken on 2013-09-27.
Similarly \file{P212-C\_B\_KS\_20130420.fits} is an observation of the field of the standard star ``P212-C'', the the second such observation during the night (``B''), in filter $K_s$, taken on 2013-04-20.
A ``set'' is defined for each observation at a different airmass during the night.  For almost all SN targets this will always be ``A'' as observations were carried out contiguously.  However, we intentionally observed standard star fields at up to three different airmasses during a night, and each of these will be assigned a different set: ``A'', ``B'', or ``C''.
The effective exposure time for the coadded image varies discretely across the image due to the steps in the dithering (see Figure~\ref{fig:weightmap}).  The exposure map image is provided as an accompanying file with the same name, but with a suffix of \file{.expmap.fits}.  These FITS images are available in the \file{stacks} subdirectory.
    \item  {\bf uncalibrated catalogs}  We provide the astrometrically calibrated but photometrically {\it uncalibrated} catalogs for each image stack that passed basic quality control.
    They are named following the image they were generated from.  E.g., \file{iPTF13dkx\_A\_H\_20130927\_photometry.fits}.  These are catalogs are available as FITS binary tables in the \file{photometry\_catalogs} directory.
    \item  {\bf } We provide summary files of our photometrically {\it uncalibrated} measurements of all 2MASS stars as \\
    \file{star\_catalogs/ALL\_2MASS\_star\_measurements.fits} and the subset of our 71 measurements of the 6 Persson standard stars as \file{Persson\_standards/ALL\_Persson\_star\_measurements.fits}.  Per-night compilations of 2MASS stars are available in \file{star\_catalogs}, while per-object compilations of the Persson standards are available in \file{Persson\_standards}.
    \item  {\bf calibrated catalogs}  We provide astrometrically and photometrically calibrated catalogs for each image stack that passed basic quality control.  They are named as calibrated photometry following the image they were generated from.  E.g., \file{iPTF13dkx\_A\_H\_20130927\_calibrated\_photometry.fits}.  These are available as FITS binary tables in the \file{catalogs} subdirectory.
	\item  {\bf star lightcurves}  We provide as individual files the lightcurves of all 2MASS stars in these images.  These lightcurve files are in the ``Wstd'' format used in ESSENCE project \citet{Miknaitis07}, and have names of the form \file{SN2012et\_2MASS23423292+2705046.Wstd.dat}.  The first token ``SN2012et'' denotes the field, while the second denotes the target by its 2MASS star name, ``2MASS23423292+2705046'' which also encodes the 2MASS RA, Dec.  These text files are available in the \file{lightcurves} subdirectory.
    \item  {\bf SN lightcurves}  We provide lightcurves of all the supernovae that we determined to be sufficiently separated from any significant contamination from their host galaxy.  These files are named in the style of ``SN2012et\_SN2012et.Wstd.dat'', for the field ``SN2012et'' and the target ``SN2012et''.  While this is somewhat redundant, it provides a consistent scheme between the stars and the supernova associated with the same field.  We also provide a version of the SN lightcurves in the ``SNPY'' format suitable for import into the SNooPy light-curve fitting framework.  These light-curve files contain the same information as presented here in Table~\ref{table:lightcurves}.  These text files are available in the \file{lightcurves} subdirectory.

\end{itemize}
\section{Discussion and Future Work}\label{sec:discussion}

The present sample of \numLc~\sneia\ presented here provide a first step toward (1) further quantifying the standard NIR luminosity of \sneia\ and (2) investigating correlations of \sneia\ luminosity with host-galaxy properties.
Adding together the 3\% uncertainty in peak flux from host-galaxy contamination, 0.03~mag field-to-field calibration uncertainty, and an overall system 0.02~mag average calibration to 2MASS in quadrature results in a total systematic uncertainty of $\sim$0.05~mag.\footnote{While the host-galaxy template uncertainty is a constant in {\em flux} and the calibration is a constant in {\em magnitude}, we combine them here into a total in magnitude for the purpose of estimation of the size of the total effect.}
For comparison, the distance modulus uncertainty introduced by a peculiar velocity dispersion of 150~km/s is $\sim(0.05, 0.02)$~mag at $z=(0.02, 0.05)$.  Thus, while the lightcurve will improve in quality with host-galaxy subtraction and in number by including all of the supernovae currently excluded due to excessive host-galaxy contamination, the current data presented here are useful to begin studies on the standardness of \sneia\ in the NIR.

Results on correlations of NIR luminosity with host-galaxy properties including these \numLc~\snia lightcurves were first presented in \citet{Ponder17} and will be further developed in \citet{Ponder18}.

The full goals of the SweetSpot program are to:
\begin{enumerate}
\item Quantify the standard brightness of \sneia\ in the NIR by populating the NIR Hubble diagram in a regime ($0.02<z<0.08$) less affected by peculiar motions than previous NIR work.
\item Determine if current optical Hubble residual correlations with host-galaxy properties are also present in the NIR.
\item Improve our understanding of the intrinsic optical and NIR colors of \sneia.
\item Improve our understanding of dust in the host galaxies of \sneia.
\end{enumerate}

(1) Our current systematic uncertainty of $\sim0.05$~mag is on the same order as the uncertainty due to typical peculiar velocities at $z=0.02$.  To increase the quantitative power for these studies we will need to reduced the lightcurve extraction and calibration uncertainties to 0.02~mag.  To achieve these goals, in the next data release we will present the full set of images, including host-galaxy reference images, and perform image subtraction to remove the 3\% uncertainty from host-galaxy contamination.  We will also be able to generate lightcurves for all of the supernovae, instead of just the \numLc\ \sneia\ presented here, which will be a roughly four-fold increase in the number of supernovae.  We will improve the calibration of the fields by using all of the WIYN+WHIRC observed stars, instead of just the 2MASS stars in each field, to aid in the translation of the photometric system across the sky.  Finally, we will improve our statistical uncertainty in the photometry with PSF modeling and extraction of photometry from the subtracted images.

An additional key current need in the field of NIR \snia\ cosmology is the development of
an updated model of K-corrections based on the latest NIR spectroscopic data
and accurate calibration between the various NIR photometric systems in use.

(2) We have preliminary results in \citet{Ponder17}, and the quantitative strength of these explorations will be improved as we improve the quality and quantity of the basic photometry as outlined above.

(3), (4) Using these NIR lightcurves to improve our understanding color and dust will require combination with optical lightcurves.  In collaboration with Peter Milne, we have obtained optical photometry with Super-LOTIS\footnote{\url{http://slotis.kpno.noao.edu/LOTIS/}} for half of the \sneia\ presented here.  We will present these data in a future paper.  Many of the \sneia\ we have presented here in this paper have also been
observed by others who have obtained optical photometric and spectroscopic data.
We intend that the release of these present NIR data will prompt further work
with joint analyses of optical and NIR properties of these \sneia.
\section{Conclusion}\label{sec:conclusion}

SweetSpot has completed its original 3-year program as an NOAO survey to image nearby \sneia\ in the NIR using WHIRC at the WIYN 3.5m Observatory at Kitt Peak.
This paper represents the first substantive data release from this survey.
We presented \numSnObservations\ processed and calibrated images for \numSne\ \sneia\ along with accompanying photometric catalogs.
We also presented \numLc\ SN lightcurves from 2011B--2013B, which includes the first 3 semesters (2012B, 2013A, 2013B) of the formal NOAO 2012B-0500 and also includes a re-analysis of the supernovae observed in the pilot survey (2011B, 2012A).

The second and full data release (DR2) is planned for Fall 2018 and will include data from the remaining 3 semesters of SweetSpot.
DR2 will also provide subtraction of host-galaxy template images to provide lightcurves for all SweetSpot \sneia.
The final Data Release 3 will feature technical improvements in instrument signature removal and stacking, improved calibration of the WIYN+WHIRC system, and a compilation of complementary optical lightcurves.

\acknowledgments

{\bf Author Contributions:}
Anja Weyant led the observational program and contributed to the image processing and light curve.
Michael Wood-Vasey is PI of SweetSpot and developed the bulk of the 2D image pipeline (based on the work of Ralf Kotulla) and also calibrated the observations and light curves.
Richard Joyce provided expertise in processing of the WHIRC data and instrument support/trouble-shooting in the early stages of the survey.
Lori Allen observed several nights at the WIYN 0.9-m in support of this program, and was the project contact for ensuring this NOAO 2012B-0500 program was appropriately scheduled.
Saurabh Jha was one of the developers of the Sweetspot project.
Peter Garnavich was one of the developers of the Sweetspot project and advised on target selection.
Tom Matheson was one of the developers of the Sweetspot project and contributed to science priority and observation planning and presented the status of the project at annual NOAO reviews.
Kara Ponder observed several nights at WIYN and oversaw image quality inspection.
Jessica Kroboth did the bulk of the processed image inspection.

We acknowledge the observing contributions of Nabila Jahan and Brandon Patel who each participated in a night of observing.  We thank Emily Yaruss and Daniel Perrefort for their assistance measuring the host-galaxy r-band brightness in pre-explosion images.

Based on observations at Kitt Peak National Observatory, National Optical Astronomy Observatory (NOAO Prop. IDs: 2011B-0482 and 2012B-0500; PI: Wood-Vasey), which is operated by the Association of Universities for Research in Astronomy (AURA) under a cooperative agreement with the National Science Foundation.

NOAO time on WIYN under single-semester program 2011B-0482 and NOAO Survey program 2012B-0500.

A.W. and M.W.-V. were supported in part by the US National Science Foundation under Grants AST-1028162 and AST-1311862.
A.W. additionally acknowledges support from PITT PACC and the Zaccheus Daniel Foundation.

We thank the staff of KPNO and the WIYN telescope and engineering staff for their efforts that enabled these observations.
In particular we thank the WIYN Observing Assistants who have participated in our program: Kristen Reetz, Malanka Riabokin, Karen Butler, Jennifer Power, Christian Soto, David Summers, Doug Williams.
We thank Diane Harmer for scheduling the 36 nights of WIYN time presented here within the constraints of cadence to observe supernovae and other WIYN programs.

We express our thanks to the hard-working groups who discover and provide rapid spectroscopic classification of \sneia\ that have allowed our follow-up program to be viable.
The ``Latest Supernovae'' website\footnote{\url{http://www.rochesterastronomy.org/supernova.html}} maintained by David Bishop was helpful in planning and executing these observations.

We thank the Tohono O'odham Nation for leasing their mountain to allow for astronomical research.

This work was supported in part by National Science Foundation Grant No. PHYS-1066293 and the hospitality of the Aspen Center for Physics.
In particular, we thank the Aspen Center for Physics for hosting the 2010 summer workshop on ``Taking Supernova Cosmology into the Next Decade'' where the original discussions that led to the SweetSpot survey took place, and the 2015 summer workshop on ``The Dynamic Universe: Understanding ExaScale Astronomical Synoptic Surveys'' where this manuscript took full form.

This research has made use of the NASA/IPAC Extragalactic Database (NED) which is operated by the Jet Propulsion Laboratory, California Institute of Technology, under contract with the National Aeronautics and Space Administration.

This publication makes use of data products from the Two Micron All Sky Survey, which is a joint project of the University of Massachusetts and the Infrared Processing and Analysis Center/California Institute of Technology, funded by the National Aeronautics and Space Administration and the National Science Foundation.

IRAF is distributed by the National Optical Astronomy Observatory, which is operated by the Association of Universities for Research in Astronomy (AURA) under a cooperative agreement with the National Science Foundation.

PyRAF is a product of the Space Telescope Science Institute, which is operated by AURA for NASA.

This research has made use of the VizieR catalog access tool, CDS, Strasbourg, France.

\facility{WIYN}

\software{IRAF, Python\footnote{\url{http://python.org}}, NumPy\footnote{\url{http://www.numpy.org}}, SciPy\footnote{\url{http://www.scipy.org}}, matplotlib~\citep{Hunter07}\footnote{\url{http://matplotlib.org}}, AstroPy~\citep{Astropy13}\footnote{\url{http://www.astropy.org}}, IDL\footnote{\url{http://www.harrisgeospatial.com/ProductsandSolutions/GeospatialProducts/IDL.aspx}}}

\bibliographystyle{aasjournal}
\bibliography{dr1}

\clearpage
\clearpage

\begin{deluxetable}{lrr}
\tabletypesize{\footnotesize}
\tablewidth{0pt}
\tablecolumns{3}
\tablecaption{Summary of awarded nights of NOAO time\label{table:nights_awarded}}

\tablehead{\colhead{Semester}         &
           \colhead{Awarded Nights}   &
           \colhead{Usable Nights\tablenotemark{a}}
}
\startdata
2011B & 7.0  & 3.5 \\
2012A & 1.0  & 1.0   \\
2012B & 8.0  & 8.0   \\
2013A & 10.5 & 9.0   \\ 
2013B & 10.5 & 8.0   \\ 
\enddata
\tablenotetext{a}{Effective total nights during which we obtained on-sky data.  The majority of the difference is due to bad weather, with some contribution from instrument failures. 
}
\end{deluxetable}
\begin{deluxetable}{rrrl}
\tablecaption{Nights Observed\label{table:nights_observed}}
\tablehead{\colhead{Date\tablenotemark{a}} & \multicolumn{2}{c}{\# Observations} & \colhead{Photometric} \\ \colhead{} & \colhead{SN\tablenotemark{b}} & \colhead{Persson\tablenotemark{c}} & \colhead{}}
\startdata
2011-10-25 & 12 & 0 & No \\
2011-11-15 & 16 & 0 & Yes \\
2011-11-21 & 8 & 0 & No \\
2011-12-08 & 25 & 0 & No \\
2012-01-08 & 20 & 0 & Yes \\
2012-04-02 & 21 & 0 & Yes \\
2012-09-25 & 12 & 0 & No \\
2012-10-01 & 11 & 0 & No \\
2012-10-07 & 12 & 0 & No \\
2012-10-28 & 17 & 0 & No \\
2012-11-02 & 21 & 2 & Yes \\
2012-11-22 & 21 & 9 & Yes \\
2012-11-25 & 22 & 17 & Yes \\
2012-11-30 & 22 & 9 & No \\
2013-03-21 & 1 & 0 & No \\
2013-03-26 & 5 & 2 & No \\
2013-04-20 & 15 & 8 & No \\
2013-04-25 & 15 & 4 & No \\
2013-04-29 & 11 & 5 & No \\
2013-05-19 & 10 & 5 & Yes \\
2013-05-22 & 6 & 6 & No \\
2013-05-23 & 12 & 9 & No \\
2013-05-28 & 12 & 12 & Yes \\
2013-06-13 & 14 & 6 & Yes \\
2013-06-17 & 15 & 6 & Yes \\
2013-09-25 & 10 & 10 & Yes \\
2013-09-27 & 11 & 18 & No \\
2013-10-15 & 19 & 21 & Yes \\
2013-10-18 & 10 & 6 & Yes \\
2013-10-19 & 7 & 7 & Yes \\
2013-10-20 & 18 & 16 & Yes \\
2013-10-25 & 5 & 0 & Yes \\
2013-11-11 & 17 & 15 & No \\
2013-11-20 & 10 & 1 & No \\
2013-12-09 & 15 & 12 & No \\
2013-12-13 & 15 & 11 & Yes
\enddata

\tablenotetext{a}{YYYY-MM-DD of local evening at KPNO (MST).}
\tablenotetext{b}{Each SN-filter combination is counted separately.}
\tablenotetext{c}{Each standard-filter-airmass observation is counted separately.}

\end{deluxetable}
\begin{deluxetable}{lllllllll}
\tabletypesize{\footnotesize}
\tablewidth{0pt}
\tablecaption{SN~Ia Properties from the Literature\label{table:snsummary}}
\tablehead{\colhead{Name} & \colhead{RA} & \colhead{Dec} & \colhead{Spec.} & \colhead{SN $z_{\rm helio}$} & \colhead{Classifier/} & \colhead{Discoverer\tablenotemark{a}} & \colhead{est.\tablenotemark{b} $T_{\rm B max}$} & \colhead{Lightcurve}  \\
\colhead{} & \multicolumn{2}{c}{J2000} & \colhead{Type} & \colhead{} & \colhead{Redshift\tablenotemark{a}} & \colhead{} & \colhead{MJD} & \colhead{}}
\startdata
CSS121006:232854+085452 & 23:28:54.48 & $+$08:54:51.6 & Ia & 0.08 & D12b & D12c & 56207 & Yes \\
CSS121009:011101-172841 & 01:11:01.091 & $-$17:28:42.28 & Ia & 0.06 & D12b & D12c & 56220 & Yes \\
CSS121114:090202+101800 & 09:02:02.420 & $+$10:18:00.31 & Ia & 0.042 & G12 & D12a & 56252 & Yes \\
CSS130218:092354+385837 & 09:23:54.52 & $+$38:58:36.8 & Ia & 0.05 & D13b & D13a & 56361 & No \\
CSS130317:082848+293031 & 08:28:47.784 & $+$29:30:30.85 & Ia & 0.08 & D13b & D13b & 56370 & No \\
LSQ12fhs & 22:52:23.423 & $-$20:36:53.35 & 91T & 0.033 & CO12 & CO12 & 56215 & Yes \\
LSQ12fmx & 03:12:52.144 & $-$00:12:12.55 & Ia & 0.067 & BN12 & BN12 & 56227 & Yes \\
LSQ12fuk & 04:58:15.927 & $-$16:17:58.14 & Ia & 0.02 & H12 & H12 & 56232 & No \\
LSQ12gef & 01:40:33.751 & $+$18:30:36.74 & Ia & 0.065 & CL12 & CL12 & 56247 & No \\
LSQ13cmt & 01:02:41.95 & $-$21:52:23.4 & Ia & 0.057 & CT13 & CT13 & 56575 & Yes \\
LSQ13crf & 03:10:50.263 & $+$01:25:19.19 & Ia & 0.06 & Tu13 & Tu13 & 56596 & No \\
LSQ13cwp & 04:03:50.662 & $-$02:39:18.57 & Ia & 0.067 & WK13 & WK13 & 56606 & Yes \\
PS1-12bwh & 07:09:24.29 & $+$39:06:15.8 & 02cx & 0.025 & W12 & W12 & 56222 & No \\
PS1-13dkh & 03:11:16.09 & $+$15:42:58.3 & 91T & 0.03 & Fo13 & Fo13 & 56572 & Yes \\  
PSNJ07250042+2347030 & 07:25:00.408 & $+$23:47:03.15 & Ia & 0.018 & BT12 & BT12 & 56218 & No \\ 
PTF11moy & 17:06:18.06 & $+$38:32:19.9 & Ia & 0.06 & GY11a & GY11a & 55824 & Yes \\
PTF11mty & 21:34:05.20 & $+$10:25:24.6 & Ia & 0.077 & GY11d & GY11d & 55835 & No \\
PTF11owc & 09:16:10.05 & $+$49:37:30.7 & Ia & 0.07 & GY11c & GY11c & 55867 & No \\
PTF11qmo & 10:06:49.748 & $-$07:41:12.39 & Ia & 0.055 & GY11b & GY11b & 55894 & Yes \\
PTF11qpc & 12:20:05.457 & $+$09:24:12.38 & Ia & 0.091 & GY11b & GY11b & 55902 & No \\
PTF11qri & 12:47:06.276 & $-$06:19:49.46 & Ia & 0.055 & GY11b & GY11b & 55897 & Yes \\
PTF11qzq & 07:19:27.311 & $+$54:13:48.84 & Ia & 0.06 & GY11b & GY11b & 55905 & No \\
PTF12iiq & 02:50:07.786 & $-$00:15:54.04 & Ia & 0.029 & GY12 & GY12 & 56179 & No \\
PTF12ikt & 01:14:43.13 & $+$00:17:07.1 & Ia & 0.045 & GY12 & GY12 & 56182 & No \\
PTF13asv & 16:22:43.19 & $+$18:57:35.0 & Ia & 0.035 & C13d & C13d & 56423 & Yes \\ 
PTF13ayw & 15:39:33.479 & $+$32:05:38.33 & Ia & 0.0538 & C13d & C13d & 56430 & No \\
PTF13dad & 01:48:08.39 & $+$37:33:29.1 & Ia & 0.086 & C13e & C13e & 56547 & Yes \\
PTF13ddg & 00:47:50.83 & $+$31:49:17.5 & Ia & 0.084 & C13e & C13e & 56545 & Yes \\
SN 2011fe & 14:03:05.80 & $+$54:16:25.3 & Ia & 0.008 & Nu11 & Nu11 & 55814 & Yes \\
SN 2011fs & 22:17:19.509 & $+$35:34:49.94 & Ia & 0.021 & BM11; To11b & J11a & 55833 & Yes \\
SN 2011gf & 21:12:24.302 & $-$07:48:52.04 & Ia & 0.027 & V11; M11 & D11 & 55827 & Yes \\
SN 2011gy & 03:29:35.319 & $+$40:52:02.93 & Ia & 0.017 & Ta11; Z11a & J11b & 55865 & No \\
SN 2011ha & 03:57:40.87 & $+$10:09:55.2 & Ia & 0.094 & O11 & Li11 & 55840 & No \\
SN 2011hb & 23:27:55.508 & $+$08:46:45.45 & Ia & 0.029 & MB11a & Ho11a & 55872 & No \\
SN 2011hk & 02:18:45.801 & $-$06:38:30.45 & 91bg & 0.018 & MB11b & N11 & 55864 & No \\  
SN 2011ho & 11:44:13.060 & $+$33:30:59.89 & Ia & 0.032 & Fo11b; MB11c & Ho11b & 55865 & No \\
SN 2011hr & 08:54:46.056 & $+$39:32:16.02 & 91T & 0.013 & Z11b & Na11 & 55883 & No \\
SN 2011io & 23:02:47.617 & $+$08:48:08.12 & Ia & 0.04 & Fr11 & Ba11 & 55894 & Yes \\
SN 2011iu & 23:51:02.342 & $+$46:43:21.55 & Ia & 0.046 & MB11d & Cox11 & 55894 & No \\
SN 2011iy & 13:08:58.38 & $-$15:31:04.0 & Ia & 0.00427 & CJ11; Y11 & I11 & 55896 & Yes \\
SN 2011jh & 12:47:14.42 & $-$10:03:47.2 & Ia & 0.00779 & To11a & NN11 & 55929 & Yes \\
SN 2012bh & 12:13:37.325 & $+$46:29:00.34 & Ia & 0.02524 & CK12 & CK12 & 56016 & Yes \\
SN 2012bm & 13:05:45.621 & $+$46:27:52.39 & Ia & 0.0248 & Ca12a & P12 & 56021 & No \\
SN 2012bo & 12:50:45.215 & $-$14:16:07.69 & Ia & 0.02543 & Ca12b; DT12 & I12 & 56021 & Yes \\
SN 2012bp & 16:18:12.451 & $+$36:28:51.87 & Ia & 0.02829 & Pa12 & Li12b & 56013 & Yes \\
SN 2012cg & 12:27:12.826 & $+$09:25:12.93 & Ia & 0.00145 & Ce12; M12 & Ka12 & 56018 & No \\
SN 2012em & 22:44:01.66 & $+$15:51:49.3 & 91bg & 0.038 & Be12; TD12 & Mi12 & 56181 & Yes \\
SN 2012et & 23:42:38.747 & $+$27:05:31.02 & Ia & 0.025 & De12 & R12a & 56186 & No \\
SN 2012fb & 01:50:51.238 & $+$33:08:24.17 & Ia & 0.038 & CW12 & Zh12 & 56195 & No \\
SN 2012fk & 02:30:52.081 & $+$22:28:46.02 & Ia & 0.035 & To12 & Li12a & 56223 & Yes \\
SN 2012fm & 06:56:13.39 & $+$84:04:50.2 & Ia & 0.014 & Za12 & Bel12 & 56228 & No \\
SN 2012fr & 03:33:35.99 & $-$36:07:37.7 & Ia & 0.005 & CS12; B12 & Kl12 & 56240 & No \\
SN 2012gm & 23:17:37.055 & $+$14:00:08.89 & Ia & 0.015 & O12 & R12b & 56260 & No \\
SN 2012go & 22:41:51.848 & $+$34:58:07.39 & Ia & 0.03 & Ke12; Ko12 & Ar12 & 56252 & No \\
SN 2013ar & 08:37:45.02 & $+$49:28:32.2 & Ia & 0.06 & ZJ13; To13c & Zh13c & 56374 & Yes \\
SN 2013be & 12:36:27.638 & $+$11:45:27.96 & Ia & 0.06585 & S13 & Zh13b & 56398 & No \\
SN 2013bo & 13:17:29.19 & $+$42:44:29.6 & Ia & 0.036 & To13b & D13c & 56393 & No \\
SN 2013bq & 13:04:08.45 & $+$43:54:07.7 & Ia & 0.06 & To13d & D13d & 56406 & No \\
SN 2013bs & 17:17:22.03 & $+$41:04:00.2 & Ia & 0.0276 & To13e & D13e & 56403 & Yes \\
SN 2013bt & 14:21:15.185 & $+$61:34:15.42 & Ia & 0.0364 & To13f & D13f & 56408 & Yes \\
SN 2013cb & 11:35:01.736 & $+$16:07:14.91 & Ia & 0.0541 & To13g; Gao13 & Zh13d & 56424 & No \\
SN 2013ck & 15:24:29.105 & $+$48:32:54.53 & Ia & 0.049 & To13a & D13g & 56424 & No \\
SN 2013cs & 13:15:14.839 & $-$17:57:55.24 & Ia & 0.00924 & Y13 & Ho13a & 56436 & Yes \\
SN 2013da & 13:45:36.338 & $-$07:19:33.69 & Ia & 0.0216 & Pr13a & Pr13a & 56449 & No \\
SN 2013fj & 22:15:28.480 & $+$15:34:03.19 & Ia & 0.03357 & ZT13 & Cia13 & 56545 & No \\
SN 2013fn & 21:00:23.673 & $-$14:29:52.42 & Ia & 0.027 & Mo13; MC13 & Ho13b & 56572 & Yes \\
SN 2013fw & 21:13:44.763 & $+$13:34:33.33 & Ia & 0.01695 & HL13 & J13 & 56603 & No \\
SNhunt175 & 15:19:24.978 & $+$20:54:01.61 & Ia & 0.0409 & Pr13b & Pr13b & 56369 & No \\  
SNhunt206 & 01:58:42.759 & $+$08:20:39.51 & Ia & 0.027 & Ke13 & Ke13 & 56548 & No \\ 
iPTF13dge & 05:03:35.091 & $+$01:34:17.03 & Ia & 0.01584 & C13b & C13b & 56555 & No \\
iPTF13dkj & 23:08:50.737 & $+$20:04:08.59 & Ia & 0.03 & C13a & C13a & 56555 & No \\
iPTF13dkl & 23:44:57.999 & $+$03:23:40.07 & Ia & 0.04 & C13a & C13a & 56556 & Yes \\
iPTF13dkx & 01:20:53.115 & $+$03:20:23.65 & Ia & 0.03 & C13a & C13a & 56561 & Yes \\
iPTF13ebh & 02:21:59.98 & $+$33:16:13.7 & Ia & 0.01327 & C13c & C13c & 56619 & Yes \\
\enddata
\tablenotetext{a}{SN classifier/redshift and discovery references are:
Ar12 \citet{Arbour12},
B12 \citet{Buil12},
BM11 \citet{Balam11}, \\
BN12 \citet{Benitez12},
BT12 \citet{Buton12},
Ba11 \citet{Balanutsa11b}, \\
Be12 \citet{Benetti12}, 
Bel12 \citet{Belligoli12},
C13a \citet{Cao13a},
C13b \citet{Cao13b},
C13c \citet{Cao13c}, \\
C13d \citet{Cao13d},
C13e \citet{Cao13e},
CJ11 \citet{Chen11},
CK12 \citet{Chornock12}, \\
CL12 \citet{Cellier-Holzem12},
CO12  \citet{Copin12},
CS12 \citet{Childress12},
CT13 \citet{Chen13}, \\
CW12 \citet{Caldwell12},
Ca12a \citet{Cappellaro12a},
Ca12b \citet{Cappellaro12b},
Ce12 \citet{Cenko12}, \\
Cia13 \citet{Ciabattari13},
Cox11 \citet{Cox11},
D11 \citet{Drake11},
D12a \citet{Drake12a},
D12b \citet{Drake12b}, \\
D12c \citet{Drake12c},
D13a \citet{Drake13a},
D13b \citet{Drake13b},
D13c \citet{Drake13c}, \\
D13d \citet{Drake13d},
D13e \citet{Drake13e},
D13f \citet{Drake13f},
D13g \citet{Drake13g},
DT12 \citet{Drout12}, \\
De12 \citet{Dennefeld12},
Fo11b \citet{Foley11b},
Fo13 \citet{Foley13},
Fr11 \citet{Fraser11}, \\
G12 \citet{Graham12},
GY11a \citet{Gal-Yam11a},
GY11b \citet{Gal-Yam11b},
GY11c \citet{Gal-Yam11c}, \\
GY11d \citet{Gal-Yam11d},
GY12 \citet{Gal-Yam12},
Gao13 \citet{Gao13},
H12 \citet{Hadjiyska12}, \\
HL13 \citet{Howell13},
Ho11a \citet{Howerton11a},
Ho11b \citet{Howerton11b},
Ho13a \citet{Howerton13a}, \\
Ho13b \citet{Howerton13b},
I11 \citet{Itagaki11},
I12 \citet{Itagaki12},
J11a \citet{Jin11a},
J11b \citet{Jin11b}, \\
J13 \citet{Jin13},
Ka12 \citet{Kandrashoff12},
Ke12 \citet{Kankare12},
Ke13 \citet{Kankare13}, \\
Kl12 \citet{Klotz12},
Ko12 \citet{Kotak12},
Li11 \citet{Lipunov11},
Li12a \citet{Lipunov12a}, \\
Li12b \citet{Lipunov12b},
M11 \citet{Marion11},
M12 \citet{Marion12},
MB11a \citet{MarionBerlind11a}, \\
MB11b \citet{MarionBerlind11b},
MB11c \citet{MarionBerlind11c},
MB11d \citet{MarionBerlind11d},
MC13 \citet{Milisavljevic13}, \\
Mi12 \citet{Mikuz12},
Mo13 \citet{Mo13},
N11 \citet{Nakano11a},
NN11 \citet{Nakano11b},
Na11 \citet{Nayak11}, \\
Nu11 \citet{Nugent11},
O11 \citet{Ochner11},
O12 \citet{Ochner12},
P12 \citet{Puckett12},
Pa12 \citet{Pastorello12}, \\
Pr13a \citet{Prieto13a},
Pr13b \citet{Prieto13b},
R12a \citet{Rich12a},
R12b \citet{Rich12b},
S13 \citet{Silverman13}, \\
TD12 \citet{Taddia12},
Ta11 \citet{Taubenberger11},
To11a \citet{Tomasella11a},
To11b \citet{Tomasella11b}, \\
To12 \citet{Tomasella12},
To13a \citet{Tomasella13a},
To13b \citet{Tomasella13b},
To13c \citet{Tomasella13c}, \\
To13d \citet{Tomasella13d}, 
To13e \citet{Tomasella13e},
To13f \citet{Tomasella13f},
To13g \citet{Tomasella13g}, \\
Tu13 \citet{Turatto13},
V11 \citet{Valenti11},
W12 \citet{Wright12},
WK13 \citet{Walker13}, \\
Y11 \citet{Yamanaka11},
Y13 \citet{Yamanaka13},
Z11a \citet{Zhang11a},
Z11b \citet{Zhang11b}, \\
ZJ13 \citet{ZhangJ13},
ZT13 \citet{Zanutta13},
Za12 \citet{Zaggia12},
Zh12 \citet{Zheng12}, \\
Zh13b \citet{Zhang13b},
Zh13c \citet{Zhang13c},
Zh13d \citet{Zhang13d},
%Ba11a \citet{Balanutsa11a},
}
\tablenotetext{b}{Time of maximum in the $B$-band reported in CBET/ATel.}
\end{deluxetable}

\clearpage
% \floattable
\begin{deluxetable}{llcccccc}
\tabletypesize{\scriptsize}
\tablewidth{0pt}
\tablecolumns{10}
\tablecaption{Persson Photometric Standard Stars\label{table:persson_standards}}
\tablehead{
  \colhead{Name}     		&
  \colhead{2MASS ID} &
  \multicolumn{3}{c}{Persson Catalog Magnitudes} & 
  \multicolumn{3}{c}{2MASS Catalog Magnitudes}  \\
  \colhead{}&
  \colhead{} &
  \colhead{$m_J$ [mag]}      &
  \colhead{$m_H$ [mag]}     &
  \colhead{$m_{\Ks}$ [mag]} &
  \colhead{$m_J$ [mag]}      &
  \colhead{$m_H$ [mag]}     &
  \colhead{$m_{\Ks}$ [mag]}  
}
\startdata
P212-C  &  10062887$+$4101245 & 11.993 $\pm$ 0.006 &  11.729 $\pm$ 0.005 &  11.697 $\pm$ 0.007 &  11.939 $\pm$ 0.014 &  11.747 $\pm$ 0.014 &  11.662 $\pm$ 0.016 \\ 
S791-C  &  13172933$-$0532383 & 11.661 $\pm$ 0.008 &  11.310 $\pm$ 0.007 &  11.267 $\pm$ 0.008 &  11.677 $\pm$ 0.025 &  11.318 $\pm$ 0.022 &  11.275 $\pm$ 0.023 \\
P330-E  &  16313382$+$3008465 & 11.816 $\pm$ 0.007 &  11.479 $\pm$ 0.005 &  11.429 $\pm$ 0.006 &  11.781 $\pm$ 0.018 &  11.453 $\pm$ 0.018 &  11.432 $\pm$ 0.020 \\
P525-E  &  00242846$+$0749005 & 11.622 $\pm$ 0.005 &  11.298 $\pm$ 0.005 &  11.223 $\pm$ 0.005 &  11.642 $\pm$ 0.025 &  11.297 $\pm$ 0.021 &  11.224 $\pm$ 0.021 \\
P161-D  &  07005180$+$4829231 & 11.680 $\pm$ 0.006 &  11.408 $\pm$ 0.006 &  11.352 $\pm$ 0.006 &  11.682 $\pm$ 0.017 &  11.385 $\pm$ 0.015 &  11.326 $\pm$ 0.016 \\
S840-F  &  05423214$+$0009019 & 11.426 $\pm$ 0.009 &  11.148 $\pm$ 0.009 &  11.058 $\pm$ 0.008 &  11.441 $\pm$ 0.021 &  11.125 $\pm$ 0.020 &  11.070 $\pm$ 0.019 \\
\enddata
\end{deluxetable}

\begin{deluxetable}{lll}
\tabletypesize{\footnotesize}
\tablewidth{0pt}
\tablecaption{Stack Inspection\label{table:stackinspection}}
\tablehead{\colhead{imageName} & \colhead{stackUsable} & \colhead{wcsGood}}
\startdata
PTF11moy\_A\_H\_20111025.fits & Y & Y \\
PTF11moy\_A\_J\_20111025.fits & Y & Y \\
PTF11mty\_A\_H\_20111025.fits & Y & Y \\
PTF11mty\_A\_J\_20111025.fits & Y & Y \\
SN2011fe\_A\_H\_20111025.fits & Y & Y \\
SN2011fe\_A\_J\_20111025.fits & Y & Y \\
SN2011fe\_A\_KS\_20111025.fits & Y & Y \\
SN2011fs\_A\_H\_20111025.fits & Y & Y \\
SN2011fs\_A\_J\_20111025.fits & Y & Y \\
SN2011fs\_A\_KS\_20111025.fits & Y & Y
\enddata
 97.3\% of the stacks are usable.
 97.2\% of the stacks have good WCS.
 96.2\% of the stacks are usable and have good WCS.
Note that there a few stacks that are not usable but have good WCS:
e.g., the stack may have only one constituent raw image
and so have many masked regions, but has a good WCS solution. (This table is a stub to show the structure.  See the online version for the full table.)
\end{deluxetable}
\begin{deluxetable}{lrrrr}
\tabletypesize{\footnotesize}
\tablewidth{0pt}
\tablecolumns{5}
\tablecaption{Photometric Calibration Terms\label{table:transform}}
\tablehead{
  \colhead{} &
  \multicolumn{1}{c}{2011B--2013A} &
  \multicolumn{1}{c}{2013B} &
  \colhead{} &
  \colhead{} 
\\
  \colhead{Filter}     &
  \colhead{zeropoint}  &
  \colhead{zeropoint}  &
  \colhead{$k$}        &
  \colhead{$c$}        
\\
  \colhead{}              &
  \colhead{[mag]}         &
  \colhead{[mag]}         &
  \colhead{[mag/airmass]} &
  \colhead{[mag/mag]}     
}
% The joint coefficients for filter: #star, #nobs (zptps_old, zptps_new, k, C):
% J :  19  56 (22.5964, 23.2272, $+0.0812$, $+0.0365$)
% Ks:  16  60 (22.1259, 22.5772, $+0.0094$, $+0.0434$)
%   [where k is from the Persson standard star joint fit]
% H : 239 611 (22.7300, 23.3014, $+0.0470$, $+0.0635$)
\startdata
$J$   & 22.5801 $\pm$ 0.01 & 23.1998 $\pm$ 0.01 &  0.0812 $\pm$ 0.02  & $+$0.2846 $\pm$ 0.03 \\
$H$   & 22.7300 $\pm$ 0.01 & 23.3014 $\pm$ 0.01 &  0.0470 $\pm$ 0.02  & $+$0.0635 $\pm$ 0.03 \\
$K_s$ & 22.2076 $\pm$ 0.01 & 22.5684 $\pm$ 0.01 &  0.0094 $\pm$ 0.02  & $+$0.1144 $\pm$ 0.03 \\
\enddata
\end{deluxetable}

\clearpage
\begin{deluxetable}{llrrrrrrrrrrrr}
\tabletypesize{\scriptsize}
\tablewidth{0pt}
\tablecaption{2MASS Calibration Stars\label{table:calibrationstars}}
\tablehead{\colhead{2MASS ID} & \colhead{Field} & \multicolumn{6}{c}{WHIRC} & \multicolumn{6}{c}{2MASS} \\ \colhead{} & \colhead{} & \colhead{$J$} & \colhead{$\sigma_J$} & \colhead{$H$} & \colhead{$\sigma_H$} & \colhead{$K_s$} & \colhead{$\sigma_{K_s}$} & \colhead{$J$} & \colhead{$\sigma_J$} & \colhead{$H$} & \colhead{$\sigma_H$} & \colhead{$K_s$} & \colhead{$\sigma_{K_s}$}\\ \colhead{ } & \colhead{ } & \colhead{$\mathrm{mag}$} & \colhead{$\mathrm{mag}$} & \colhead{$\mathrm{mag}$} & \colhead{$\mathrm{mag}$} & \colhead{$\mathrm{mag}$} & \colhead{$\mathrm{mag}$} & \colhead{$\mathrm{mag}$} & \colhead{$\mathrm{mag}$} & \colhead{$\mathrm{mag}$} & \colhead{$\mathrm{mag}$} & \colhead{$\mathrm{mag}$} & \colhead{$\mathrm{mag}$}}
\startdata
23285863$+$0854063 & CSS121006:232854+085452 & \nodata & \nodata & 15.137 & 0.011 & \nodata & \nodata & 15.469 & 0.063 & 15.022 & 0.088 & 14.849 & 0.112 \\
01105481$-$1728484 & CSS121009:011101-172841 & 13.813 & 0.015 & 13.537 & 0.015 & \nodata & \nodata & 14.252 & 0.028 & 13.591 & 0.034 & 13.480 & 0.045 \\
09015664$+$1017118 & CSS121114:090202+101800 & 15.327 & 0.010 & 14.802 & 0.010 & \nodata & \nodata & 15.322 & 0.054 & 14.818 & 0.054 & 14.679 & 0.105 \\
09240185$+$3858360 & CSS130218:092354+385837 & \nodata & \nodata & 14.612 & 0.011 & \nodata & \nodata & 15.258 & 0.047 & 14.706 & 0.070 & 14.445 & 0.069 \\
08285459$+$2929165 & CSS130317:082848+293031 & 12.956 & 0.011 & 12.157 & 0.010 & \nodata & \nodata & 12.768 & 0.023 & 12.127 & 0.019 & 11.989 & 0.022 \\
22521836$-$2038146 & LSQ12fhs & 15.039 & 0.018 & 14.354 & 0.017 & \nodata & \nodata & 14.962 & 0.041 & 14.322 & 0.046 & 14.213 & 0.060 \\
03125316$-$0012084 & LSQ12fmx & 14.596 & 0.011 & 14.295 & 0.011 & \nodata & \nodata & 14.510 & 0.032 & 14.236 & 0.032 & 14.066 & 0.065 \\
04581563$-$1617462 & LSQ12fuk & 12.845 & 0.015 & 12.302 & 0.015 & \nodata & \nodata & 12.899 & 0.024 & 12.277 & 0.023 & 12.075 & 0.024 \\
01403113$+$1831584 & LSQ12gef & 12.576 & 0.010 & 11.999 & 0.010 & \nodata & \nodata & 12.818 & 0.024 & 12.128 & 0.026 & 11.937 & 0.020 \\
01025011$-$2151228 & LSQ13cmt & 14.921 & 0.017 & 14.416 & 0.017 & \nodata & \nodata & 15.683 & 0.059 & 15.102 & 0.057 & 14.882 & 0.104 \\
03105792$+$0125138 & LSQ13crf & 12.707 & 0.011 & 12.373 & 0.011 & \nodata & \nodata & 12.717 & 0.024 & 12.310 & 0.029 & 12.201 & 0.021 \\
04035391$-$0238401 & LSQ13cwp & 11.779 & 0.011 & 11.494 & 0.011 & \nodata & \nodata & 11.775 & 0.023 & 11.491 & 0.024 & 11.386 & 0.023 \\
07092861$+$3906536 & PS1-12bwh & 14.751 & 0.010 & 14.407 & 0.010 & \nodata & \nodata & 14.682 & 0.039 & 14.308 & 0.049 & 14.329 & 0.062 \\
03111139$+$1542476 & PS1-13dkh & 12.760 & 0.010 & 12.211 & 0.010 & \nodata & \nodata & 12.792 & 0.023 & 12.212 & 0.028 & 12.083 & 0.022 \\
07245876$+$2348557 & PSNJ07250042+2347030 & 14.417 & 0.010 & 14.003 & 0.010 & \nodata & \nodata & 14.395 & 0.030 & 13.968 & 0.041 & 13.915 & 0.037 \\
17061982$+$3832029 & PTF11moy & 15.725 & 0.017 & 15.254 & 0.021 & \nodata & \nodata & 15.726 & 0.067 & 15.165 & 0.102 & 15.068 & 0.122 \\
21340153$+$1025244 & PTF11mty & 15.047 & 0.015 & 14.678 & 0.018 & \nodata & \nodata & 15.118 & 0.047 & 14.667 & 0.063 & 14.565 & 0.090 \\
09161694$+$4938008 & PTF11owc & 15.439 & 0.010 & 14.886 & 0.011 & \nodata & \nodata & 15.736 & 0.065 & 15.073 & 0.068 & 14.843 & 0.087 \\
10064651$-$0740348 & PTF11qmo & 11.832 & 0.012 & 11.639 & 0.013 & \nodata & \nodata & 12.035 & 0.022 & 11.791 & 0.022 & 11.737 & 0.024 \\
12201145$+$0923443 & PTF11qpc & \nodata & \nodata & 12.432 & 0.010 & \nodata & \nodata & 12.531 & 0.021 & 12.239 & 0.025 & 12.187 & 0.026 \\
12470715$-$0620106 & PTF11qri & 15.256 & 0.015 & 14.707 & 0.017 & \nodata & \nodata & 15.017 & 0.029 & 14.673 & 0.060 & 14.757 & 0.096 \\
07192539$+$5413348 & PTF11qzq & 14.056 & 0.010 & 13.676 & 0.010 & \nodata & \nodata & 14.170 & 0.028 & 13.691 & 0.029 & 13.751 & 0.037 \\
02500494$-$0015196 & PTF12iiq & 10.666 & 0.012 & 10.586 & 0.011 & \nodata & \nodata & 10.507 & 0.028 & 10.070 & 0.022 & 9.911 & 0.022 \\
01143707$+$0015215 & PTF12ikt & 11.269 & 0.011 & 11.059 & 0.011 & \nodata & \nodata & 11.255 & 0.027 & 10.874 & 0.031 & 10.786 & 0.022 \\
16224201$+$1856258 & PTF13asv & \nodata & \nodata & 14.982 & 0.011 & \nodata & \nodata & 15.718 & 0.064 & 14.923 & 0.062 & 14.818 & 0.080 \\
15393516$+$3207091 & PTF13ayw & 15.755 & 0.014 & 15.096 & 0.012 & \nodata & \nodata & 15.988 & 0.070 & 15.345 & 0.102 & 14.990 & 0.103 \\
01480916$+$3734125 & PTF13dad & 20.982 & 0.556 & 13.663 & 0.010 & \nodata & \nodata & 13.915 & 0.024 & 13.652 & 0.026 & 13.609 & 0.030 \\
00475085$+$3151019 & PTF13ddg & \nodata & \nodata & 11.877 & 0.013 & \nodata & \nodata & 11.972 & 0.024 & 11.704 & 0.031 & 11.698 & 0.025 \\
14025710$+$5416408 & SN 2011fe & 14.708 & 0.039 & 14.590 & 0.034 & 14.519 & 0.053 & 15.060 & 0.038 & 14.642 & 0.072 & 14.844 & 0.108 \\
22171903$+$3535266 & SN 2011fs & 14.838 & 0.014 & 14.205 & 0.013 & 14.003 & 0.016 & 14.819 & 0.035 & 14.203 & 0.045 & 14.106 & 0.066 \\
21122821$-$0747132 & SN 2011gf & 15.641 & 0.019 & 15.080 & 0.022 & \nodata & \nodata & 15.712 & 0.067 & 15.015 & 0.101 & 14.790 & 0.103 \\
03293746$+$4052433 & SN 2011gy & 14.630 & 0.016 & 14.062 & 0.013 & \nodata & \nodata & 14.771 & 0.032 & 14.075 & 0.035 & 13.950 & 0.034 \\
03573953$+$1008196 & SN 2011ha & 15.440 & 0.011 & 14.914 & 0.011 & \nodata & \nodata & 15.543 & 0.056 & 14.989 & 0.075 & 14.808 & 0.084 \\
23275381$+$0847589 & SN 2011hb & 12.784 & 0.018 & 12.432 & 0.015 & \nodata & \nodata & 12.871 & 0.024 & 12.412 & 0.024 & 12.301 & 0.026 \\
02184937$-$0637528 & SN 2011hk & 14.934 & 0.015 & 14.283 & 0.014 & \nodata & \nodata & 15.022 & 0.045 & 14.408 & 0.047 & 14.257 & 0.059 \\
11440746$+$3330234 & SN 2011ho & \nodata & \nodata & 10.800 & 0.010 & \nodata & \nodata & 12.622 & 0.021 & 12.035 & 0.022 & 11.792 & 0.018 \\
08544574$+$3933348 & SN 2011hr & 15.065 & 0.010 & 14.537 & 0.010 & \nodata & \nodata & 15.126 & 0.038 & 14.589 & 0.053 & 14.249 & 0.058 \\
23024227$+$0848225 & SN 2011io & 16.017 & 0.012 & 15.620 & 0.011 & \nodata & \nodata & 15.732 & 0.070 & 15.163 & 0.090 & 14.966 & 0.128 \\
23505722$+$4643151 & SN 2011iu & 15.204 & 0.012 & 14.578 & 0.011 & \nodata & \nodata & 15.323 & 0.051 & 14.685 & 0.055 & 14.467 & 0.078 \\
13085417$-$1532148 & SN 2011iy & 14.030 & 0.016 & 13.535 & 0.015 & \nodata & \nodata & 14.129 & 0.026 & 13.601 & 0.027 & 13.510 & 0.037 \\
12472073$-$1002493 & SN 2011jh & 12.356 & 0.013 & 12.086 & 0.012 & \nodata & \nodata & 12.353 & 0.023 & 12.024 & 0.022 & 11.972 & 0.024 \\
12133551$+$4627088 & SN 2012bh & 14.834 & 0.010 & 14.597 & 0.010 & \nodata & \nodata & 15.110 & 0.040 & 14.511 & 0.057 & 14.464 & 0.071 \\
13054793$+$4629189 & SN 2012bm & 14.393 & 0.014 & 14.722 & 0.013 & \nodata & \nodata & 15.271 & 0.042 & 14.625 & 0.059 & 14.226 & 0.048 \\
12504489$-$1414194 & SN 2012bo & 15.745 & 0.021 & 15.274 & 0.024 & \nodata & \nodata & 15.606 & 0.055 & 15.193 & 0.087 & 15.117 & 0.150 \\
16181090$+$3629229 & SN 2012bp & 15.503 & 0.011 & 15.296 & 0.010 & \nodata & \nodata & 15.588 & 0.062 & 15.147 & 0.085 & 15.203 & 0.151 \\
12271948$+$0924165 & SN 2012cg & 14.285 & 0.010 & 13.506 & 0.010 & \nodata & \nodata & 14.664 & 0.041 & 14.106 & 0.053 & 13.784 & 0.058 \\
22435933$+$1550085 & SN 2012em & 14.414 & 0.010 & 13.818 & 0.010 & \nodata & \nodata & 14.414 & 0.029 & 13.807 & 0.037 & 13.634 & 0.044 \\
23424027$+$2706047 & SN 2012et & 15.519 & 0.011 & 15.002 & 0.011 & \nodata & \nodata & 15.395 & 0.059 & 14.883 & 0.088 & 14.752 & 0.096 \\
01505417$+$3308549 & SN 2012fb & 14.555 & 0.010 & 13.978 & 0.010 & \nodata & \nodata & 14.500 & 0.029 & 13.869 & 0.036 & 13.576 & 0.036 \\
02305565$+$2227192 & SN 2012fk & 15.810 & 0.010 & 15.486 & 0.010 & \nodata & \nodata & 15.942 & 0.081 & 15.245 & 0.080 & 15.250 & 0.159 \\
06552328$+$8405019 & SN 2012fm & 12.432 & 0.016 & 12.073 & 0.016 & \nodata & \nodata & 12.507 & 0.022 & 12.085 & 0.030 & 12.007 & 0.023 \\
03333369$-$3607144 & SN 2012fr & 12.192 & 0.036 & 11.851 & 0.037 & \nodata & \nodata & 12.148 & 0.024 & 11.773 & 0.027 & 11.673 & 0.026 \\
23173291$+$1359237 & SN 2012gm & 11.667 & 0.010 & 11.155 & 0.010 & 11.262 & 0.010 & 11.741 & 0.022 & 11.159 & 0.021 & 11.060 & 0.020 \\
22414990$+$3458008 & SN 2012go & 14.143 & 0.010 & 13.923 & 0.010 & \nodata & \nodata & 14.114 & 0.025 & 13.673 & 0.017 & 13.735 & 0.050 \\
08374071$+$4927483 & SN 2013ar & 15.767 & 0.010 & 14.866 & 0.010 & \nodata & \nodata & 15.415 & 0.043 & 14.693 & 0.061 & 14.518 & 0.068 \\
12362858$+$1144499 & SN 2013be & 12.299 & 0.010 & 12.052 & 0.010 & \nodata & \nodata & 12.586 & 0.022 & 12.291 & 0.028 & 12.232 & 0.027 \\
13173089$+$4244190 & SN 2013bo & 14.866 & 0.011 & 14.312 & 0.010 & \nodata & \nodata & 14.927 & 0.037 & 14.411 & 0.049 & 14.229 & 0.060 \\
13041752$+$4354337 & SN 2013bq & 14.627 & 0.010 & 13.753 & 0.010 & \nodata & \nodata & 15.179 & 0.047 & 14.592 & 0.054 & 14.395 & 0.067 \\
17171754$+$4105044 & SN 2013bs & 10.749 & 0.010 & 10.485 & 0.010 & \nodata & \nodata & 10.852 & 0.022 & 10.533 & 0.016 & 10.472 & 0.018 \\
14211406$+$6134348 & SN 2013bt & 13.904 & 0.012 & 13.298 & 0.011 & \nodata & \nodata & 13.968 & 0.026 & 13.359 & 0.026 & 13.172 & 0.028 \\
11350065$+$1608296 & SN 2013cb & 15.265 & 0.012 & 14.859 & 0.010 & \nodata & \nodata & 15.635 & 0.057 & 15.064 & 0.080 & 14.814 & 0.081 \\
15242515$+$4834230 & SN 2013ck & 15.166 & 0.011 & 14.862 & 0.010 & \nodata & \nodata & 15.356 & 0.047 & 14.818 & 0.068 & 14.592 & 0.106 \\
13151954$-$1757077 & SN 2013cs & 14.275 & 0.016 & 13.718 & 0.015 & \nodata & \nodata & 14.342 & 0.032 & 13.736 & 0.034 & 13.526 & 0.046 \\
13454012$-$0718110 & SN 2013da & 15.488 & 0.013 & 15.017 & 0.012 & \nodata & \nodata & 15.266 & 0.046 & 14.919 & 0.068 & 14.814 & 0.128 \\
22152384$+$1534390 & SN 2013fj & 15.182 & 0.012 & 14.823 & 0.013 & \nodata & \nodata & 15.253 & 0.046 & 14.762 & 0.050 & 14.758 & 0.134 \\
21002395$-$1431048 & SN 2013fn & 15.920 & 0.014 & 15.501 & 0.014 & 15.412 & 0.016 & 16.007 & 0.069 & 15.396 & 0.086 & 15.377 & 0.138 \\
21134413$+$1336121 & SN 2013fw & 13.034 & 0.010 & 12.720 & 0.010 & 14.043 & 0.010 & 13.007 & 0.025 & 12.633 & 0.026 & 12.576 & 0.025 \\
15192067$+$2053326 & SNhunt175 & 13.327 & 0.012 & 12.892 & 0.011 & \nodata & \nodata & 13.428 & 0.024 & 13.007 & 0.022 & 12.867 & 0.025 \\
01583809$+$0820382 & SNhunt206 & 15.056 & 0.012 & 14.475 & 0.011 & \nodata & \nodata & 15.189 & 0.042 & 14.522 & 0.052 & 14.351 & 0.081 \\
05033130$+$0133109 & iPTF13dge & 16.395 & 0.011 & 15.638 & 0.011 & 15.070 & 0.012 & 15.571 & 0.100 & 14.756 & 0.101 & 14.264 & 0.101 \\
23084927$+$2003397 & iPTF13dkj & 13.940 & 0.010 & 13.345 & 0.010 & \nodata & \nodata & 13.921 & 0.025 & 13.370 & 0.026 & 13.045 & 0.026 \\
23445822$+$0325271 & iPTF13dkl & 14.479 & 0.012 & 14.227 & 0.012 & \nodata & \nodata & 14.940 & 0.039 & 14.280 & 0.058 & 14.183 & 0.063 \\
01205658$+$0322007 & iPTF13dkx & 11.398 & 0.011 & 11.083 & 0.011 & \nodata & \nodata & 11.358 & 0.024 & 10.996 & 0.026 & 10.929 & 0.024 \\
02215630$+$3316108 & iPTF13ebh & 15.885 & 0.011 & 15.473 & 0.010 & 15.278 & 0.012 & 15.937 & 0.067 & 15.501 & 0.122 & 15.331 & 0.156 \\
\enddata
\end{deluxetable}

\clearpage
\begin{deluxetable}{lrrlrrrr}
\tabletypesize{\footnotesize}
\tablewidth{0pt}
\tablecaption{SweetSpot Lightcurves\label{table:lightcurves}}
\tablehead{\colhead{Name} & \colhead{mjd} & \colhead{approxPhase} & \colhead{filter\tablenotemark{a}} & \colhead{mag} & \colhead{magErr} & \colhead{flux\tablenotemark{b}} & \colhead{fluxErr}\\ \colhead{ } & \colhead{$\mathrm{d}$} & \colhead{$\mathrm{d}$} & \colhead{ } & \colhead{$\mathrm{mag}$} & \colhead{$\mathrm{mag}$} & \colhead{ } & \colhead{ }}
\startdata
CSS121009:011101-172841 & 56234.27359 & 14.27359 & WHIRCJ & 18.97756 & 0.05157 & 256.4340 & 12.1698 \\
CSS121009:011101-172841 & 56229.31361 & 9.31361 & WHIRCH & 18.64361 & 0.04781 & 348.7850 & 15.3479 \\
CSS121009:011101-172841 & 56229.28541 & 9.28541 & WHIRCJ & 19.27324 & 0.07101 & 195.3014 & 12.7551 \\
CSS121009:011101-172841 & 56234.22657 & 14.22657 & WHIRCH & 18.53778 & 0.04162 & 384.4939 & 14.7330 \\
CSS121114:090202+101800 & 56254.50847 & 2.50847 & WHIRCH & 18.08766 & 0.02963 & 582.0156 & 15.8819 \\
CSS121114:090202+101800 & 56257.45088 & 5.45088 & WHIRCH & 18.29991 & 0.02746 & 478.6713 & 12.1028 \\
CSS121114:090202+101800 & 56262.48385 & 10.48385 & WHIRCH & 18.39446 & 0.03869 & 438.7490 & 15.6268 \\
CSS121114:090202+101800 & 56254.52481 & 2.52481 & WHIRCJ & 18.39741 & 0.02063 & 437.5576 & 8.3115 \\
CSS121114:090202+101800 & 56257.46387 & 5.46387 & WHIRCJ & 18.84659 & 0.02922 & 289.3104 & 7.7855 \\
CSS121114:090202+101800 & 56262.49573 & 10.49573 & WHIRCJ & 19.18741 & 0.04400 & 211.3655 & 8.5610
\enddata

\tablenotetext{a}{
The WIYN+WHIRC $K_s$ filter is called ``WHIRCK'' in \code{SNooPy}'s system transmission database.
We adopt the SNooPy filter names here for future convenience.
}
\tablenotemark{b}{
Fluxes are given with respect to a zeropoint of 25: ${\rm mag}=-2.5\log_{10}({\rm flux}) + 25$.
}
(This table is a stub to show the structure.  See the online version for the full table.)
\end{deluxetable}

\clearpage
\begin{deluxetable}{lrrrr}
\tabletypesize{\footnotesize}
\tablewidth{0pt}
\tablecaption{SweetSpot DR1 Summary\label{table:dr1}}
\tablehead{\colhead{Category} & \colhead{Total} & \colhead{$J$} & \colhead{$H$} & \colhead{$K_s$}}
\startdata
Total Open-Shutter Time [s] & 743091 & & & \\
Total Sky Target Open-Shutter Time [s] & 707467 & 391913 & 289382 & 25214 \\
Total Science Frames & 14356 & 5553 & 7509 & 1080 \\
Total Science Frames that are Standard Stars & 2220 & 720 & 813 & 687 \\
 & & & & \\
Fields Observed\tablenotemark{a} & \numObjects & \numObjectsJ & \numObjectsH & \numObjectsKs \\
2MASS Stars Observed & \numTwomassLc & \numTwomassLcJ & \numTwomassLcH & \numTwomassLcKs \\
Supernovae Observed\tablenotemark{a} & \numSne & \numSneJ & \numSneH & \numSneKs \\
Supernovae Observed $\leq 20$~day past $B$-max & \numSnObjectsPreTwentyDaysBMax & \numSnObjectsPreTwentyDaysBMaxJ & \numSnObjectsPreTwentyDaysBMaxH & \numSnObjectsPreTwentyDaysBMaxKs \\
Supernovae Observed $0$--$20$~days past $B$-max & \numSnObjectsPlateau & \numSnObjectsPlateauJ & \numSnObjectsPlateauH & \numSnObjectsPlateauKs \\
Supernovae Observed $\leq 0$~day past $B$-max & \numSnObjectsPreBMax & \numSnObjectsPreBMaxJ & \numSnObjectsPreBMaxH  & \numSnObjectsPreBMaxKs  \\
& & & & \\
Observations & \numObjObservations & \numObjObservationsJ & \numObjObservationsH & \numObjObservationsKs \\
Average \# Observations/Observed Object & \avgObjObservations & \avgObjObservationsJ & \avgObjObservationsH & \avgObjObservationsKs \\
Observations of Supernovae & \numSnObservations & \numSnObservationsJ & \numSnObservationsH & \numSnObservationsKs \\
Average \# Observations/Observed Supernovae & \avgSnObservations & \avgSnObservationsJ & \avgSnObservationsH & \avgSnObservationsKs \\
Median \# Observations/Observed Supernovae & \medSnObservations & \medSnObservationsJ & \medSnObservationsH & \medSnObservationsKs \\
& & & & \\
Supernova Lightcurves & \numLc & \numLcJ & \numLcH & \numLcKs \\
Supernova Lightcurve Points & \numLcPoints & \numLcPointsJ & \numLcPointsH & \numLcPointsKs \\
Supernova Lightcurve Average \# Points/Observed Supernovae & \avgLcPoints & \avgLcPointsJ & \avgLcPointsH & \avgLcPointsKs \\
Supernova Lightcurve Median \# Points/Observed Supernovae & \medLcPoints & \medLcPointsJ & \medLcPointsH & \medLcPointsKs \\
Supernovae Lightcurve Points $\leq 20$~day past $B$-max & \numLcPointsPreTwentyDaysBMax & \numLcPointsPreTwentyDaysBMaxJ & \numLcPointsPreTwentyDaysBMaxH & \numLcPointsPreTwentyDaysBMaxKs \\
Supernovae Lightcurve Points $10$--$20$ days past $B$-max & \numLcPointsPlateau & \numLcPointsPlateauJ & \numLcPointsPlateauH & \numLcPointsPlateauKs \\
Supernovae Lightcurve Points $\leq 0$~day past $B$-max & \numLcPointsPreBMax & \numLcPointsPreBMaxJ & \numLcPointsPreBMaxH & \numLcPointsPreBMaxKs \\
\enddata

\tablenotetext{a}{The difference between ``Fields Observed'' and ``Supernovae Observed'' is the six standard Persson standard star fields that were observed.}

\end{deluxetable}

\clearpage
\begin{deluxetable}{llllrrl}
\tablecaption{Image Information\label{table:imageinfo}}
\tablehead{\colhead{Name} & \colhead{Set} & \colhead{Filter} & \colhead{Date\tablenotemark{a}} & \colhead{approxMjd} & \colhead{approxPhase} & \colhead{imageName}\\ \colhead{ } & \colhead{ } & \colhead{ } & \colhead{ } & \colhead{ } & \colhead{$\mathrm{d}$} & \colhead{ }}
\startdata
CSS121006:232854+085452 & A & H & 2012-10-28 & 56228 & 21 & CSS121006:232854+085452\_A\_H\_20121028.fits \\
CSS121006:232854+085452 & A & H & 2012-11-02 & 56233 & 26 & CSS121006:232854+085452\_A\_H\_20121102.fits \\
CSS121009:011101-172841 & A & H & 2012-10-28 & 56228 & 8 & CSS121009:011101-172841\_A\_H\_20121028.fits \\
CSS121009:011101-172841 & A & H & 2012-11-02 & 56233 & 13 & CSS121009:011101-172841\_A\_H\_20121102.fits \\
CSS121009:011101-172841 & A & J & 2012-10-28 & 56228 & 8 & CSS121009:011101-172841\_A\_J\_20121028.fits \\
CSS121009:011101-172841 & A & J & 2012-11-02 & 56233 & 13 & CSS121009:011101-172841\_A\_J\_20121102.fits \\
CSS121114:090202+101800 & A & H & 2012-11-22 & 56253 & 1 & CSS121114:090202+101800\_A\_H\_20121122.fits \\
CSS121114:090202+101800 & A & H & 2012-11-25 & 56256 & 4 & CSS121114:090202+101800\_A\_H\_20121125.fits \\
CSS121114:090202+101800 & A & H & 2012-11-30 & 56261 & 9 & CSS121114:090202+101800\_A\_H\_20121130.fits \\
CSS121114:090202+101800 & A & J & 2012-11-22 & 56253 & 1 & CSS121114:090202+101800\_A\_J\_20121122.fits
\enddata
(This table is a stub to show the structure.  See the online version for the full table.)
\tablenotetext{a}{YYYY-MM-DD of local evening at KPNO (MST).}

\end{deluxetable}

\end{document}